\documentclass{aastex}
\usepackage{spr-astr-addons}
\usepackage{url}
\usepackage[colorlinks=true,citecolor=blue,linkcolor=blue]{hyperref}
\usepackage{xcolor}
\usepackage{amsmath}

\RequirePackage{color}

\begin{document}

\title{Identifying Galactic Sources of High-Energy Neutrinos}

\slugcomment{Not to appear in Nonlearned J., 45.}
%% Running heads
\shorttitle{Galactic Neutrinos}
\shortauthors{Ali Kheirandish}

\author{Ali Kheirandish\altaffilmark{1,2}}
\affil{kheirandish@psu.edu}
%\email{\emaila{}}

\altaffiltext{1}{Dept. of Physics, The Pennsylvania State University, University Park, PA 16802, USA}
\altaffiltext{2}{Institute for Gravitation and the Cosmos, The Pennsylvania State University, University Park, PA 16802, USA}

\begin{abstract}
High-energy neutrinos present the ultimate signature for a cosmic ray accelerator. Galactic sources responsible for acceleration of cosmic rays up to the knee in cosmic ray spectrum will provide a guaranteed, albeit subdominant, contribution to the high-energy cosmic neutrino flux. In this review, we discuss the the prospects for identification of high-energy neutrinos from sources of the very high energy gamma ray emission in the Milky Way. We present the status of the search for point-like and extended emission from these sources, and describe how the results of these studies indicate that neutrino telescopes are closing in on identifying Galactic sources of high-energy neutrinos.
\end{abstract}


\section{Introduction}
Galactic sources constitute one of the main components of the high-energy emission observed at Earth. The most manifest component of the high-energy radiation from the Galaxy is the measured gamma-ray emission from sources in the Galactic plane. Today, the observed high-energy gamma-rays from Galactic sources reach 100 TeV, indicating that Galactic sources are capable of accelerating particles to very high energies. While the cosmic rays spectrum extend to energies far beyond these energies, Galactic sources has been understood as the principal contributors up to the ``knee" in the cosmic ray spectrum.

The origin of the very high energy cosmic rays has been a longstanding challenge of particle astrophysics, and despite the success of gamma-ray astrophysics in identifying Galactic and extragalactic sources of high-energy gamma rays, it falls short of revealing whether these sources are the origin of the cosmic rays because accelerated electrons will emit high-energy gamma rays. However, if gamma rays are pionic in origin, i.e. produced in hadronic interactions, then high-energy neutrinos are inevitably accompanying them and should reach Earth simultaneously. Therefore, high-energy neutrinos can unveil the origin of the cosmic rays and provide the smoking gun for the hadronic interactions at a source.

The rationale for searching for Galactic sources of high-energy cosmic neutrinos relies on the search for PeVatrons. Galactic cosmic rays are thought to reach energies of at least several PeV, the {\em knee} in cosmic ray spectrum. The interaction of cosmic rays with energies greater than PeV with the dense environments in the Galaxy leads to production of charged and neutral meson that would eventually decay to high-energy neutrinos, as well as gamma-rays, and should be seen in neutrino detectors.

In the context of multimessenger connection, identification of the source of high-energy neutrinos and very high energy cosmic rays can benefit from the information brought by other channels of observation. For Galactic source, since they are in our neighborhood, the very high energy gamma rays that should accompany the high-energy neutrinos would not be affected by the extragalactic background light (EBL) and therefore, they can be seen in similar energy ranges that neutrino telescopes operate effectively.

The observation of high-energy cosmic neutrinos in the IceCube Neutrino Observatory \citep{Aartsen2013e, Aartsen:2014gkd} opened the new era of neutrino astronomy and multimessenger astrophysics. Today, for the first time since the discovery of high-energy cosmic neutrinos, the neutrino sky seen by IceCube is showing evidences for anisotropies. The first evidence for a neutrino emission from a source was found through a multimeesnger observation, revealing a blazar as a source of a neutrino with energies $\sim$ 300 TeV \citep{IceCube:2018dnn} and uncovering a flaring activity of the source in the IceCube archival data \citep{IceCube:2018cha}. In addition, the results of the 10 year time-integrated search for neutrino sources \citep{Aartsen:2019fau} has revealed more sources with positive fluctuation and indicate that neutrino telescopes are getting close to identify sources of high-energy neutrinos. 
Possible neutrino emission from the Galactic plane appears as a subdominant component to the observed cosmic neutrino flux. However, recent studies imply that identification of the Galactic source of high-energy neutrinos is likely in near future.

A cubic kilometer detector like IceCube, and in near future KM3NeT \citep{Adrian-Martinez:2016fdl}, is sensitive to generic cosmic ray sources with an energy density in high-energy neutrinos comparable to their energy density in cosmic rays and their associated pionic gamma rays. Galactic cosmic rays reaching energies of few PeV meet this condition.

The goal of this review is to discuss the current status of identifying Galactic sources of high-energy neutrinos which is tied to one of the central questions in particle astrophysics: where is the birthplace of cosmic rays? We first present an overview of the high-energy neutrino observations and discuss recent developments in flux measurements and search for the origin of high-energy neutrinos. After a brief discussion of Galactic cosmic rays in Section \ref{sec:gcr}, we will talk about the identification of potential sources of high-energy neutrinos in the Milky Way based on gamma-ray observations and discuss the prospect for their observations in Sections \ref{sec:sources} and \ref{sec:diff}. We will highlight analyses aimed at identifying the correlation of high-energy neutrinos with these objects and argue how their findings indicate that identifying Galactic component of cosmic neutrino flux is likely in near future.

%%%%%%%%%%%%%%%%%%%%%%%%%%%%%%%%%%%%
\section{High-Energy Cosmic Neutrinos}\label{sec:henu}
Soon after their discovery, neutrinos emerged as ideal astronomical messengers thanks to their feeble interaction and immunity to the interstellar magnetic fields. Their ability to traverse the universe unscathed, granted them as particles that can point back to the origin of the cosmic rays. However, the weak interacting nature of neutrinos made their observation difficult. Early prediction of the neutrino flux from astrophysical sources necessitated instrumenting immense detectors for their observation. Theoretical predictions for a high-energy neutrino emission concentrated on sources that were thought capable of accelerating cosmic rays to very high energies, such as active galactic nuclei (AGN), gamma ray bursts (GRBs), and supernova remnants (SNR) as well as the interaction of ultra high energy cosmic rays (UHECR) with the cosmic background radiation. These predictions relied on the anticipation that cosmic accelerators may produce cosmic rays, gamma rays, and neutrinos with similar energies and indicated that cubic kilometer detectors were required for observation of high-energy neutrino flux from these sources.

Secondary charged particles produced by neutrino interaction with a nucleus emit Cherenkov light when they travel faster that the speed of light in a medium. Therefore, development of large scale neutrino detectors focused on using highly transparent natural water or ice that were instrumented with photomultipliers to detect the Cherenkov lights. DUMAND\footnote{Deep Underwater Muon And Neutrino Detector Project}, NT200, AMANDA\footnote{Antarctic Muon And Neutrino Detector Array}, and ANTARES experiments pursued this idea and presented the proof of concept for building the first cubic kilometer detector: IceCube\footnote{For a comprehensive history of neutrino astronomy and development of neutrino detectors see \citet{Spiering:2012xe}.}. 

The IceCube Neutrino Observatory transformed one cubic kilometer of Antarctic ice into a Cherenkov detector and serves as world's principal neutrino telescope. The IceCube detector consists of 51600 digital optical modules (DOMs) deployed below the depth of 14500 meters, that contain photomultiplier tubes (PMTs) as well as the digitizing electronics that capture PMT light signal \citep{Abbasi:2010vc, Abbasi:2008aa, Aartsen:2016nxy}. The information recorded by DOMs are the basis for reconstruction of the energy, direction, and localization of particle interactions in the ice.

IceCube was completed in 2010 and one year after completion, it succeeded in observing an extraterrestrial flux of high-energy neutrinos \citep{Aartsen2013e}. Since then, IceCube has continuously been measuring the high-energy cosmic neutrino flux and the excess of neutrinos  (over atmospheric neutrino background) has been confirmed in multiple channels of observation \citep{Aartsen:2014gkd, Aartsen2015b, Kopper:2015vzf, Aartsen:2016xlq, Haack:2017dxi, Schneider:2019ayi, Stettner:2019tok, Aartsen:2020aqd}. 

IceCube detects neutrinos with energies greater than 10 GeV. Atmospheric neutrinos and muons produced by cosmic rays interaction in the atmosphere are  the main backgrounds for the search of high-energy cosmic neutrinos. On average, IceCube observes $10^{6}$ atmospheric neutrinos and $10^{11}$ atmospheric muons each year \citep{Halzen:2010yj}. The background neutrinos spectra falls sharply with increasing energies, and at energies beyond 100 TeV the astrophysical flux becomes the dominant component.

In order to identify the cosmic neutrino flux, two principle methods were incorporated to distinguish signal from background. The first method relies upon the observation of muon neutrinos that interact primarily outside the instrumented volume. Charged current interaction of high-energy muon neutrinos produces a kilometer-long muon track that passes through the detector. In this method, Earth is used as a filter to reduce the background of cosmic ray muons. As a result, at high energies ($\gtrsim$100 TeV) the background is severely suppressed. Using this method, IceCube has identified an excess of high-energy neutrino flux at energies beyond 100 TeV that can not be explained by the atmospheric flux. Analyzing 10 years of data, IceCube has observed a flux of high-energy muon neutrinos that is well described by simple power-law with a spectral index of 2.28 \citep{Stettner:2019tok}.

The alternative method for identification of astrophysical neutrino flux uses a {\em veto} technique to remove the background, thus excluding events that did not interact in an inner fiducial volume of the detector. This event selection benefits from detection of neutrinos from every direction in the sky and all neutrino flavors can be identified. This method succeeded in revealing the first evidence for cosmic neutrinos \citep{Aartsen2013e}, and today, 7.5 years of high-energy starting events (HESE) revealed an excess of astrophysical flux of high-energy neutrinos with statistical significance of $\sim 8 \sigma$, rejecting atmospheric origin \citep{Schneider:2019ayi}. 
The measured flux in HESE is described by a simple power-law with an index of 2.88.

These two method of separating cosmic neutrinos from atmospheric backgrounds have complementary advantages. The HESE selection observes an all-sky sample of events that do not trace any accompanying muons from an atmospheric shower, and therefore, are highly unlikely to be of atmospheric origin. In addition to starting muon tracks, the sample includes showers produced by charged-current interactions of electron and tau neutrinos, and neutral current interactions of neutrinos of all flavors. Showers have typically an angular uncertainty of $10-15$ degrees and benefit from an energy resolution better than 15\%, since the deposited energy is contained inside the detector \citep{Aartsen:2013vja}. On the other hand, the through-going muons provide an angular resolution of $\leq 0.4$ degree at high energies. 

 \begin{figure*}[ht]
 \includegraphics[width=\textwidth]{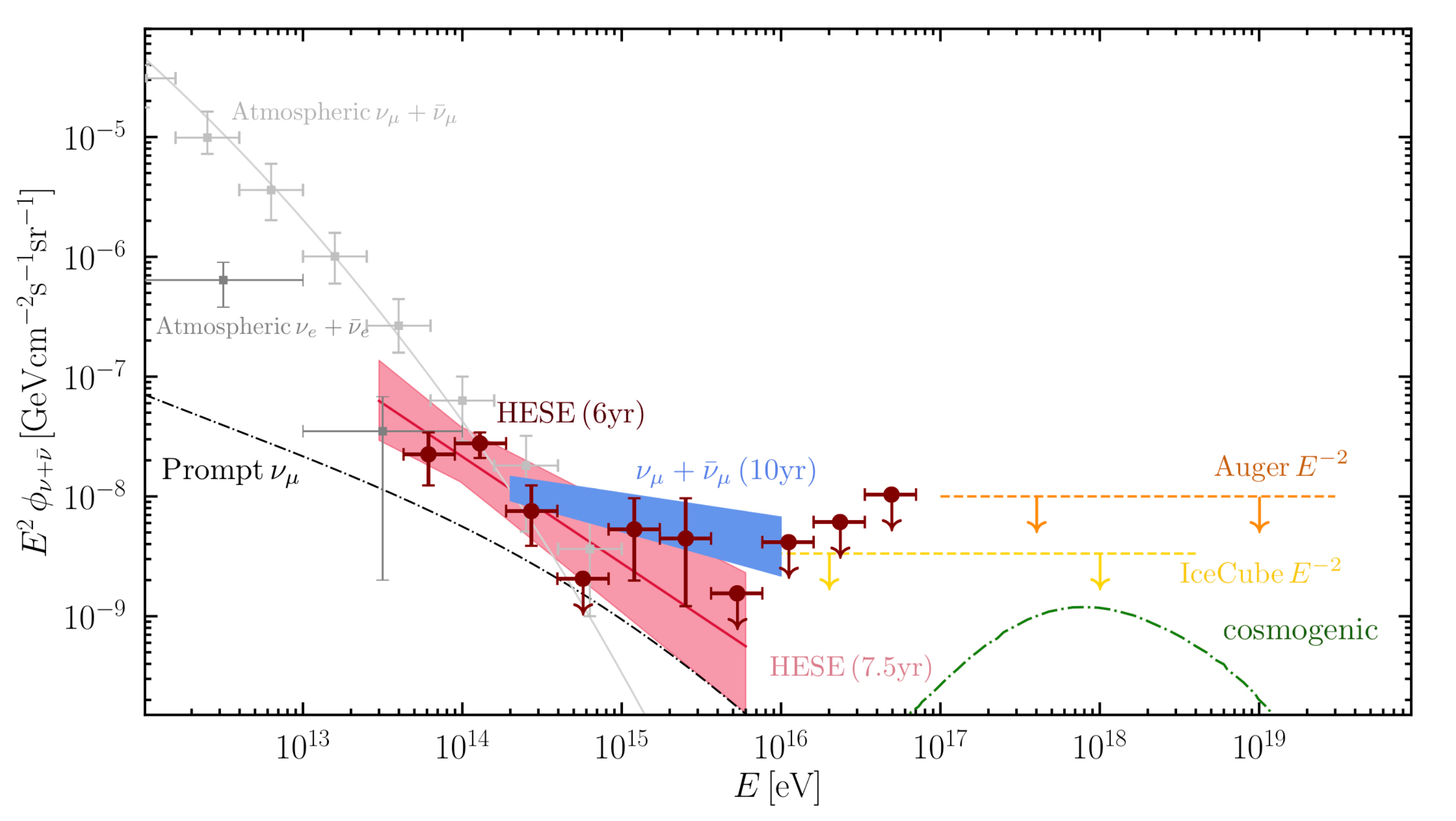}
 \caption{The flux of high-energy neutrinos (per flavor) above 1 TeV.  At low energies, IceCube's measured atmospheric muon and electron neutrinos \citep{Aartsen:2014qna,Aartsen:2015xup} are shown in light and dark grey, respectively. The predicted prompt flux \citep{Enberg:2008te} expected from atmospheric charm production in cosmic ray showers is shown in black. The red line shows the fitted spectra for HESE 7.5 analysis and the shaded region shows the uncertainties \citep{Schneider:2019ayi}. The dark red points show differential data points in HESE 6 year analysis. The results for 10 year muon neutrino flux measurement is shown in blue shaded area \citep{Stettner:2019tok}. At higher energies, the predicted flux of cosmogenic neutrinos \citep{Ahlers:2012rz} shown in green, as well as the upper limit on the extremely high energy flux of neutrinos from IceCube \citep{Aartsen:2016ngq} and Pierre Auger Observatory \citep{Aab:2015kma}.} 
 \label{fig:nu-spec}
 \end{figure*}

Figure \ref{fig:nu-spec} showes a summary of high-energy cosmic neutrino flux measurements in each method. At lower energies, the atmospheric muon and electron neutrino fluxes are shown, which fall rapidly with and index of $\sim 3.7$. At $\gtrsim 10$ TeV, the fitted spectrum for contained events in 7.5 year HESE measurement is shown, which becomes the dominant component. Note that the atmospheric muon and electron neutrino fluxes here are before applying the {\em veto}. Taking advantage of this technique, the atmospheric component is severely suppressed in the HESE analysis. The high-energy muon sample from the Northern hemisphere measures the astrophysical flux at relatively higher energies, as it relies on the growing neutrino-nucleus cross section to shield the atmospheric flux via Earth. The cosmic neutrino flux in this analysis is effectively measured above 100 TeV. As could be seen in the figure, the measurements are compatible within common energy ranges (for discussion on the compatibility of the measurements, see \citet{Aartsen:2020aqd}). In the meantime, different spectral indices found in these analyses hint at possible spectral features in the global spectrum of the cosmic neutrinos. While the proof of such features in high-energy neutrino spectrum is contingent on more statistical and systematic improvements in measurement of the high-energy neutrino flux, the high level of neutrino flux at 10 TeV has a major physical impact: it could be symptomatic of sources and their properties.

 Since both charged and neutral pions are produced in astrophysical beam dumps, the high-energy neutrino production is inevitably accompanied by high-energy gamma rays. If the source is nearby, gamma rays will arrive at Earth. However, gamma rays from distant source will suffer absorption and cascading in EBL, which will degraded their energy and eventually contribute to the isotropic gamma-ray background (IGRB) measured by the {\em Fermi} satellite \citep{Ackermann:2014usa}. Consequently, neutrino-gamma connection imposes strong constraint on the source emission and the origin of high-energy cosmic neutrinos. Most notably, if hadronuclear ($pp$) interactions were the dominant channel of neutrino production at PeV energies, the neutrino flux cannot have a spectrum softer than $E^{-2.2}$ as the accompanying gamma ray flux will overshoot the IGRB measurements \citep{Murase2013b}. Furthermore, sources responsible for the low-energy excess in IceCube data, at 10 TeV, should be opaque to gamma-ray emission in order to avoid violating the IGRB limits \citep{Murase:2015xka, Capanema:2020rjj}.
 
Another potential origin of lower energy excess in IceCube data is considered to be the Galactic component of the neutrino flux. The starting event selection is dominated by events from the Southern hemisphere where there is a greater exposure to the Galactic plane compared to the Northern Sky.
Galactic sources are believed to contribute mostly to the flux below 100 TeV energies, and this speculation is supported by the other measurement lowering the threshold in starting event analysis \citep{Aartsen:2014muf} as well as the search with cascade data \citep{Aartsen:2020aqd}.

An extensive effort has been placed in order to identify the origin of high-energy cosmic neutrinos. These effort include time-integrated searches for steady emission \citep{Aartsen:2014cva, Aartsen:2016oji, Aartsen:2018ywr, Aartsen:2019fau} as well as time-dependent search for transients \citep{Aartsen:2014aqy, Aartsen:2015wto,  Aartsen:2019wbt}. In addition, extended sources and cross correlation with potential astrophysical sources has been done in the form of stacking searches. 

The consistency of the arrival direction of astrophysical neutrinos with isotropic distribution implies that the flux of high-energy neutrinos are predominantly extragalactic in origin. After almost ten years of operation, the first evidence for neutrino emission from a source was identified through a multimessenger campaign that was initiated by a 290 TeV neutrino in September 2017 \citep{2017GCN.21916....1K}. A flaring gamma-ray blazar, TXS 0506+056, was identified as the source of the high-energy neutrino \citep{IceCube:2018dnn} and remarkably, an archival search in IceCube data revealed a neutrino flare in 2014-15 that dominated the neutrino flux from the direction of TXS 0506+056 in a ten year period \citep{IceCube:2018cha}. This observation presented the first evidence for a neutrino source. Later, signatures of anisotropy has emerged in 10 year analysis of IceCube muon neutrino data \citep{Aartsen:2019wbt}. Most notably, a nearby seyfert galaxy, NGC 1068, has been identified as the most significant source in the pre-identified list in the search and surprisingly, the all-sky untriggered search finds a hot spot compatible with this source. This source, and  three other sources were found with a pre-trial significance of $4\sigma$. Overall, the anisotropy is observed at a 3$\sigma$ level in the search. 

In summary, the origin of the majority of the IceCube cosmic neutrino flux is still a mystery. The recent developments has offered compelling evidences for the origin of high-energy neutrinos and they are driven by the extragalactic component of the flux. This is not a surprise as the cosmic neutrino flux is predominantly extragalactic. However, the Galactic contribution to the astrophysical neutrino flux cannot be excluded and several hints in the neutrino data may point toward possible identification of Galactic sources of cosmic neutrinos in near future. In fact, Galactic sources were thought as guaranteed contributors to the high-energy neutrino flux. In the next section, we will discuss how Galactic cosmic rays support this argument and later in Section \ref{sec:sources}, we will discuss these hints in addition to the strategies for targeting and identifying sources of high-energy neutrinos in the Milky Way.

%%%%%%%%%%%%%%%%%%%%%%%%%%%%%%%%%%%%
\section{Galactic cosmic ray accelerators}\label{sec:gcr}

 \begin{figure}[t]
 \includegraphics[width=\columnwidth]{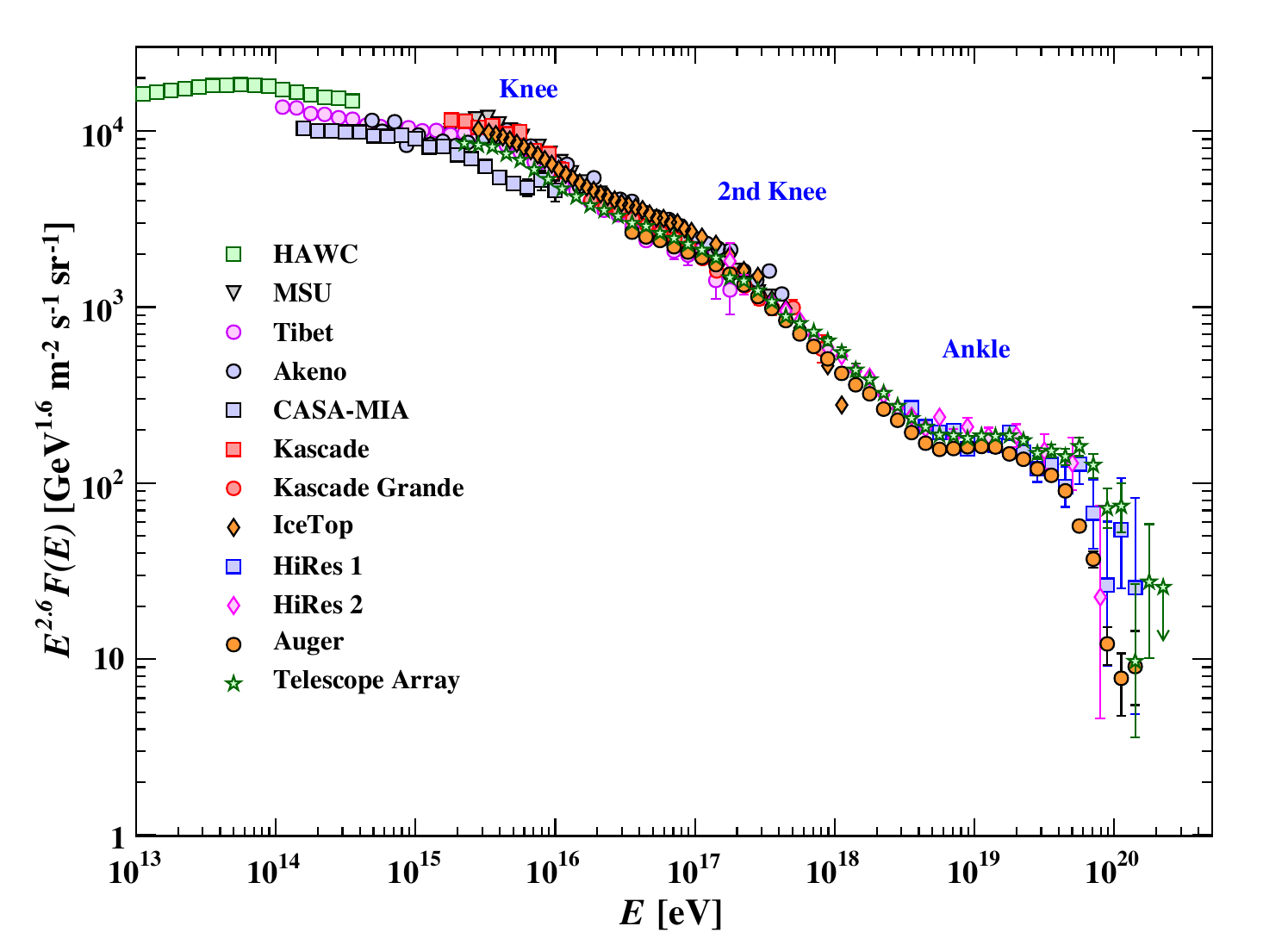}
 \caption{Energy-weighted cosmic ray flux at energies above 10 TeV and the three prominent features of the cosmic ray spectrum: the {\em knee}, the {\em second knee}, and the {\em ankle}. The {\em knee} refers to the steepening of the spectrum at energies above $\sim$ 3 PeV and Galactic cosmic rays are believed to be the dominant contributors up to this energy. The {\em second knee} has been identified around $\sim 300$ PeV and marks steepening of the flux. The {\em ankle} is referred to the flattening feature of the spectrum, at energies beyond EeV. Figure from Particle Data Group \citep{Tanabashi:2018oca}.} 
 \label{fig:cr-spec}
 \end{figure}

Cosmic ray spectrum spans a wide range of energies, over 14 orders of magnitude. Despite the increasing knowledge on the cosmic ray spectrum since their discovery more than a hundred year ago, the principle puzzle in cosmic ray physics remains unanswered: Where do cosmic rays originate from? 

One prominent feature in cosmic ray spectrum is the steepening of the spectrum above $\sim$ 3 PeV, see Figure  \ref{fig:cr-spec}. This break energy is known as the {\em cosmic ray knee}. The location of the {\em knee}, and other features in the cosmic ray spectrum, offer important clues about the sources of cosmic rays.

{The necessary condition for a source to be considered as a particle accelerator up to a certain energy is to satisfy the Hillas criterion \citep{1984ARA&A..22..425H}. That is, the gyro-radius of the particle shall not excess the size of the astrophysical object. Otherwise, it cannot contain the particle. The other relevant parameters determining the maximum energy an accelerator can achieve are the acceleration mechanism and energy losses at the source. Galactic sources such as supernova remnants can meet Hillas' condition and reach energies beyond the cosmic ray {\em knee} if the magnetic filed is amplified compared to the average interstellar medium's average value. Possible Galactic wind termination shocks can also accelerate cosmic rays beyond the {\em knee} in Milky Way, see \citep{Bustard:2016swa} for details. 

 Supernova remnant was among the very early suggestions as the source of cosmic rays up to energies of several PeVs \citep{BaadeAndZwicky}. 
Galactic cosmic rays posses an energy density of $\sim 10^{-12}\, \rm erg s^{-1}$ and their average containment time in the Galaxy is three million years. In order to maintain a steady energy density for the Galactic cosmic rays, a source should typically accelerate particles with a power of $10^{41} \, \rm erg s^{-1}$. This can be easily met when 10\% of the power generated by supernova remnants ($10^{51} \, \rm erg s^{-1}$) is released to the Galaxy every 30 years. This coincidence is what motivated supernova remnants as the major source responsible cosmic ray acceleration and production of Galactic cosmic rays. For a recent review on Galactic cosmic rays see \citet{BeckerTjus:2020xzg}.

Cosmic rays interaction with matter or radiation at the source, or during their propagation in Galactic environment, would result in secondary particles that can be used to pinpoint the origin of the cosmic rays. Gamma rays and neutrinos are promptly produced from decay of pions. For Galactic sources, the hadronuclear ($pp$) interaction is dominant and the photohadronic component is negligible. The threshold for production of pions in hadronuclear interaction leads to a feature in gamma-ray flux known as {\em pion bump}, appearing at $\sim$ 70 MeV. Since accelerated electron-positron pair can emit high-energy gamma rays, such feature is crucial to find a direct evidence for hadronic interactions in gamma-ray observations. Interestingly, the Fermi Large Area Telescope ({\em Fermi}-LAT) has identified this feature in gamma-ray spectra of supernova remnants IC 443, W 44, and W 51 \citep{Ackermann:2013wqa, Jogler:2015ddc}.

Although identification of the {\em pion bump} is an intriguing evidence for the presence of hadrons in supernova remnants, it does not suffice to identify the origin of very high energy cosmic rays in the Milky Way. The energy at which the feature is present is quite below the energy associated to very high energy cosmic rays. Hence, even if very high energy gamma rays were identified from a source the leptonic versus hadronic origin of them cannot be disentangled. 
Neutrinos on the other hand  can only originate from decay of pion, and therefore, provide the smoking gun for hadronic interactions. Neutrinos are another secondary particle that stem from cosmic ray interactions and they would accompany any pionic high-energy gamma ray emission. Therefore, neutrinos present the only unambiguous signature for cosmic ray sources in the Galaxy, and indeed anywhere in the universe. 

%%%%%%%%%%%%%%%%%%%%%%%%%%%%%%%%%%%%
\section{Identifying Galactic Sources of High-Energy Neutrinos}\label{sec:sources}
A cubic kilometer neutrino detector like IceCube has sufficient sensitivity to probe for a cosmic ray source that releases a similar energy density in neutrinos, very high energy gamma rays, and cosmic rays \citep{Gaisser1997, AlvarezMuniz:2002yr}. Galactic sources might be satisfying this condition. 

The search for Galactic neutrino sources is centered on the search for PeVatrons, sources with the required energetics to produce cosmic rays, at least up to the {\em knee} in cosmic ray spectrum. Supernova remnants has been the primary candidates. Accelerating cosmic rays beyond PeV energies, generic PeVatrons should emit very high energy gamma rays whose spectrum extend to several hundred TeV without a cut off. 

Here, we focus on hadronuclear interaction ($pp$). The photohadronic ($p\gamma$) interactions are generally considered to be negligible for Galactic sources. Hadronuclear interactions will result in production of charged and neutral pions:
\begin{eqnarray}
pp \longrightarrow N_\pi [\pi^+  + \pi^-  + \pi^0] + X,
\end{eqnarray}

where $N_\pi$ is the pion multiplicity. High-energy flux of neutrinos is produced by the immediate decay of charged pions 
\begin{eqnarray}
\pi^+  \longrightarrow \mu^+ + \bar\nu_\mu% ~ \& ~ \pi^-  \longrightarrow \mu^- + \nu_\mu
\end{eqnarray}
followed by $\mu^+ \longrightarrow e^+ + \nu_e +\bar\nu_\mu$. Simultaneously, high-energy gamma rays are produced via neutral pions decay:
\begin{eqnarray}
\pi^0  \longrightarrow \gamma + \gamma.
\end{eqnarray}

High-energy neutrinos and gamma rays carry on average 1/4 and 1/2 of the parent pions energies, respectively. The pion's energy is related to proton's energy via the average inelasticity {\em per pion}, $\kappa_\pi$, such that $E_\pi=\kappa_\pi E_p$. For $pp$ interactions, $\kappa_\pi \simeq 0.2$ \citep{Kelner:2006tc}. Neutrinos would carry between  3-5\% of the parent cosmic ray energy. The efficiency of a beam dump to produce pions is expressed as
\begin{eqnarray}
f_\pi = 1-\exp(-\kappa \sigma_{pp} n l),
\end{eqnarray}

where $\kappa$ is the inelastisity for $pp$ interaction, $\sigma$ is the interaction cross section, $n$ is the target density, and $l$ is the target size. Using the source production rate function $Q_p$ in units of $\rm GeV^{-1}\,s^{-1}$,  one can find the total neutrino emission rate as 
\begin{eqnarray}
E_\nu^2 Q_\nu = \frac{3}{4} f_\pi \frac{K_\pi}{K_\pi+1} E_p^2 Q_p \biggr\rvert_{E_p = 4E_\nu/\kappa_\pi},
\end{eqnarray}

where the factor 3 takes into account the number of neutrinos produced per pion and $K_\pi$ is the fraction of charged to neutral pions and equals to 2 for $pp$ interactions. 

The gamma-ray and neutrino flux at the source are connected considering the number of charged and neutral pions:
\begin{equation}\label{eq:GAMMAtoNU}
\frac{1}{3}\sum_{\alpha}E^2_\nu Q_{\nu_\alpha}(E_\nu) \simeq \frac{K_\pi}{4} E^2_\gamma Q_\gamma(E_\gamma)\biggr\rvert_{E_\gamma = 2E_\nu}\,.
\end{equation}

Interaction with the background light or dense environments alters the gamma-ray flux before arriving at Earth. In order to apply the above relation to the gamma-ray and neutrino fluxes detected at Earth, one should correct the gamma-ray flux by multiplying the right hand side of equation \ref{eq:GAMMAtoNU} by $\exp\big(\tau(z, E)\big)$. For extragalactic sources, this correction is essential as EBL absorption would attenuate the gamma-ray flux at very high energies. For Galactic source, however, the effect is negligible.

Neutrino oscillations will modify the original flavor ratio of the high-energy neutrinos. Therefore, when considering the flux at Earth for each flavor, the flux of neutrinos is equally distributed between the three flavors of neutrinos.

In the context of multimessenger connection, identification of potential Galactic cosmic ray, and cosmic neutrino, sources rely upon identification of very high energy gamma ray emitters. The associated gamma rays produced with high-energy neutrino could lay out the whereabouts of potential sources in the Galaxy.

The first very high energy survey of the Galaxy was performed by the Milagro collaboration \citep{Abdo:2006fq}. This survey revealed the brightest gamma-ray emitters in the Galactic plane and pointed towards possible sites of cosmic ray acceleration in the Galaxy. Milagro was a water Cherenkov telescope with a wide field of view. The success of Milagro in identifying bright gamma-ray spots in the Milky Way, which are the most luminous source in the plane after Crab, was followed by observations by imaging air Cherenkov telescopes (IACTs) such as HESS\footnote{High Energy Stereoscopic System}, VERITAS\footnote{Very Energetic Radiation Imaging Telescope Array System}, and MAGIC\footnote{Major Atmospheric Gamma-Ray Imaging Cherenkov} which provided a throve of data on the gamma-ray emission from the Milky Way. Today, with operation of the HAWC\footnote{High Altitude Water Cherenkov} gamma-ray observatory, as the successor of Milagro, we have an unprecedented sensitivity to very high energy gamma ray emitters in the Milky Way. 

The initial Milagro survey of the Galaxy presented the first view of the Galaxy at 10 TeV and identified a handful of TeV emitters. Today, there are more than 100 sources identified with TeV emission in the Milky Way. Pulsar wind nebula (PWN), supernova remnant (SNR), binaries, molecular clouds, and Shell SNR are different type of sources with identified classes. However, the majority of detected sources are yet classified as {\em unidentified}. The TeVCat\footnote{tevcat.uchicago.edu} data base provides a present-day summary of the detected sources with TeV emission \citep{Wakely:2007qpa}.

Six sources with extended emission were identified in the initial survey of the Milagro collaboration \citep{Abdo:2006fq} as potential cosmic ray accelerators in the Galactic plane. These prominent sources were found to have a hard spectrum that extended to very high energies without showing signatures of energy cut-off. This behavior made them ideal candidates for long sought PeVatrons in the Galaxy. Three of these sources -- MGRO J1908+06, MGRO J2019+37, and MGRO J2031+41 -- were most significant sources after Crab. The two other, MGRO J2043+36 and MGRO J2032+37, were candidate sources with relatively high level of gamma-ray emission. Finally, MGRO J1852+01 was found below the threshold despite its large flux.

Four of these sources were located in the Cygnus region, a star forming zone in the Milky Way with high level of activity and young stars which could provide substantial requirements for particle acceleration and interaction. The other two sources were close to inner galaxy and depicted a large flux in TeV energies. They were located at low declination, which is the area in the sky IceCube is most sensitive to. Therefore, provided that the emission was hadronic, they had the highest likelihood of observation in IceCube.

The early predictions based on the measured fluxes and extensions reported by the Milagro collaboration determined that IceCube should be able to see neutrinos from these sources after five years of operation \citep{Beacom:2007yu, Halzen:2008zj, GonzalezGarcia:2009jc}. Follow-up observations by Milagro reported a low-energy cut-off in the spectrum of these sources \citep{Abdo:2012jg} which reduced their chance of observation in IceCube \citep{Gonzalez-Garcia:2013iha}.

The prospects for observation of these sources in IceCube is entangled with the uncertainties in the spectrum and the extension of the sources. Sources with hard spectrum, close to $E^{-2}$, are more likely to be identified in the neutrino sources search as the astrophysical signal in neutrino data can be distinguished from the soft atmospheric background. In addition, neutrino source search become less sensitive with larger extensions as the number of background in the direction of source would enhance by the larger extension of the source. It should be noted that understanding the true extension is necessary to obtain the optimal sensitivity in the search for extended sources \citep{Aartsen:2014cva}.

The discrepancy in measurement of spectrum and extension of the Milagro sources emerged as IACT experiments performed follow up observation of these sources. The new information brought by HESS, VERITAS, MAGIC, and ARGO-YBJ observatories showed tensions and discrepancies in the spectrum and extension of these sources. While the likelihood of identifying these objects as neutrino sources highly depends on the true spectra and extensions, the clear picture is hard to assess as each experiment faces systematic limitations. The wide-field water Cherenkov telescopes like Milagro are suited to measure gamma-ray flux from extended sources while IACTs are pointing observatories and are limited in exploiting extended regions. Their small field of view prevents them from constructing background from off-source regions for extended sources. On the other hand, while IACTs benefit from good energy resolution, and hence more accurate spectral measurements, their sensitivity declines for energies beyond 10 TeV. Here, we focus on the prospects for observation of MGRO J1908+06 as one of the preeminent candidate sources, for detailed discussion on the rest of sources see \citet{Halzen:2016seh}.

{The high-energy gamma ray emission from MGRO J1908+06 was reported by air-shower detectors (EAS) like the Milagro experiment.~\citep{Abdo:2007ad,Abdo:2009ku,Smith:2010yn}, and the ARGO-YBJ \citep{ARGO-YBJ:2012goa}. The high-energy gamma ray flux has been also detected also by IACTs. Figure \ref{fig:1908} compares the flux measurements by different observatories.  The HESS collaboration measures a flux systematically lower than the Milagro and ARGO-YBJ data \citep{Aharonian:2009je}, which finds a  hard spectrum with no evidence of a cut-off for energies $<20$~TeV. With better angular resolution, it could be that HESS detects the flux from a point source that cannot be resolved by the Milagro and ARGO-YBJ observation. MGRO J1908+06 has also been observed by VERITAS \citep{Aliu:2014xra}, and the measured flux is of the same order as the one reported by HESS. Finally, the value recently reported by HAWC points towards a similar normalization \citep{Abeysekara:2015qba}. 

 \begin{figure}[t]
 \includegraphics[width=\columnwidth]{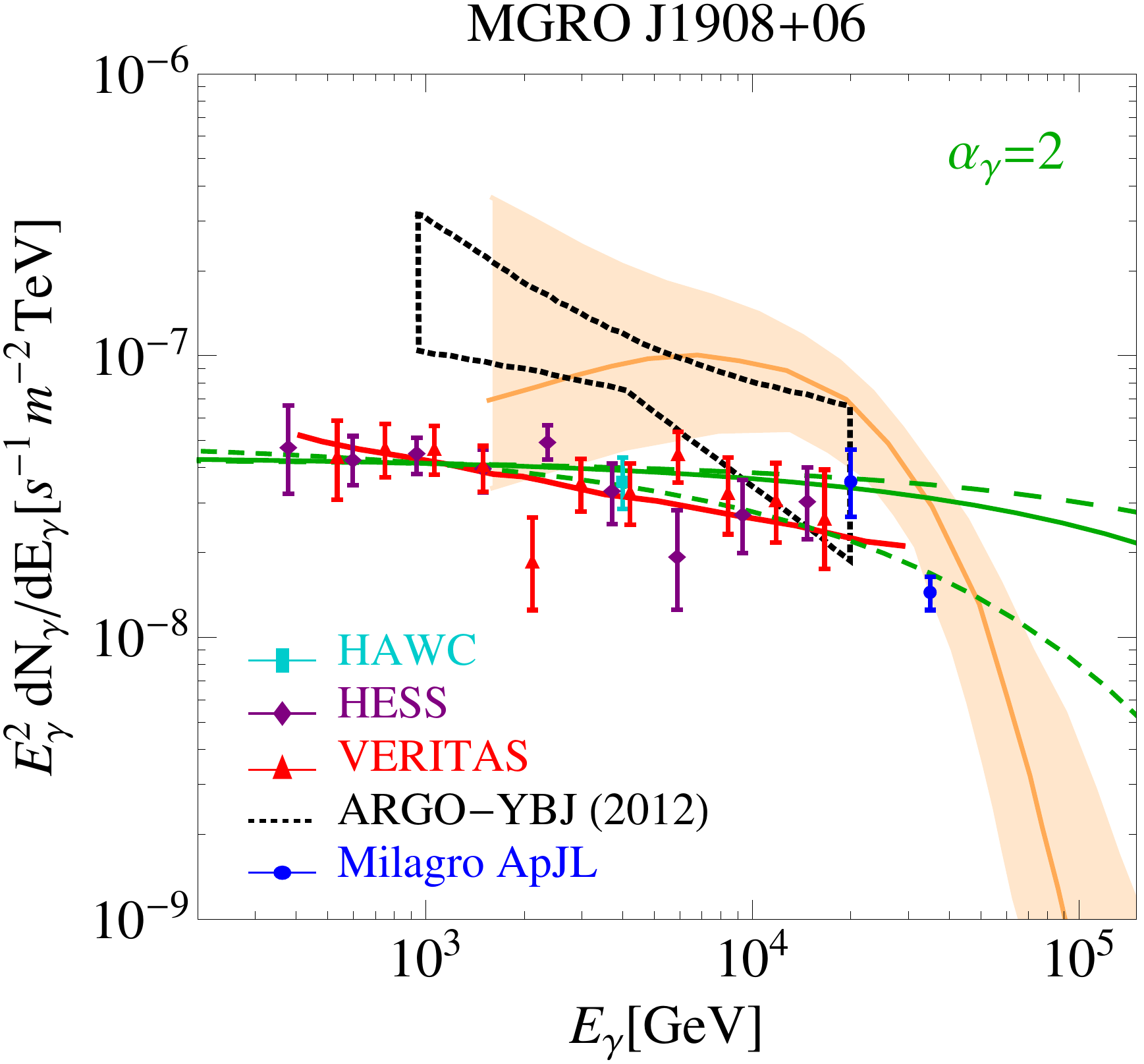}
 \caption{Measured gamma-ray flux of MGRO J1908+06 from different experiments. We show in purple data points measured flux by HESS~\citep{Aharonian:2009je}, in red the flux from VERITAS~\citep{Aliu:2014rha}, and in cyan the flux from HAWC~\citep{Abeysekara:2015qba}. The blue data points sow the previous measurements by Milagro~\citep{Abdo:2007ad,Abdo:2009ku}, and the orange line and the shaded orange region show the best fit and the $1\sigma$ band as reported by the Milagro collaboration \citep{Smith:2010yn} . The dotted area identifies the ARGO-YBJ $1\sigma$ band~\citep{ARGO-YBJ:2012goa}. The green lines show the spectra obtained considering $\alpha_\gamma=2$ and fixing the normalization to the best fit reported by HESS, the allowed the cut-off energy vary: $E_{\rm cut, \gamma} =30,~300, {\rm and}~800$~TeV (short-dashed, solid, and long-dashed lines, in green). Figure from \citep{Halzen:2016seh}.} 
 \label{fig:1908}
 \end{figure}
 
 \begin{figure}[ht]
 \includegraphics[width=\columnwidth]{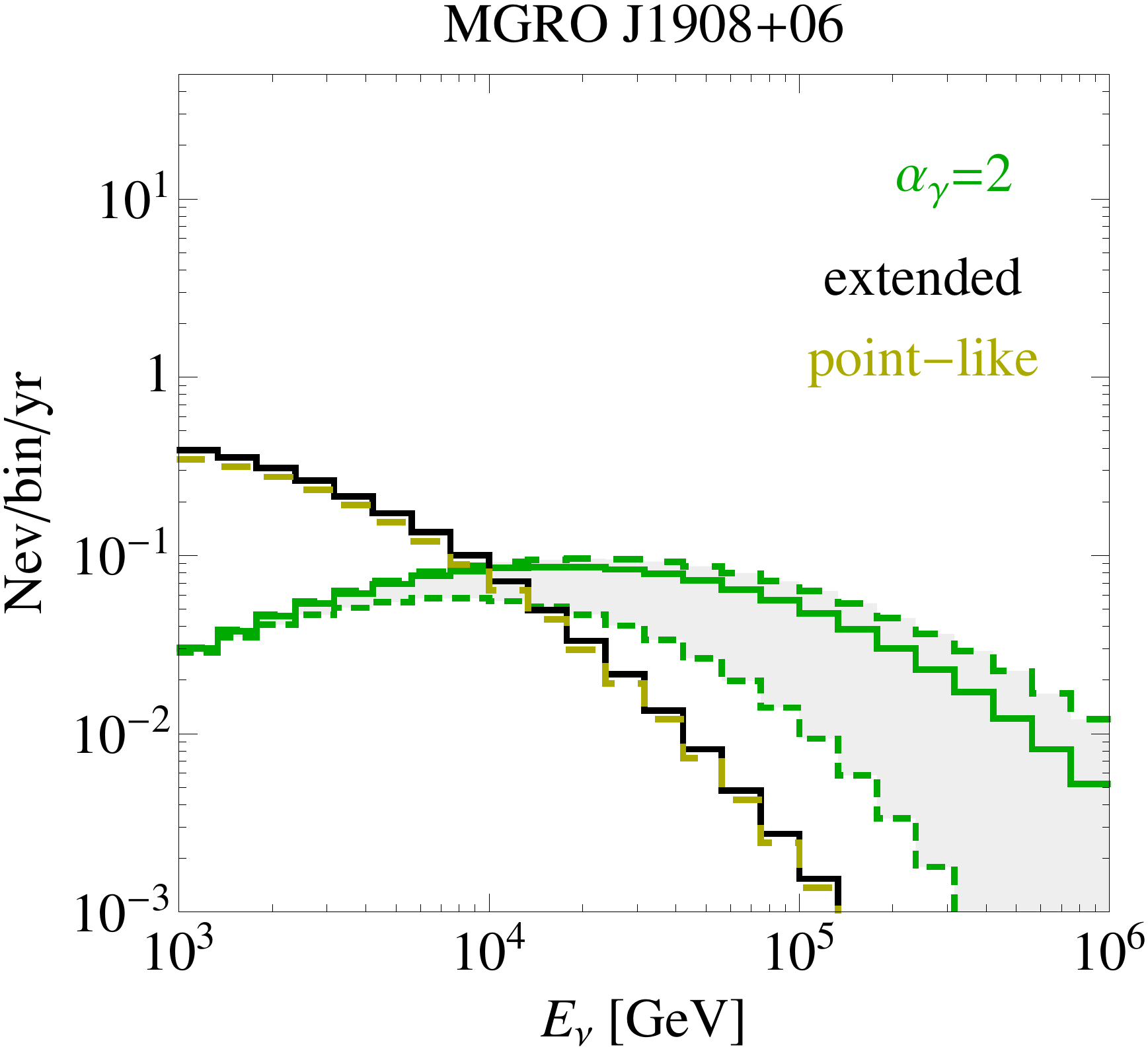}
 \caption{ Event distribution from MGRO 1908+06 for the spectra parametrized by HESS measurement.The grey band encodes the uncertainty on the cut-off energy. The black (gold dashed) line shows the background from atmospheric neutrinos for extended (point-like) sources. Figure from \citep{Halzen:2016seh}.} 
 \label{fig:1908events}
 \end{figure}

In addition to different fluxes, the reported extension differs depending on the experiment. The ARGO-YBJ finds an extension of $0.5^\circ$ \citep{Bartoli:2012tj} while HESS finds an extension of $0.34^\circ$ \citep{Aharonian:2009je}. The measured extension in VERITAS observation is $0.44^\circ$ \citep{Aliu:2014xra}.

MGRO J1908+06 is currently classified as an unidentified source. {\em Fermi}-LAT observes the pulsar PSR J1907+0602 within the extension reported by the Milagro collaboration \citep{Abdo:2010ht}. On the other hand, the large size, the measured hard spectrum in TeV photon that persist with distance from the pulsar are not characteristic of a PWN scenario \citep{Aliu:2014xra} and MGRO J1908+06 is perhaps consistent with a SNR. 

Multimessenger relation between gamma-ray and neutrino production rates can be used to evaluate the expected high-energy neutrino flux from MGRO J1908+06. Here, we adopt the following parameterization of the gamma-ray flux

\begin{equation}
\frac{dN_{\gamma}(E_\gamma)}{dE_\gamma}
=k_{\gamma}
 \left(\frac{E_\gamma}{\rm TeV}\right)^{-\alpha_{\gamma}} 
\exp\left(-\sqrt{\frac{E_\gamma}{E_{cut,\gamma}}}\right).
\label{eq:cutoff}
\end{equation}

This parameterization corresponds to the production rate from a source with cosmic ray injection rate that is described by a power-law with an exponential cut-off \citep{Kappes:2006fg}.} The green lines in Figure \ref{fig:1908} show the spectra obtained within this parameterization considering $\alpha_\gamma=2$ and fixing the normalization to the best fit reported by HESS, and allowing the cut-off energy to vary: $E_{\rm cut, \gamma} =30,~300, {\rm and}~800$~TeV.  

In Figure \ref{fig:1908events} we show the expected number of events when the neutrino flux evaluated from spectrum reported by HESS, considering energy cut-off for the spectrum ranging from 30 to 800 TeV. The expected number of background events for both extended and point source assumption are shown for comparison. Here, we are assuming  $0.34^\circ$ extension reported by HESS. The corresponding p-value for observation of neutrino emission in the direction of MGRO J1908+06 is shown in Figure \ref{fig:1908pvals}. Different assumptions on the high-energy cut-off for the spectrum and the extension of the source result in different likelihood for identification of MGRO J1908+06. The most optimistic scenario occurs when the source is point-like and the cut-off is beyond several hundreds of TeV. Overall, provided that the gamma-ray flux from MGRO J1908+06 extends to energies beyond 100 TeV, using the flux reported by HESS we anticipate a $3\sigma$ observation in about 10 to 15 years of IceCube data.

\begin{figure}[t]
 \includegraphics[width=\columnwidth]{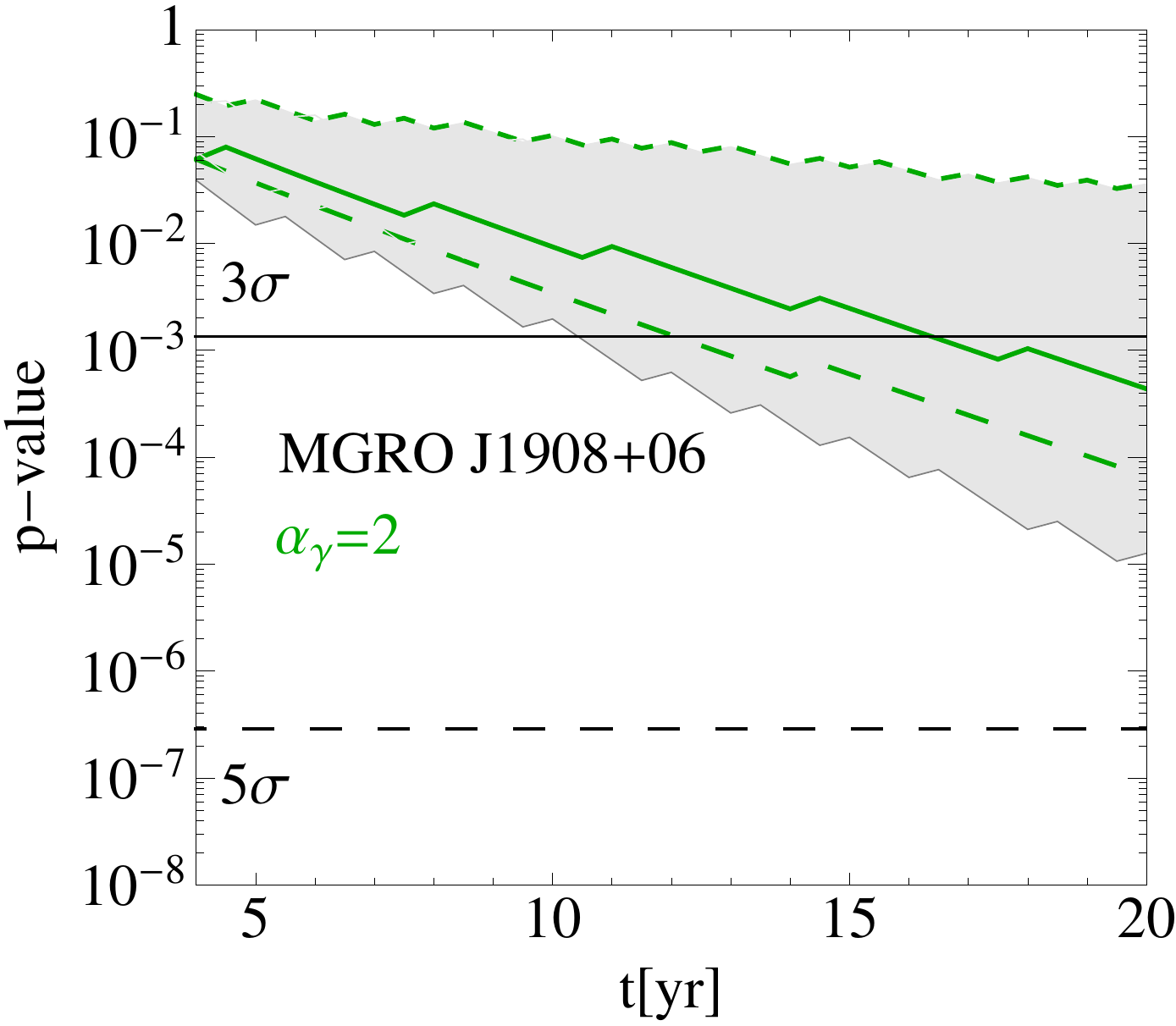}
 \caption{The expected p-value for MGRO J1908+06 as a function of time for different spectral assumptions compatible with the values reported by HESS \citep{Aharonian:2009je}, as shown in Fig.~\ref{fig:1908events}. The grey band encodes the uncertainty due to different values of  $E_{cut,\gamma}$, and morphology. Green lines show the case of extended source.  Figure from \citep{Halzen:2016seh}.} 
 \label{fig:1908pvals}
 \end{figure}
 
 \begin{figure}[ht]
 \includegraphics[width=0.9\columnwidth]{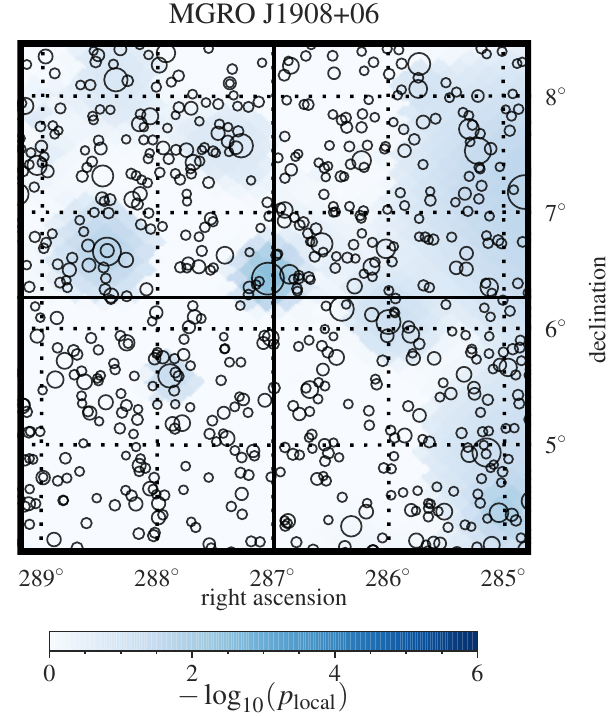}
 \caption{P-value map of neutrino excesses near MGRO 1908+06 in point source search with eight years of IceCube up-going muons. Figure from \citep{Aartsen:2018ywr}.} 
 \label{fig:8r1908}
 \end{figure}

MGRO 1908+06 has been included in IceCube's triggered search for point-like and extended sources \citep{Aartsen:2014cva, Aartsen:2016oji, Aartsen:2018ywr, Aartsen:2019fau}. Recent search for steady sources in the muon neutrino flux with 8 years of IceCube of data \citep{Aartsen:2018ywr} found this source as the second warmest source in the catalog, with a pre-trial p-value of 0.0088. The fitted flux upper limit is consistent with the expectations discussed above. Figure \ref{fig:8r1908} shows the p-value of the neutrino excess and neutrino events found by the 8 year IceCube search in the vicinity of MGRO J1908+06. It should be noted that in this search the source is assumed to be without extension, and as mentioned earlier the likelihood for observation depends on the size of the source. 

It is worth mentioning that recent study of PeVatron candidates in the HESS Galactic plane survey data \citep{H.E.S.S.:2018zkf} obtains a lower bound on cut-off energy of MGRO J1908+06 and supports the idea of this source as PeVatron \citep{Spengler:2019sde}. A major development regarding MGRO J1908+06 and the rest of the sources identified in Milagro survey has brought by HAWC very high energy survey of the Galaxy. We will come back to this later.

Besides MGRO J1908+06 and MGRO 2019+37 that are monitored by IceCube steady source searches, IceCube has searched for collective neutrino emission from the six initially identified sources, aka Milagro 6, via stacking analysis. Stacking likelihood searches \citep{Achterberg:2006ik} probe for correlation of a collection of sources, or a catalog, to a sample of neutrino events. This type of search benefits from a better sensitivity for a lower value of flux per source. The most recent results for the Milagro 6 are reported in \citep{Aartsen:2017ujz}. 

The progress of very high energy gamma ray astrophysics through the last decades has disclosed more than 100 sources with TeV emission in the Galactic plane. The majority of the sources that have identified partners are classified as pulsar wind nebula. Given that they carry a significant fraction of the high-energy emission from the Galaxy, their potential contribution to the neutrino flux is important. 

Confined inside supernova remnants, pulsar wind nebulae are bright diffuse nebulae whose emission is powered by pulsar winds induced by the highly spinning and magnetized pulsars in the center. The nonthermal high-energy emission from a pulsar wind nebula is generally believed to arise from relativistic electron-positron pairs. Those are considered as the primary components of the pulsar winds that are powered by the rotational energy of the central pulsars. In this leptonic scenario, the dominant process for the low-frequency emission in radio, optical, and X-ray is the synchrotron emission of relativistic pairs. Meanwhile, the inverse Compton scattering (ICS) of synchrotron photons becomes dominant at high frequencies, producing TeV emission. This scenario can explain the spectrum from radio to TeV \citep{Kargaltsev:2010jy}.  However, the presence of hadrons in the pulsar wind cannot be excluded yet,  neither by theory nor observation. 

The hadronic scenario for high-energy emission and particle acceleration in pulsar wind nebulae was first discussed in the context of the very high energy gamma ray emission from the Crab Nebula.  There, protons accelerated in the outer gap of the pulsar interacting with the nebula \citep{Cheng:1990au} and heavy nuclei accelerated in the pulsar magnetosphere interacting with soft photons \citep{Bednarek:1997cn}. Later, high-energy neutrino emission from pulsar wind nebulae were studied for particle acceleration in the termination shocks which is then followed by hadronuclear or photohdadronic interactions in the source region, see e.g. \citep{ Guetta:2002hv,Amato:2003kw,Bednarek:2003cv,Lemoine:2014ala,DiPalma:2016yfy}. As a result, a neutrino flux is expected from cosmic rays interaction with the dense environment, see \citep{Amato:2006ts} for details. 

 Presence or absence of hadrons (ions) in pulsar winds have important consequences as any evidence for their existence in pulsar winds would provide significants clues about the mechanism of acceleration in these sources. Minor contamination of ions at the termination shock would result in significant amount of energy contents released in hadrons \citep{Amato:2013fua}. This is particularly important in the context of theoretical modeling of the pulsar wind, especially an obstacle known as $\sigma$ problem. The parameter $\sigma$ is defined as the ratio of the wind Poynting flux to its kinetic energy flux. Theoretical modeling of pulsar magnetospheres and wind indicate large $\sigma$ values. However, the magnetohydrodynamic simulations cannot match shock size and expansion speed at same time. This conflicting scenario could be resolved if the majority of the pulsar winds energy is carried by hadrons and, in addition, can explain how efficient acceleration of leptons is obtained in the termination shocks \citep{DiPalma:2016yfy}.
 
Pulsar wind nebulae have been explored as potential sources of high-energy neutrinos. The searches for neutrino emission from pulsar wind nebulae aims at identification of any excess in the direction of individual sources \citep{Aartsen:2013uuv, Aartsen:2014cva, Aartsen:2016oji} as well as looking for cumulative emission in a stacking searches \citep{Aartsen:2017ujz, Liu:2019iga}. So far, neither of these tests have yielded an evidence for correlation of pulsar wind nebulae with neutrinos arrival directions. 

The obtained upper limits on the neutrino flux from individual sources can constrain the maximal hadronic emission from few of these source, which can be incorporated to revisit the role of hadrons and discuss the prospects for the observations for future detectors such as KM3NeT. Especially, since a good number of these sources are located in the Southern hemisphere, see \citep{DiPalma:2016yfy}.

Meanwhile, the stacking analyses have aimed at identifying correlation of the arrival direction of neutrinos with catalogs of pulsar wind nebulae under specifics assumptions. The stacking analysis presented in \citep{Aartsen:2017ujz} focused on catalogs of 9 supernova remnants with associated pulsar wind nebulae that were observed by gamma-ray observatories. 

In a recent study, IceCube has examined correlation of high-energy neutrinos with 35 TeV pulsar wind nebulae \citep{Liu:2019iga}. In this search, sources identified as pulsar wind nebula with gamma-ray emission at energies higher than 1 TeV were studied for potential correlation with muon neutrino data. The selection of sources were based on the high-energy gamma-ray observations of HAWC, HESS, MAGIC, and VERITAS. The associated pulsars of these pulsar wind nebulae were identifies listed in the Australia Telescope National Facility (ATNF) catalog \citep{Manchester:2004bp}.

 In order to test the correlation of IceCube neutrinos to these sources four different hypotheses were considered. Each scenario provided a weighting scheme for the stacking analysis that enables testing the validity of the hypothesis. In the simplest scheme, all the sources are treated equally. Thus, no preference is given to any source. Other schemes examine correlation with major probability of the pulsar wind nebula such as its level of high-energy gamma ray emission, pulsar's spin frequency as a measure of how energetic the pulsar is, and finally the age of the pulsar. The latter assumes that younger pulsars are more efficient neutrino emitters. 

In the absence of any significant correlation found under any of these assumption, upper limits were imposed on the total neutrino flux from these TeV emitters in the Galaxy. The corresponding upper limit on the gamma-ray flux could be found via multimessenger relation for the neutrino and gamma-ray flux. Figure \ref{fig:pwn} summarizes the results. With better sensitivities at higher energies, these stacking analyses find upper limits at the level of the total observed high-energy gamma-ray emission indicating that neutrino flux measurements getting close to determine the feasibility of models of hadronic emission from pulsar wind nebulae.

In a separate study \citep{Aartsen:2019fau}, another hypothesis was tested. Using 10 years of IceCube data, 33 pulsar wind nebulae were studied to examine the hypothesis that wether the neutrino emission is correlated with the integral of the extrapolated gamma-ray flux above 10 TeV.  While, this search also did not find any evidence of correlation, it addresses an essential question regarding the contribution of pulsar wind nebulae to the very high energy side of IceCube's cosmic neutrino observations.

These analyses constrain the high-energy neutrino emission from pulsar wind nebulae, mainly at energies beyond 10 TeV.  At lower energies, the constraints  are weaker, as the sensitivity of point sources study decreases. It is worth mentioning that a good fraction of identified TeV pulsar wind nebulae are locating in the Southern hemisphere, including the bright source Vela X. 
 We will discuss in Section \ref{sec:summary} how future experiments positioned in the Northern hemisphere, and improved analyses will provide better measurements of possible neutrino emission from pulsar wind nebulae. 

 \begin{figure}[t]
 \includegraphics[width=\columnwidth]{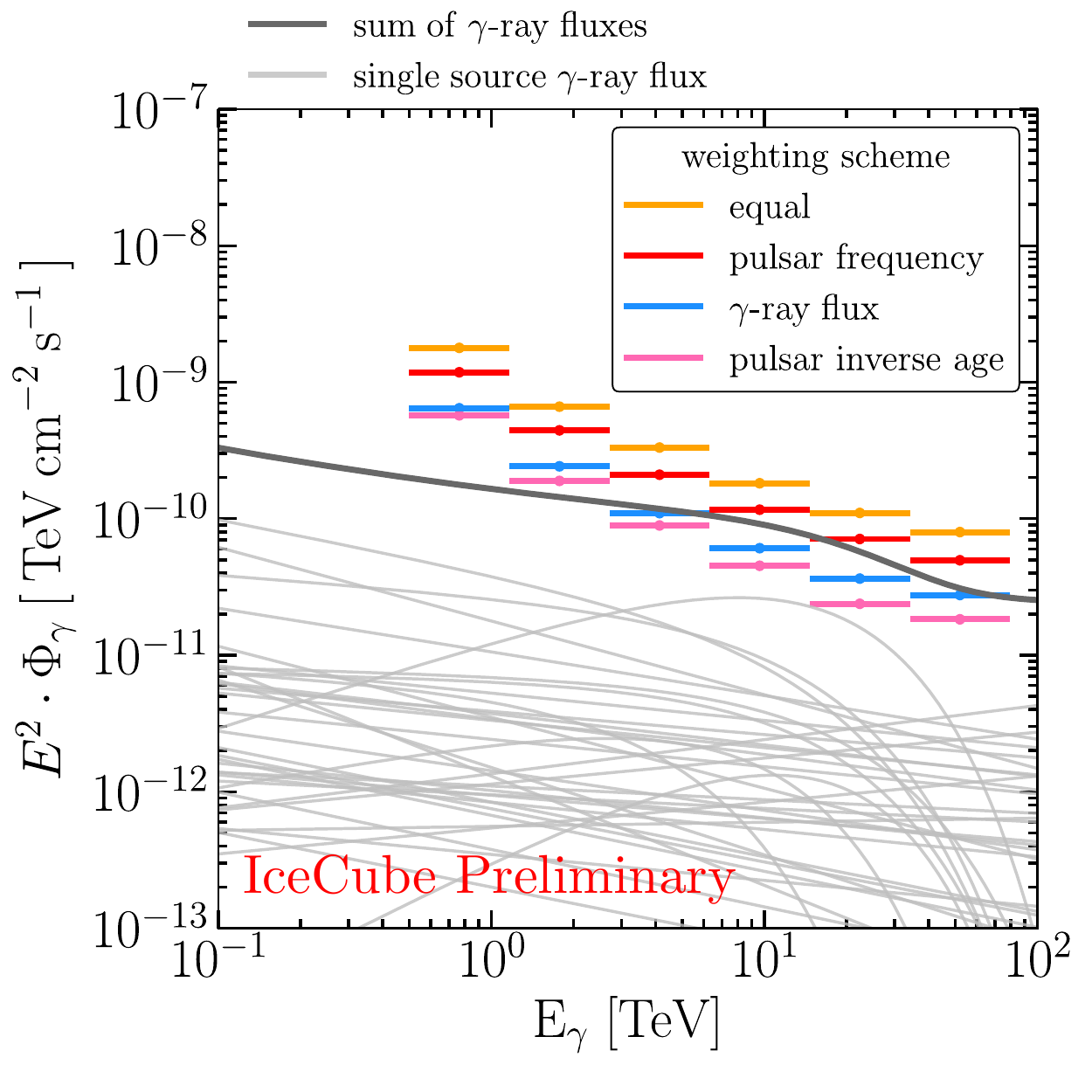}
 \caption{Differential upper limits on the hadronic component of the gamma-ray emission from TeV pulsar wind nebulae. The light grey lines show the observed gamma-ray spectra of the sources while the dark grey line presents the sum of fluxes. Red, orange, magenta and blue steps show the differential upper limit on the hadronic gamma-ray emission. The upper limits are obtained by converting 90\% CL differential upper limit on the neutrino flux, and each color corresponds to a given weighting method. 
 Figure from \citep{Liu:2019iga}.} 
 \label{fig:pwn}
 \end{figure}
  
 Multimessenger identification of potential neutrino sources in the Galaxy based on gamma-ray observations would highly benefit from a better access to very high energies, beyond 10 TeV which is the typical sensitivity of IACTs. Leptonic scenarios are less likely to produce a spectrum that would extend to energies beyond 100 TeV. HAWC gamma-ray observatory as the most sensitive operating ground based wide-field telescope has provided an unprecedented view of the Galaxy in recent years. In a major development, HAWC's very high energy survey of the Galactic plane has revealed more than 20 brand new sources to the list of TeV gamma-ray emitters in the Galaxy. Currently, most of these newly resolved sources are classified as {\em unidentified}. 
Another advantage offered by HAWC's access to higher energies is that it probes an energy ranges that is most relevant for IceCube.  Finally, IceCube is most sensitive to the sources in the Northern Hemisphere that is explored by HAWC.  

  \begin{figure}[ht]
 \includegraphics[width=\columnwidth]{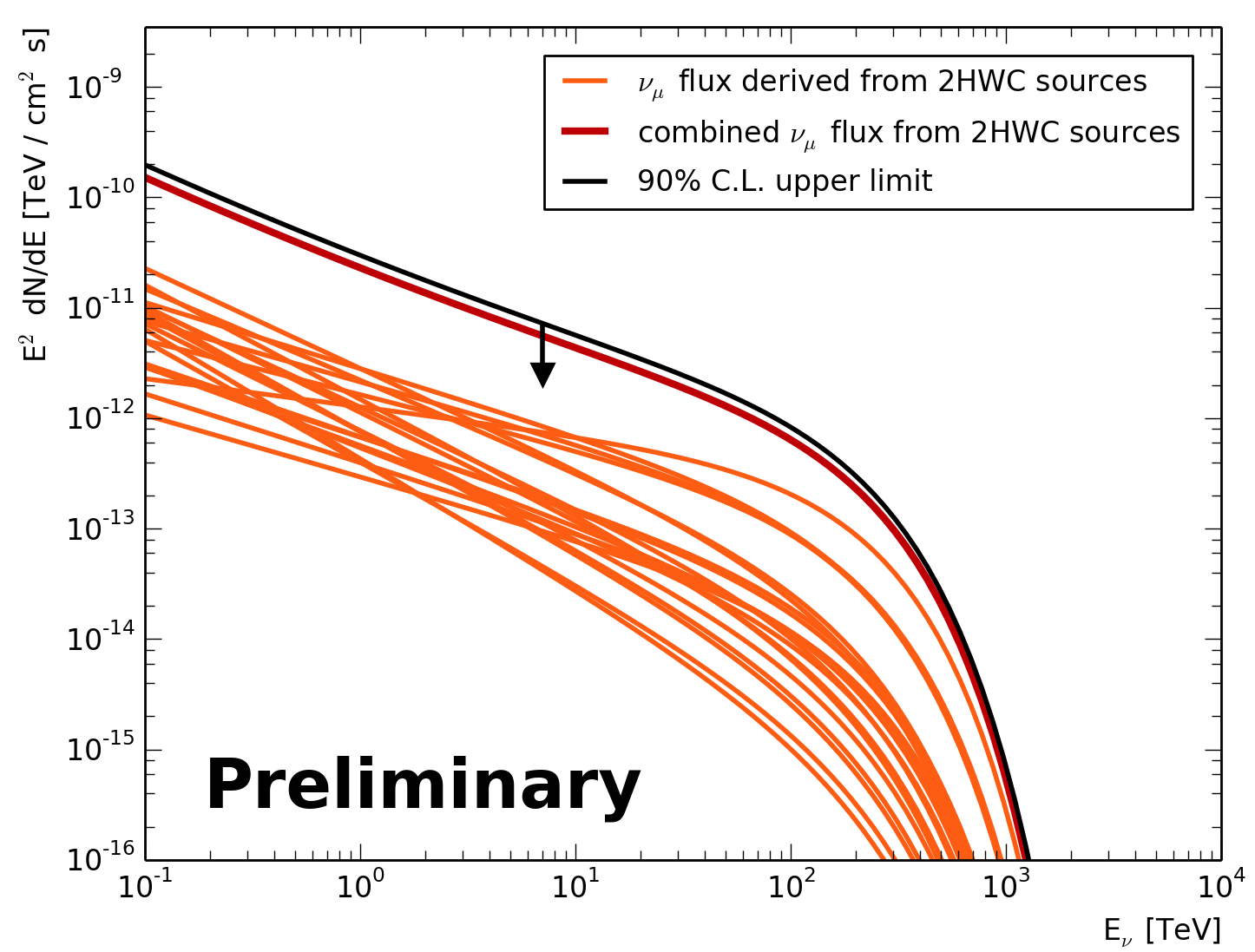}
 \caption{Upper limit (90\% C.L.) on the flux of muon neutrinos (black) for the stacking search of sources in 2HWC catalog (PWN excluded). The projected muon neutrino fluxes (thin orange) represent the expected flux from each source assuming that the high-energy gamma ray flux measured by HAWC is produced in hadronuclear interactions. The combined flux (red) shows sum of the individual fluxes. Figure from \citep{Kheirandish:2019bke}.}
 \label{fig:hwcstack}
 \end{figure}

In order to find the common sources of very high energy gamma rays and high-energy neutrinos, the IceCube and HAWC collaborations designed and performed a joint analysis \citep{Kheirandish:2019bke}. This analysis examined the correlation of neutrinos and 2HWC sources \citep{Abeysekara:2017hyn} as well as correlation of the high-energy neutrinos arrival direction with the morphology of the very high energy gamma ray emission. The former stacking analysis excluded the sources previously identified as PWN, thus the analysis was mainly focused on the {\em unidentified} sources. The latter analysis incorporated the full morphology of the gamma-ray emission from the plane visible in the Northern Sky. Additionally, special regions were singled out for the correlation study. The selection of these regions were motivated by the early findings of Milagro. Cygnus region, area surrounding the MGRO 1908+06, and the inner Galaxy that included candidate source MGRO J1852+01 (now confirmed source in HAWC) are the three preselected regions.

The findings of these four analyses, using 8 years of IceCube data, constrains the contribution of hadronic interactions to the gamma-ray emission observed in Cygnus region and the Galactic plane visible in the Northern Sky. On the other hand, the upper limits on the neutrino flux from the other searches is above the estimated flux based on the gamma-ray emission. The upper limit for the stacking analysis is shown in Figure \ref{fig:hwcstack}. This could indicate that the current data cannot establish sufficient sensitivity for identification of the emission from background. More years of data and improved selections will be required. This is in accordance with the expectations for the likelihood of identification of MGRO J1908+06. 

Let us know turn into to another special region in the search: the Cygnus complex. The joint IceCube-HAWC analysis found an under-fluctuation of data for this region that led to a stringent constraint on the hadronic component of the high-energy emission from this complex.

The Cygnus region is star forming region with a lot of stellar activity. 4 out of 6 sources initially identified in the Milagro observations belonged to this region. Because of the highly extended emission from this region, identification of the sources in this area has been challenging for the IACT experiments such as VERITAS. The region where MGRO J2031+41 was identified is particularly interesting, as it coincides with the localization of the Cygnus cocoon. Due to the uncertainties associated with the origin of the flux of the Cygnus cocoon and $\gamma$-Cygni, a complete picture of high-energy emission from MGRO J2031+41 is missing. The combined analysis of ARGO-YBJ and Fermi data finds a hard spectrum \citep{Argo:2014tqa} , which under point-like assumption for the emission, 
 indicates the observation might be likely in IceCube over the course of $\sim 15$ years \citep{Halzen:2016seh}.

{The Cygnus~X complex has been considered as a promising nearby source of very high energy cosmic rays and neutrinos, see e.g.\citep{Ackermann:2011lfa, Aharonian:2002ij,Tchernin:2013wfa,Gonzalez-Garcia:2013iha,Nierstenhoefer:2015gta, Guenduez:2017qrw, Aharonian:2018oau}. This complex contains massive molecular clouds, populous associates of massive young stars, and luminous H\textsc{II} regions \citep{1981A&A...101...39B, 1992ApJS...81..267L, 2015ApJ...811...85K}.  
Moreover, observation of a hard gamma-ray spectrum, in combination with dense molecular clouds \citep{2012A&A...541A..79G, 2016A&A...591A..40S} and a large number of young OB stars \citep{2015MNRAS.449..741W} support the idea of Cygnus~X as a plausible source of cosmic rays and high-energy astrophysical neutrinos.

 High-energy gamma ray emission from this region was tentatively detected by EGRET\footnote{Energetic Gamma Ray Experiment Telescope} \citep{1996A&AS..120C.423C}. Later, hard gamma-ray emission from the region was observed by HEGRA\footnote{High-Energy-Gamma-Ray Astronomy} \citep{Aharonian:2002ij} which was followed by confirmation of TeV emission by the Milagro collaboration \citep{2007ApJ...658L..33A}. More recent observation has been conducted by MAGIC \citep{Albert:2008yk}, \textit{Fermi} \citep{Ackermann:2011lfa, FermiLAT:2011lax}, ARGO-YBJ \citep{Bartoli:2014irw}, and VERITAS \citep{Aliu:2014rha} collaborations which helped obtaining a better understanding of point-like and extended emission in this region. 

 An important feature revealed in the high-energy gamma ray emission from the Cygnus~X region is an extended excess of hard emission, referred to as the Cygnus Cocoon, that was first detected by the \textit{Fermi} collaboration \citep{Ackermann:2011lfa}. Extended emission has been also reported at TeV energies by Milagro, ARGO-YBJ, VERITAS, and HAWC that is partially coincident with the Cygnus Cocoon. However, the true nature of the Cocoon is still unrevealed. 
 
In the following, we will discuss the potential neutrino flux corresponding to the hard gamma-ray emission from the “Cygnus Cocoon”. We adopt a single-zone model of cosmic ray interactions that is primarily used in the central molecular zones of starburst galaxies \citep{Yoast-Hull:2013wwa}. In this model, at lower energies ($E \sim 10$ MeV) bremsstrahlung scattering is the principal contributor to the gamma-ray spectrum. However, pionic photons contributions grow with increase in energy and become dominate at $\sim 10\, $--$ 100$ GeV.  Here, high-energy gamma ray emission due to inverse Compton scattering is negligible due to the steepness of the cosmic ray electrons at very high energies.

The comparison of the combined gamma-ray spectrum for Cygnus~X with observations from \textit{Fermi}, Milagro, ARGO-YBJ, and HAWC are shown in Figure \ref{fig:gammacyg}. The spectral fits provided in the 3FGL are only valid between 100 MeV and 300 GeV.  However, for the Cocoon, we extrapolated the spectral fit to higher energies to compare with TeV energy gamma-ray observations.  Finally, we include only the off-pulse emission for the 3 brightest pulsars. 
See \citet{Yoast-Hull:2017gaj} for details. 

 \begin{figure}[t]
 \includegraphics[width=1.0\columnwidth]{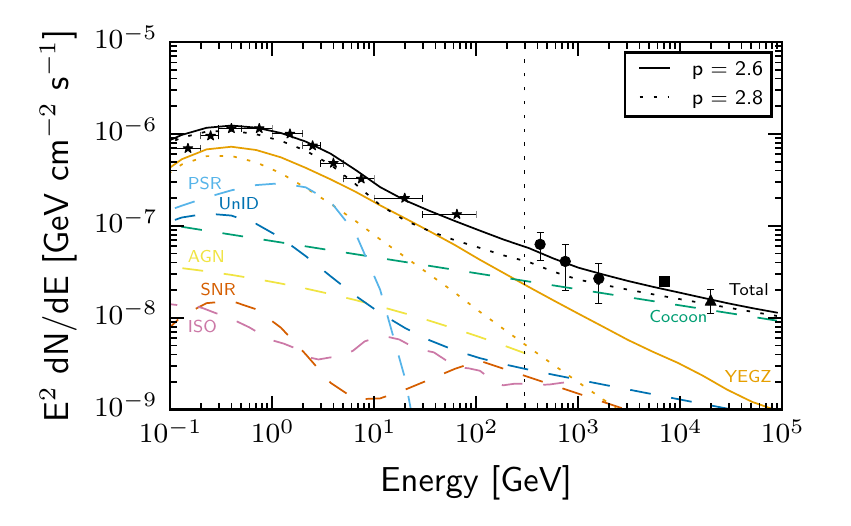}
 \caption{Observed and modeled gamma-ray fluxes from Cygnus X region. Different modeling components include gamma-ray emission from YEGZ models, pulsars (PSRs), active galactic nuclei (AGN), supernova remnants (SNRs), unidentified sources (UnID), the isotropic gamma-ray background (ISO), and the Cygnus Cocoon. Observational points show data from \textit{Fermi} (black stars) \citep{FermiLAT:2011lax}, ARGO-YBJ (black circles) \citep{Bartoli:2015era}, HAWC (black square) \citep{Abeysekara:2017hyn}, and Milagro (black triangle) \citep{2007ApJ...658L..33A}. The vertical dotted black line identifies 300 GeV beyond which the spectrum for Fermi 3FGL sources are extrapolated.  Figure from \citep{Yoast-Hull:2017gaj}.} 
 \label{fig:gammacyg}
 \end{figure}

\begin{figure}[t]
 \includegraphics[width=1.0\columnwidth]{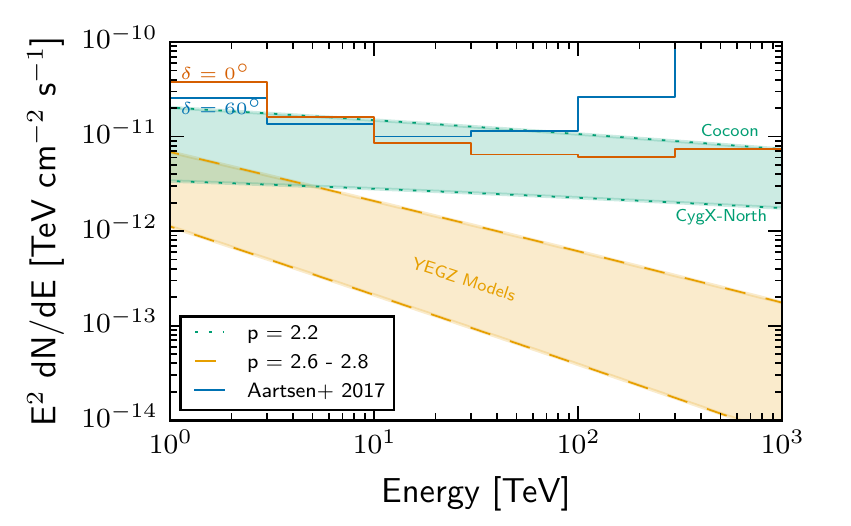}
 \caption{Expected neutrino flux from Cygnus region. The neutrino spectra from the soft, diffuse YEGZ models ($p = 2.6 - 2.8$), the Cygnus Cocoon, and the CygX-North molecular cloud complex, along with the point source differential discovery potential for IceCube \citep{Aartsen:2016oji}. Note that the IceCube sensitivity to extended sources is naturally lower than that for point sources, and thus this plot represents the most optimistic case for detection.  Figure from \citep{Yoast-Hull:2017gaj}.} 
 \label{fig:cocoonnu}
 \end{figure}

 Assuming that the Cocoon is a single source and is dominated by gamma-rays from neutral pion decay, single-zone YEGZ interaction model \citep{Yoast-Hull:2013wwa} finds a neutrino flux slightly larger than IceCube's differential discovery potential at 1 PeV\citep{Aartsen:2016oji}. 
 Therefore, the possibility of detecting neutrinos from the Cocoon is feasible, provided the cosmic ray spectrum is hadronic and extends to PeV energies without steepening.
 
While the contribution of the hadronic processes to the observed high-energy emission is not clear, 
  let us turn our focus to a subregion within the Cocoon that surrounds a large molecular gas cloud complex: CygX-North.  
   This sub-region is centered on $(l = 81.5^{\circ},~b = 0.5^{\circ})$, left of Cyg OB2 (see Figure 7 in \cite{2006A&A...458..855S}) and its emission is expected to be dominated by hadronic interactions, most likely from an unresolved source, either a supernova remnant or a pulsar wind nebula.
   
   Assuming a predominant hadronic component for the high-energy emission from CygX-North, 
   we estimate the neutrino flux from this subregion by adapting the cosmic ray spectrum to the high-energy gamma rays associated to this region.  The 
 resulting neutrino flux falls below IceCube's discovery potential. 
 This implies that the likelihood of observing CygX-North in IceCube is slim. The summary of expected neutrino fluxes from Cygnus cocoon and CygX-North is shown in Figure \ref{fig:cocoonnu}.
 
 In summary, the prospects for identification of neutrino emission from the Cygnus X region depends on the slope of the cosmic ray spectrum.  Soft cosmic ray population in Cygnus X leads to a neutrino flux that is several orders of magnitude below the current IceCube sensitivity. However, assuming a hard cosmic ray population equivalent to that required for the Cocoon will generate a neutrino flux that is potentially detectable by IceCube. 

Assuming a leptohadronic scenario for the high-energy emission from Cygnus region and incorporating the broad data set from radio, MeV (COMPTEL), GeV (Fermi), TeV (Argo) and tens of TeV (Milagro) energies, shows that diffusion loss plays a significant role in Cygnus X \citep{Guenduez:2017qrw}. The fit to the broadband observations can describe the spectrum up to TeV. However, the very high energy gamma ray emission reported by Milagro is underestimated. This transport model with a broad multiwavelength fit predicts a neutrino flux which approaches the sensitivity of IceCube at very high energies, nominally $> 50$ TeV, and expects the flux in the Cygnus X region to suffice for IceCube to measure a significant neutrino flux in the next decade. Nevertheless, this model is rather pessimistic as no strong individual source contribute to the flux. 

The upper limit obtained in the joint IceCube-HAWC analysis indicates that at most 60\% of the gamma-ray flux observed by HAWC can originate from hadronic interactions. While this result limits the ability of neutrino searches to detect neutrino emission from this region, more accurate observation of the extension and emission regions within the Cygnus region, and the cocoon could change this picture. 

Finally, let us comment on the area around MGRO J1852+01, which is particularly interesting for IceCube.  We mentioned earlier that this source missed the statistical threshold in Milagro survey. HAWC observes a high intensity emission from this region and conclusively observed this source, however, the emission is mostly centered around the source 2HWC 1857+027. The large flux reported by Milagro in a $3^{\circ}\times3^{\circ}$ region was most likely coming from this source. Given the large flux and the vicinity to the horizon, this region is a sweet spot for IceCube. Interestingly, the largest excess in the joint IceCube-HAWC search is associated to the region around 2HWC 1857+027.

The results of the joint IceCube-HAWC analysis did not find any significant correlation. However, it provided new clues about the potential sources of high-energy neutrino in the Galaxy. There are two main outcomes from this analysis: first, the hadronic component of the emission from Cygnus region is constrained. Second, the results for regions in the inner Galaxy are marginal to the expectation. That is, while neutrino telescopes are not yet sensitive enough for identifying emission from those sources their observation is likely with more years of data.

In a more recent development, the HAWC collaboration reported measurement of gamma-ray flux beyond 58 TeV from four source \citep{Abeysekara:2019gov}. These sources coincide with the Milagro sources previously identified and support their speculation as PeVatrons. This unprecedented measurement helps with characterization of the very high energy flux, which is essential for tuning the multimessenger searches for neutrino sources. 
}

Another potential source of high-energy neutrinos in the Galaxy are binaries. Neutrino emission from binaries has been anticipated \citep{Torres:2006ub, Levinson:2001as, Guetta:2002hk, Aharonian:2005cx} based on the observation of high-energy gamma ray emission from microquasars. Microquasars, X-ray binaries with prominent jets, has promoted binaries as potential sites of cosmic ray accelerations. Presence of accretion and jets in microquasars creates suitable environment for particle acceleration and interaction. 

The central role of stellar wind and radiation fields in the production mechanisms of the high-energy photons is highlighted by the fact that all known gamma-ray binaries have a high-mass companion star. Similar to other sources of high-energy gamma rays, both leptonic and hadronic scenarios are proposed for the high-energy emission from binaries. In the leptonic mechanism, inverse Compton scattering of the stellar radiation off relativistic electron is thought to be responsible for generation of high-energy gamma rays. In hadronic models, however, neutral pions from hadronuclear interactions of  protons in the jets with cold protons of the stellar wind produce the high-energy gamma rays, see e.g. \citet{Romero:2003td, Romero:2005fr}. Accompanying high-energy neutrinos produced in this scenario is the fundamental feature of hadronic scenario. Moreover, cooling of the secondary particles produced in charged pions decay is another feature that stem from the inclusion of relativistic protons in the jet. This feature would alter the broadband emission from gamma-ray binaries. Studies of jets impact in the stellar medium for Cygnus X-1 suggest that they can carry a significant fraction of kinetic energy \citep{Gallo:2005tf,Heinz:2005jc}.

X-ray binaries are known for their periodic and outburst emissions which makes them good candidates for time-dependent searches. Time-dependent searches generally benefit from lower background rates and can identify correlations via few events, thus enhancing the sensitivity for identification these source. Microquasars periodic emission and X-ray flares have been used for possible temporal neutrino emission \citep{Abbasi:2011ke, Aartsen:2015wto, Adrian-Martinez:2014ito}. Microquasars were also suggested as the dominant contributors to the Galactic component of the high-energy neutrino flux \citep{Anchordoqui:2014rca}. This was partially motivated by the spatial clustering of the first sample of high-energy events near LS 5039, a known microquasar which has been historically suggested as a potential cosmic ray accelerator \citep{Aharonian:2005cx}.

The discussion presented here was mainly centered on the potential Galactic sources of high-energy neutrinos in the Northern Sky. In the future, with operation of KM3NeT the sensitivity of the point source search to explore the Southern Sky will significantly improve. Such development can help closing in on the Galactic component of the high-energy neutrino flux. It is worth noting that there are important aspects with regard to the Galactic gamma-ray emission visible in the Southern hemisphere. The HESS collaboration has announced observation of very high energy gamma ray emission near the Galactic center \citep{Abramowski:2016mir}. Such signal believed to be difficult for leptonic scenarios to account for. Observation of neutrino emission could establish this source as a Galactic PeVatron. In addition, neutrino emission from Galactic center has been discussed by exploiting Sagittarius A* (Sgr A*), a low-luminosity AGN located at the center of the Milky Way \citep{Fujita:2016yvk, Anchordoqui:2016dcp}. Observations X-ray echos have revealed a high level of activity in its past \citep{Koyama:1996sj, Ryu:2012ib}. Accelerated cosmic rays will interact with the dense environment in its vicinity and generate a flux of high-energy neutrinos.
Finally, KM3NeT is expected to be sensitive enough to observe RX J1713.7-3946, a known supernova remnant and potential neutrino source, as well as Vela X \citep{Aiello:2018usb}. It is worth mentioning that the IceCube search for point sources in the cascade data found RX J1713.7-3946 as the most significant source \citep{Aartsen:2019epb}. We should remark that while located in the Northern hemisphere, HAWC observes a small portion of the Galactic plane in the Southern hemisphere. This region includes sources with high level of gamma-ray emission. Provided that the very high energy gamma rays are hadronic, these sources are predicted to be detected by KM3NeT \citep{Niro:2019mzw} within 10 years of operation.

\section{Diffuse Galactic Neutrino Emission }\label{sec:diff}

Generation of high-energy neutrinos and gamma rays is not limited to the surrounding areas of cosmic ray sources. Interaction of accelerated cosmic rays with interstellar medium leads to production of charged and neutral pions that will decay to neutrinos and gamma rays. The dense environment of the Milky Way provides target material for interaction of very high energy cosmic rays during their propagation. Therefore, diffuse gamma-ray and neutrino emission from Galactic plane is considered as a guaranteed non-isotropic component in the high-energy sky \citep{Stecker:1978ah, Domokos:1991tt, Berezinsky:1992wr, Ingelman:1996md, Evoli:2007iy}.

The intensity of the Galactic diffuse neutrino emission is expected to vary along the plane. Gas density, distribution of sources, and cosmic rays density along the Galactic plane are the key parameters determining the intensity of the emission. 

Efforts to model the Galactic diffuse neutrino emission benefit from observation of the diffuse Galactic gamma-ray emission as the the intensity and morphology of diffuse Galactic neutrino flux can be assessed by incorporating the diffuse Galactic gamma-ray emission profile together with the local measurement of cosmic rays \citep{Ahlers:2013xia,Joshi:2013aua,Kachelriess:2014oma}.

The diffuse component of gamma-ray emission from Galactic plane has been measured by {\em Fermi}-LAT in the range 100 MeV to tens of GeV \citep{Ackermann:2012pya, Acero:2016qlg}. {\em Fermi}'s measurement has become one of the core elements in modeling of the Galactic diffuses high-energy gamma rays.  Model parameters are tuned to match the {\em Fermi} data for diffuse Galactic gamma-ray emission.

GALPROP \citep{2011CoPhC.182.1156V} is an example of successful attempts to model the diffuse Galactic gamma-ray component. In order to model the Galactic diffuse gamma-ray emission, propagation and diffusion of cosmic rays in the interstellar medium is studied. The production rate of cosmic rays is established by assuming a distribution of potential sources of cosmic rays in the Galaxy, for instance supernova remnants. In the meantime, the interstellar gas and radiation filed densities are limited by {\em Fermi}-LAT and radio observations. 

The conventional GALPROP modelings find a good agreement with the gamma-ray measurements below 10 GeV. However, they cannot accommodate the observed flux beyond 10 GeV for the inner Galaxy \citep{Ackermann:2012pya}. The KRA$_\gamma$ models~\citep{Gaggero:2015xza, Gaggero:2014xla, Gaggero:2017jts} try to resolve this issue by modifying the diffusion and introducing an exponential cutoff at a certain energy. Energy cut-offs at cutoffs at 5 and 50 {PeV} are chosen to accommodate the measurements at very high energies by KASCADE~\citep{Antoni:2005wq} and KASCADE-Grande~\citep{Apel:2013uni}.  The models are known as KRA$_\gamma^{5}$ and KRA$_\gamma^{50}$ to specify the considered energy cut-offs. Figure \ref{fig:diffnu} shows the direction and energy integrated for neutrinos predicted in KRA$_\gamma^{5}$ model.

Following IceCube's observation of high-energy neutrino flux, the contribution of the diffuse Galactic neutrino emission to the total isotropic neutrino flux has been studied \citep{Ahlers:2013xia, Ahlers:2015moa, Albert:2017oba, Aartsen:2017ujz} using a variety of assumptions and data sets including the KRA$_\gamma$ predictions. 

\begin{figure}[t]
 \includegraphics[width=\columnwidth]{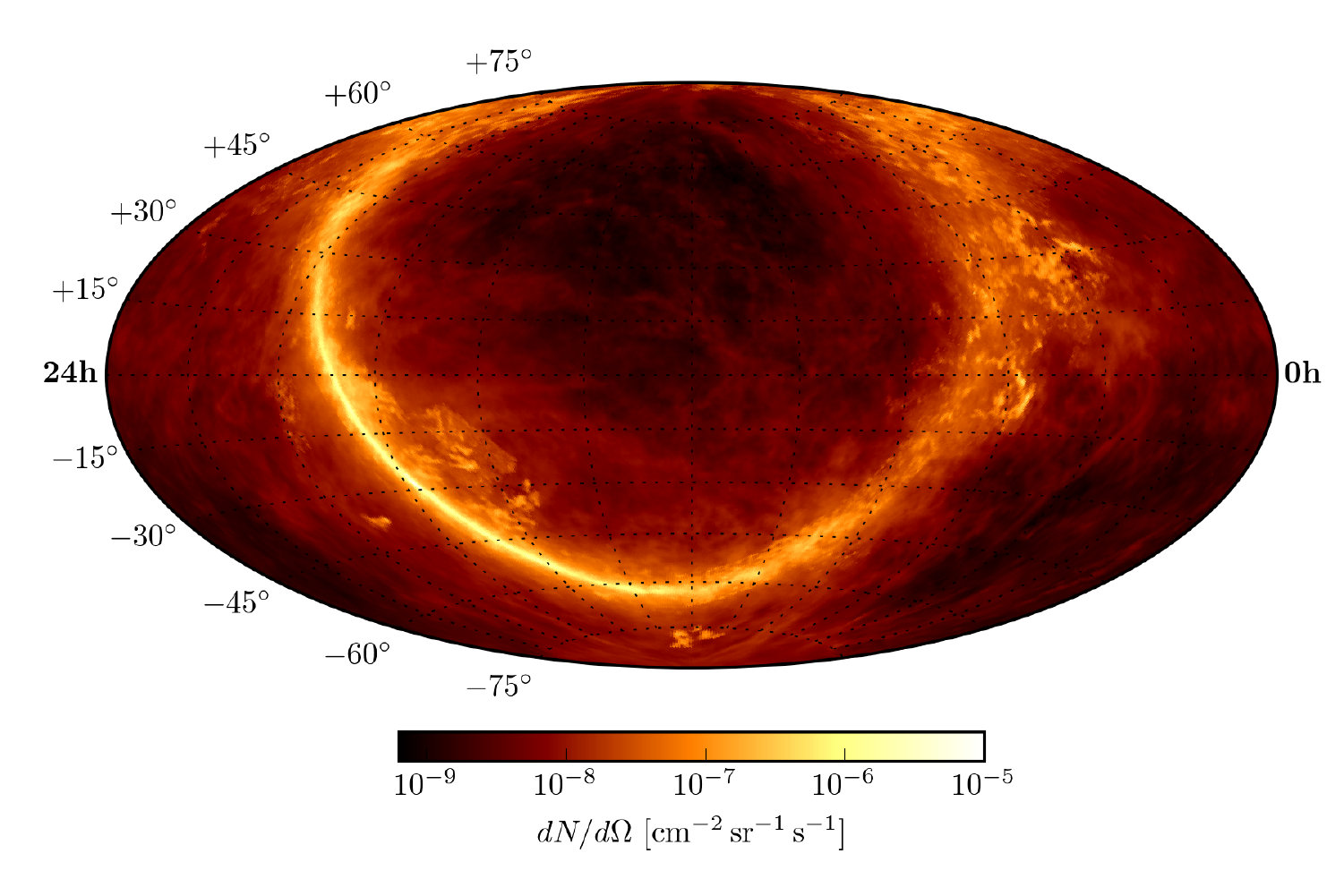}
 \caption{Diffuse gamma-ray emission from Galactic plane: 
 The color map shows energy integrated neutrino flux  from the KRA$_\gamma^5$ model~\citep{Gaggero:2015xza}, illustrated as a function of direction in equatorial coordinates.
 Figure from \citep{Albert:2018vxw}.}
 \label{fig:diffnu}
 \end{figure}
 
  \begin{figure}[ht]
 \includegraphics[width=\columnwidth]{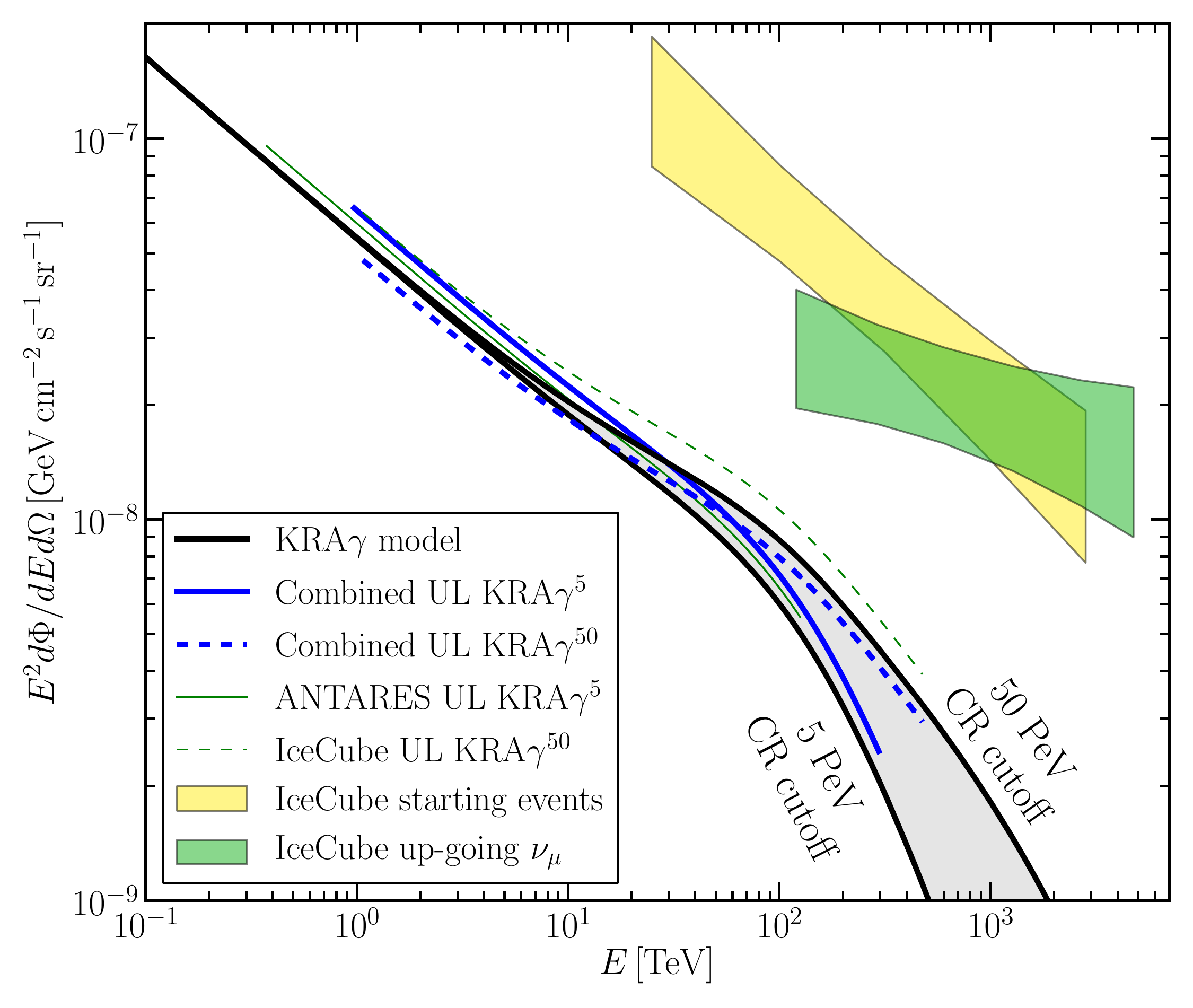}
 \caption{Upper limits on the Galactic diffuse gamma-ray emission compared to predicted neutrino flux from  KRA$_\gamma$. The total neutrino spectrum from HESE and up-going muon search has been shown for comparison. Figure from \citep{Albert:2018vxw}.}
 \label{fig:difflim}
 \end{figure}

In a recent study, the ANTARES and IceCube collaborations performed a joint search to target the diffuse Galactic component. The joint constraint from IceCube and ANTARES on the diffuse Galactic plane neutrino emission \citep{Albert:2018vxw} uses ten years of ANTARES muon track and cascade data, as well as seven years of IceCube muon track data to perform a joint likelihood test for neutrino emission. Their results impose constraints on the diffuse Galactic component hypothesis based on the KRA$_\gamma$ model. 
The results of the maximum likelihood analysis finds a non-zero component for diffuse Galactic components. However, the excess falls short of statistical significance. The upper limits from these test constraints the KRA$_\gamma^{50}$ model at $0.9\times \Phi_{\textrm{KRA}_\gamma^{50}}$. The upper limit on the KRA$_\gamma^{5}$ model, however, is above the predicted limit. Figure \ref{fig:difflim} compares the upper limit with the predicted flux from KRA$_\gamma$ models as well as the total neutrino flux measured by IceCube.

{\em Fermi} bubbles are considered as another potential site of cosmic ray diffusion and diffuse high-energy neutrino emission. The extended region assigned as {\em Fermi} bubbles extend to $\sim$ 55 degrees away from center of the Galaxy and indicates a hard gamma-ray spectrum up to $\sim$ 100 GeV \citep{Fermi-LAT:2014sfa}. It has been suggested that cosmic ray acceleration near Galactic center is the origin of gamma-ray emission \citep{Crocker:2010dg}. However, the source of high-energy emission from {\em Fermi} bubbles has not been established experimentally. Contribution of {\em Fermi} bubbles to the high-energy neutrino flux has been studied \citep{Lunardini:2013gva, Ahlers:2013xia} and IceCube and ANTARES have imposed upper limits on the neutrino emission \citep{Aartsen:2019epb, Albert:2017bdv}. 

%%%%%%%%%%%%%%%%%%%%%%%%%%%%%%%%%%%% 
\section{Summary \& Outlook}\label{sec:summary}
The discovery of high-energy neutrinos with astrophysical origin in IceCube has offered a unique view of the high-energy universe. Prior to the observation of cosmic neutrinos, our knowledge of the high-energy universe was limited to cosmic ray and gamma-ray measurements. Cosmic neutrinos reveal an unobstructed view at wavelengths where the universe is opaque to photons. Today, high-energy neutrino astronomy has demonstrated the greater than expected role of hadronic processes in the nonthermal universe. The finding that challenges our understanding of the universe. The observation of high-energy cosmic neutrinos has also facilitated the multimessenger search for the origin of cosmic rays, a century old puzzle in high-energy astrophysics.

The arrival direction of high-energy neutrinos indicate a predominant extragalactic origin. However, the contribution of the Galactic component cannot be excluded at this point. 

Galactic source are considered as {\em guaranteed} contributors to the flux of high-energy neutrinos. Sources of very high energy cosmic rays in the Galaxy are believed to be responsible for acceleration  of very high energy cosmic rays that reach energies of few PeV. Their interaction with stellar and interstellar gas and radiation would result in production of neutrinos and gamma rays. High-energy neutrinos are produced in astrophysical beam dumps that can supply enough target density for interaction of very high energy cosmic rays. The dense environment of the Milky Way meets this condition.

In this review, we discussed the prospects for identifying neutrino emission from Galactic sources and presented the status of their observation. The discussion was centered around major gamma-ray emitters in the Galaxy as they establish one pillar for multimessenger identification of the sources of cosmic rays. Our discussion regarding the current limits and recent studies was mainly based on the IceCube's measurements, as IceCube is currently the largest operating detector. The ANTARES detector in the Mediterranean is positioned in the northern sky and despite smaller effective area, can provide useful information for Galactic sources in the Southern hemisphere.

The subdominant Galactic component of the high-energy neutrino flux is estimated to contribute at most to less than 15\% of the total measured flux. However, identification of Galactic sources with strong neutrino emission is likely in near future. While the contribution to the flux at very high energies is less likely \citep{Ahlers:2015moa}, the emerging features in cosmic neutrino spectrum and the low-energy excess in IceCube flux around $\sim$ 10 TeV could suggest a relatively soft Galactic contribution dominating in the Southern sky.

In addition, the search for point-like and extended sources in the Galaxy will soon reach the required sensitivity to test optimal emission scenarios for such sources. 

In the context of multimessenger astrophysics for non-thermal hadronic emission, identification of potential sources of high-energy neutrinos in the Galaxy is tied to very high energy gamma ray observations. Prior to the observation of the high-energy neutrino flux, potential sources had been identified in the very high energy survey of the Galaxy by Milagro. The current observation is consistent with the expectations and the sensitivity of neutrino source searches are marginal to the optimal scenario for neutrino emission. 

The importance of  gamma-ray observations in identifying the sources of high-energy cosmic neutrinos is emphasized by the recent developments in the neutrino sources search. Identifying Galactic sources of high-energy neutrinos is entangled with the uncertainties and tensions found in gamma-ray observation of very high energy sources in the Milky Way. Resolving sources in populated regions, and precise measurement of the flux and extension of the source play crucial role in identifying potential sources of high-energy neutrinos in the Galaxy. Current improvement and future advancement in the observation of the very high energy emission from Galactic plane will assist this goal.

HAWC observation of the flux at energies beyond 50 TeV has supported the idea of proposed source like MGRO J1908+06 as potential source of high-energy neutrinos. The plans to establish the Southern Wide-field Gamma-ray Observatory (SWGO) \citep{Albert:2019afb} will provide a deeper survey of the Galactic plane in Southern hemisphere. Addition of Cherenkov Telescope Array (CTA) \citep{Acharya:2013sxa, Acharya:2017ttl} and and LHAASO \citep{DiSciascio:2016rgi} to the network of very high energy gamma ray observatories would provide an enhanced sensitivity that reaches to PeV regime, essential to identify the PeVatrons in the Galaxy. Based on current estimations, CTA is expected to resolve $\sim$ 100 supernova remnants \citep{Zanin:2017rao}. It is also worth mention future MeV telescopes like eAstrogram and AMEGO that will have a central role in identification of the {\em pion bump} in the broadband spectrum of Galactic sources. This is also important in the context of modeling diffuse emission, as unresolved sources with harder intrinsic cosmic ray spectra can have a significant contribution to diffuse gamma-ray and neutrino fluxes \citep{Ahlers:2013xia, Lipari:2018gzn}.
 
The search for Galactic sources of high-energy neutrino will also benefit from improvements in data selection, increased statistics, and new techniques for distinguishing signal from background. New developments in identifying starting muon tracks will improve IceCube's sensitivity to Galactic sources in the Southern sky \citep{Silva:2019fnq, Mancina:2019hsp}. Addition of cascade data sets, despite their poor angular resolution, has proved to add substantial improvements to the search. 

Future neutrino detectors will considerably improve the likelihood of finding Galactic sources of high-energy neutrinos. Development of neutrino detectors in the Northern hemisphere such as KM3NeT and Baikal  \citep{Avrorin:2019vfc} will boost sensitivity towards Galactic center and other regions of interest in the Southern Sky. Moreover, 
next-generation of IceCube neutrino observatory with five times the effective area of IceCube is expected to improve the neutrino source search sensitivity by the same order \citep{Aartsen:2014njl, Ahlers:2014ioa, Aartsen:2019swn}. Increased sensitivity to point-like and extended sources will be achieved improved angular resolution, better energy resolution, higher statistics, and reduction of the background. 

Neutrino production from cosmic ray interaction in Galactic sources and interstellar medium was the subject of this review. Another potential Galactic contributor to the high-energy neutrino flux is dark matter. Dark matter signatures present the only alternative explanation of a high-energy neutrino signal from the Galaxy. Dark matter decay and annihilation create distinctive signatures in the observed cosmic ray spectrum. The contribution of dark matter to the high-energy neutrino flux can be examined via signatures in events energy distribution and arrival direction, see e.g.  \citep{Beacom:2006tt, Murase:2012xs, Bai:2013nga, Murase:2015gea, Bhattacharya:2019ucd}. Due to the high concentration of dark matter in the center of the Galaxy, anisotropic signature is expected. Possible interaction of extragalactic component of the high-energy neutrinos with Galactic dark matter on the other hand would create a deficit in the neutrinos towards Galactic center  \citep{Arguelles:2017atb}. For recent review of the constraints on Galactic dark matter signals in neutrino data, see \citet{Arguelles:2019ouk}.

Currently, gamma-ray observations cannot differentiate hadronic and leptonic emission. However, they point to the most likely location of particle acceleration in the Galaxy and have identified potential sources of cosmic neutrinos. No strong evidence for a neutrino source in the Milky Way has been found yet but the identification of Galactic sources of high-energy neutrinos are more likely in near future. Detecting the accompanying neutrinos from a high-energy gamma ray emitter in the Milky Way would provide incontrovertible evidence for cosmic-ray acceleration.

\acknowledgments
The author would like to thank Francis Halzen, Kohta Murase, and Ibrahim Safa for comments and discussions. AK also acknowledges the IGC Postdoctoral Award. 

\sloppy


\begin{thebibliography}{169}
% BibTex style file: spr-mp.bst (nameyear,cnd), 2011-05-27
\ifx \bisbn   \undefined \def \bisbn  #1{ISBN #1}\fi
\ifx \binits  \undefined \def \binits#1{#1} \fi
\ifx \bauthor  \undefined \def \bauthor#1{#1} \fi
\ifx \batitle  \undefined \def \batitle#1{#1} \fi
\ifx \bjtitle  \undefined \def \bjtitle#1{#1}\fi
\ifx \bvolume  \undefined \def \bvolume#1{\textbf{#1}}\fi
\ifx \byear  \undefined \def \byear#1{#1} \fi
\ifx \bissue  \undefined \def \bissue#1{#1} \fi
\ifx \bfpage  \undefined \def \bfpage#1{#1} \fi
\ifx \blpage  \undefined \def \blpage #1{#1} \fi
\ifx \burl  \undefined \def \burl#1{\textsf{#1}} \fi
\ifx \doiurl  \undefined \def \doiurl#1{\textsf{#1}} \fi
\ifx \betal  \undefined \def \betal{\textit{et al.}} \fi
\ifx \binstitute  \undefined \def \binstitute#1{#1} \fi
\ifx \binstitutionaled  \undefined \def \binstitutionaled#1{#1} \fi
\ifx \bctitle  \undefined \def \bctitle#1{#1} \fi
\ifx \beditor  \undefined \def \beditor#1{#1} \fi
\ifx \bpublisher  \undefined \def \bpublisher#1{#1} \fi
\ifx \bbtitle  \undefined \def \bbtitle#1{#1} \fi
\ifx \bedition  \undefined \def \bedition#1{#1} \fi
\ifx \bseriesno  \undefined \def \bseriesno#1{#1} \fi
\ifx \blocation  \undefined \def \blocation#1{#1} \fi
\ifx \bsertitle  \undefined \def \bsertitle#1{#1} \fi
\ifx \bsnm \undefined \def \bsnm#1{#1} \fi
\ifx \bsuffix \undefined \def \bsuffix#1{#1} \fi
\ifx \bparticle \undefined \def \bparticle#1{#1} \fi
\ifx \barticle \undefined \def \barticle#1{#1} \fi
\ifx \bconfdate \undefined \def \bconfdate #1{#1} \fi
\ifx \botherref \undefined \def \botherref #1{#1} \fi
\ifx \url \undefined \def \url#1{\textsf{#1}} \fi
\ifx \bchapter \undefined \def \bchapter#1{#1} \fi
\ifx \bbook \undefined \def \bbook#1{#1} \fi
\ifx \bcomment \undefined \def \bcomment#1{#1} \fi
\ifx \oauthor \undefined \def \oauthor#1{#1} \fi
\ifx \citeauthoryear \undefined \def \citeauthoryear#1{#1} \fi
\ifx \endbibitem  \undefined \def \endbibitem {}\fi
\ifx \bconflocation  \undefined \def \bconflocation#1{#1} \fi
\ifx \arxivurl  \undefined \def \arxivurl#1{\textsf{#1}} \fi

\bibitem[\protect\citeauthoryear{Aab et~al.}{2015}]{Aab:2015kma}
\begin{barticle}
\bauthor{\bsnm{Aab}, \binits{A.}}, \betal:
\bjtitle{Phys. Rev.}
\bvolume{D91}(\bissue{9}),
\bfpage{092008}
(\byear{2015}).
\arxivurl{1504.05397}.
doi:\doiurl{10.1103/PhysRevD.91.092008}
\end{barticle}
\endbibitem

\bibitem[\protect\citeauthoryear{Aartsen et~al.}{2013a}]{Aartsen2013e}
\begin{barticle}
\bauthor{\bsnm{Aartsen}, \binits{M.G.}}, \betal:
\bjtitle{Science}
\bvolume{342},
\bfpage{1242856}
(\byear{2013}a).
\arxivurl{1311.5238}.
doi:\doiurl{10.1126/science.1242856}
\end{barticle}
\endbibitem

\bibitem[\protect\citeauthoryear{Aartsen et~al.}{2013b}]{Aartsen:2013uuv}
\begin{barticle}
\bauthor{\bsnm{Aartsen}, \binits{M.G.}}, \betal:
\bjtitle{Astrophys. J.}
\bvolume{779},
\bfpage{132}
(\byear{2013}b).
\arxivurl{1307.6669}.
doi:\doiurl{10.1088/0004-637X/779/2/132}
\end{barticle}
\endbibitem

\bibitem[\protect\citeauthoryear{Aartsen et~al.}{2014a}]{Aartsen:2013vja}
\begin{barticle}
\bauthor{\bsnm{Aartsen}, \binits{M.G.}}, \betal:
\bjtitle{JINST}
\bvolume{9},
\bfpage{03009}
(\byear{2014}a).
\arxivurl{1311.4767}.
doi:\doiurl{10.1088/1748-0221/9/03/P03009}
\end{barticle}
\endbibitem

\bibitem[\protect\citeauthoryear{Aartsen et~al.}{2014b}]{Aartsen:2014njl}
\begin{botherref}
\oauthor{\bsnm{Aartsen}, \binits{M.G.}}, et al.:
{IceCube-Gen2: A Vision for the Future of Neutrino Astronomy in Antarctica}
(2014b).
\arxivurl{1412.5106}
\end{botherref}
\endbibitem

\bibitem[\protect\citeauthoryear{Aartsen et~al.}{2014c}]{Aartsen:2014gkd}
\begin{barticle}
\bauthor{\bsnm{Aartsen}, \binits{M.G.}}, \betal:
\bjtitle{Phys.Rev.Lett.}
\bvolume{113},
\bfpage{101101}
(\byear{2014}c).
\arxivurl{1405.5303}.
doi:\doiurl{10.1103/PhysRevLett.113.101101}
\end{barticle}
\endbibitem

\bibitem[\protect\citeauthoryear{Aartsen et~al.}{2014d}]{Aartsen:2014cva}
\begin{barticle}
\bauthor{\bsnm{Aartsen}, \binits{M.G.}}, \betal:
\bjtitle{Astrophys. J.}
\bvolume{796}(\bissue{2}),
\bfpage{109}
(\byear{2014}d).
\arxivurl{1406.6757}.
doi:\doiurl{10.1088/0004-637X/796/2/109}
\end{barticle}
\endbibitem

\bibitem[\protect\citeauthoryear{Aartsen et~al.}{2015a}]{Aartsen:2014muf}
\begin{barticle}
\bauthor{\bsnm{Aartsen}, \binits{M.G.}}, \betal:
\bjtitle{Phys. Rev.}
\bvolume{D91}(\bissue{2}),
\bfpage{022001}
(\byear{2015}a).
\arxivurl{1410.1749}.
doi:\doiurl{10.1103/PhysRevD.91.022001}
\end{barticle}
\endbibitem

\bibitem[\protect\citeauthoryear{Aartsen et~al.}{2015b}]{Aartsen2015b}
\begin{barticle}
\bauthor{\bsnm{Aartsen}, \binits{M.G.}}, \betal:
\bjtitle{Phys. Rev. Lett.}
\bvolume{115}(\bissue{8}),
\bfpage{081102}
(\byear{2015}b).
\arxivurl{1507.04005}.
doi:\doiurl{10.1103/PhysRevLett.115.081102}
\end{barticle}
\endbibitem

\bibitem[\protect\citeauthoryear{Aartsen et~al.}{2015c}]{Aartsen:2015xup}
\begin{barticle}
\bauthor{\bsnm{Aartsen}, \binits{M.G.}}, \betal:
\bjtitle{Phys. Rev.}
\bvolume{D91},
\bfpage{122004}
(\byear{2015}c).
\arxivurl{1504.03753}.
doi:\doiurl{10.1103/PhysRevD.91.122004}
\end{barticle}
\endbibitem

\bibitem[\protect\citeauthoryear{Aartsen et~al.}{2015d}]{Aartsen:2014aqy}
\begin{barticle}
\bauthor{\bsnm{Aartsen}, \binits{M.G.}}, \betal:
\bjtitle{Astrophys. J.}
\bvolume{805}(\bissue{1}),
\bfpage{5}
(\byear{2015}d).
\arxivurl{1412.6510}.
doi:\doiurl{10.1088/2041-8205/805/1/L5}
\end{barticle}
\endbibitem

\bibitem[\protect\citeauthoryear{Aartsen et~al.}{2015e}]{Aartsen:2015wto}
\begin{barticle}
\bauthor{\bsnm{Aartsen}, \binits{M.G.}}, \betal:
\bjtitle{Astrophys. J.}
\bvolume{807}(\bissue{1}),
\bfpage{46}
(\byear{2015}e).
\arxivurl{1503.00598}.
doi:\doiurl{10.1088/0004-637X/807/1/46}
\end{barticle}
\endbibitem

\bibitem[\protect\citeauthoryear{Aartsen et~al.}{2016a}]{Aartsen:2016ngq}
\begin{barticle}
\bauthor{\bsnm{Aartsen}, \binits{M.G.}}, \betal:
\bjtitle{Phys. Rev. Lett.}
\bvolume{117}(\bissue{24}),
\bfpage{241101}
(\byear{2016}a).
\bcomment{[Erratum: Phys. Rev. Lett.119,no.25,259902(2017)]}.\\
\arxivurl{1607.05886}.
doi:\doiurl{10.1103/PhysRevLett.117.241101, 10.1103/PhysRevLett.119.259902}
\end{barticle}
\endbibitem

\bibitem[\protect\citeauthoryear{Aartsen et~al.}{2016b}]{Aartsen:2016xlq}
\begin{barticle}
\bauthor{\bsnm{Aartsen}, \binits{M.G.}}, \betal:
\bjtitle{Astrophys. J.}
\bvolume{833}(\bissue{1}),
\bfpage{3}
(\byear{2016}b).
\arxivurl{1607.08006}.
doi:\doiurl{10.3847/0004-637X/833/1/3}
\end{barticle}
\endbibitem

\bibitem[\protect\citeauthoryear{Aartsen et~al.}{2017a}]{Aartsen:2016oji}
\begin{barticle}
\bauthor{\bsnm{Aartsen}, \binits{M.G.}}, \betal:
\bjtitle{Astrophys. J.}
\bvolume{835}(\bissue{2}),
\bfpage{151}
(\byear{2017}a).
\arxivurl{1609.04981}.
doi:\doiurl{10.3847/1538-4357/835/2/151}
\end{barticle}
\endbibitem

\bibitem[\protect\citeauthoryear{Aartsen et~al.}{2017b}]{Aartsen:2017ujz}
\begin{barticle}
\bauthor{\bsnm{Aartsen}, \binits{M.G.}}, \betal:
\bjtitle{Astrophys. J.}
\bvolume{849}(\bissue{1}),
\bfpage{67}
(\byear{2017}b).
\arxivurl{1707.03416}.
doi:\doiurl{10.3847/1538-4357/aa8dfb}
\end{barticle}
\endbibitem

\bibitem[\protect\citeauthoryear{Aartsen et~al.}{2017c}]{Aartsen:2016nxy}
\begin{barticle}
\bauthor{\bsnm{Aartsen}, \binits{M.G.}}, \betal:
\bjtitle{JINST}
\bvolume{12}(\bissue{03}),
\bfpage{03012}
(\byear{2017}c).
\arxivurl{1612.05093}.
doi:\doiurl{10.1088/1748-0221/12/03/P03012}
\end{barticle}
\endbibitem

\bibitem[\protect\citeauthoryear{Aartsen et~al.}{2019a}]{Aartsen:2019wbt}
\begin{botherref}
\oauthor{\bsnm{Aartsen}, \binits{M.G.}}, et al.:
{A Search for MeV to TeV Neutrinos from Fast Radio Bursts with IceCube}
(2019a).
\arxivurl{1908.09997}.
doi:\doiurl{10.3847/1538-4357/ab564b}
\end{botherref}
\endbibitem

\bibitem[\protect\citeauthoryear{Aartsen et~al.}{2019b}]{Aartsen:2019swn}
\begin{botherref}
\oauthor{\bsnm{Aartsen}, \binits{M.G.}}, et al.:
{Neutrino astronomy with the next generation IceCube Neutrino Observatory}
(2019b).
\arxivurl{1911.02561}
\end{botherref}
\endbibitem

\bibitem[\protect\citeauthoryear{Aartsen et~al.}{2019c}]{Aartsen:2019epb}
\begin{barticle}
\bauthor{\bsnm{Aartsen}, \binits{M.G.}}, \betal:
\bjtitle{Astrophys. J.}
\bvolume{886},
\bfpage{12}
(\byear{2019}c).
\arxivurl{1907.06714}.
doi:\doiurl{10.3847/1538-4357/ab4ae2}
\end{barticle}
\endbibitem

\bibitem[\protect\citeauthoryear{Aartsen et~al.}{2019d}]{Aartsen:2018ywr}
\begin{barticle}
\bauthor{\bsnm{Aartsen}, \binits{M.G.}}, \betal:
\bjtitle{Eur. Phys. J.}
\bvolume{C79}(\bissue{3}),
\bfpage{234}
(\byear{2019}d).
\arxivurl{1811.07979}.
doi:\doiurl{10.1140/epjc/s10052-019-6680-0}
\end{barticle}
\endbibitem

\bibitem[\protect\citeauthoryear{Aartsen et~al.}{2020a}]{Aartsen:2020aqd}
\begin{botherref}
\oauthor{\bsnm{Aartsen}, \binits{M.G.}}, et al.:
{Characteristics of the diffuse astrophysical electron and tau neutrino flux
  with six years of IceCube high energy cascade data}
(2020a).
\arxivurl{2001.09520}
\end{botherref}
\endbibitem

\bibitem[\protect\citeauthoryear{Aartsen et~al.}{2020b}]{Aartsen:2019fau}
\begin{barticle}
\bauthor{\bsnm{Aartsen}, \binits{M.G.}}, \betal:
\bjtitle{Phys. Rev. Lett.}
\bvolume{124}(\bissue{5}),
\bfpage{051103}
(\byear{2020}b).
\arxivurl{1910.08488}.
doi:\doiurl{10.1103/PhysRevLett.124.051103}
\end{barticle}
\endbibitem

\bibitem[\protect\citeauthoryear{Aartsen et~al.}{2015}]{Aartsen:2014qna}
\begin{barticle}
\bauthor{\bsnm{Aartsen}, \binits{M.G.}}, \betal:
\bjtitle{Eur.\ Phys.\ J.\ C}
\bvolume{75}(\bissue{3}),
\bfpage{116}
(\byear{2015}).
\arxivurl{1409.4535}.
doi:\doiurl{10.1140/epjc/s10052-015-3330-z}
\end{barticle}
\endbibitem

\bibitem[\protect\citeauthoryear{Aartsen et~al.}{2018a}]{IceCube:2018dnn}
\begin{barticle}
\bauthor{\bsnm{Aartsen}, \binits{M.G.}}, \betal:
\bjtitle{Science}
\bvolume{361}(\bissue{6398}),
\bfpage{1378}
(\byear{2018}a).
doi:\doiurl{10.1126/science.aat1378}
\end{barticle}
\endbibitem

\bibitem[\protect\citeauthoryear{Aartsen et~al.}{2018b}]{IceCube:2018cha}
\begin{barticle}
\bauthor{\bsnm{Aartsen}, \binits{M.G.}}, \betal:
\bjtitle{Science}
\bvolume{361}(\bissue{6398}),
\bfpage{147}
(\byear{2018}b).
doi:\doiurl{10.1126/science.aat2890}
\end{barticle}
\endbibitem

\bibitem[\protect\citeauthoryear{Abbasi et~al.}{2009}]{Abbasi:2008aa}
\begin{barticle}
\bauthor{\bsnm{Abbasi}, \binits{R.}}, \betal:
\bjtitle{Nucl. Instrum. Meth.}
\bvolume{A601},
\bfpage{294}
(\byear{2009}).
\arxivurl{0810.4930}.
doi:\doiurl{10.1016/j.nima.2009.01.001}
\end{barticle}
\endbibitem

\bibitem[\protect\citeauthoryear{Abbasi et~al.}{2010}]{Abbasi:2010vc}
\begin{barticle}
\bauthor{\bsnm{Abbasi}, \binits{R.}}, \betal:
\bjtitle{Nucl. Instrum. Meth.}
\bvolume{A618},
\bfpage{139}
(\byear{2010}).
\arxivurl{1002.2442}.
doi:\doiurl{10.1016/j.nima.2010.03.102}
\end{barticle}
\endbibitem

\bibitem[\protect\citeauthoryear{Abbasi et~al.}{2012}]{Abbasi:2011ke}
\begin{barticle}
\bauthor{\bsnm{Abbasi}, \binits{R.}}, \betal:
\bjtitle{Astrophys. J.}
\bvolume{748},
\bfpage{118}
(\byear{2012}).
\arxivurl{1108.3023}.
doi:\doiurl{10.1088/0004-637X/748/2/118}
\end{barticle}
\endbibitem

\bibitem[\protect\citeauthoryear{Abdalla et~al.}{2018}]{H.E.S.S.:2018zkf}
\begin{barticle}
\bauthor{\bsnm{Abdalla}, \binits{H.}}, \betal:
\bjtitle{Astron. Astrophys.}
\bvolume{612},
\bfpage{1}
(\byear{2018}).
\arxivurl{1804.02432}.
doi:\doiurl{10.1051/0004-6361/201732098}
\end{barticle}
\endbibitem

\bibitem[\protect\citeauthoryear{Abdo et~al.}{2007}]{Abdo:2006fq}
\begin{barticle}
\bauthor{\bsnm{Abdo}, \binits{A.A.}}, \betal:
\bjtitle{Astrophys. J.}
\bvolume{658},
\bfpage{33}
(\byear{2007}).
\arxivurl{astro-ph/0611691}.
doi:\doiurl{10.1086/513696}
\end{barticle}
\endbibitem

\bibitem[\protect\citeauthoryear{{Abdo} et~al.}{2007}]{2007ApJ...658L..33A}
\begin{barticle}
\bauthor{\bsnm{{Abdo}}, \binits{A.A.}},
\bauthor{\bsnm{{Allen}}, \binits{B.}},
\bauthor{\bsnm{{Berley}}, \binits{D.}},
\bauthor{\bsnm{{Blaufuss}}, \binits{E.}},
\bauthor{\bsnm{{Casanova}}, \binits{S.}},
\bauthor{\bsnm{{Chen}}, \binits{C.}},
\bauthor{\bsnm{{Coyne}}, \binits{D.G.}},
\bauthor{\bsnm{{Delay}}, \binits{R.S.}},
\bauthor{\bsnm{{Dingus}}, \binits{B.L.}},
\bauthor{\bsnm{{Ellsworth}}, \binits{R.W.}},
\bauthor{\bsnm{{Fleysher}}, \binits{L.}},
\bauthor{\bsnm{{Fleysher}}, \binits{R.}},
\bauthor{\bsnm{{Gebauer}}, \binits{I.}},
\bauthor{\bsnm{{Gonzalez}}, \binits{M.M.}},
\bauthor{\bsnm{{Goodman}}, \binits{J.A.}},
\bauthor{\bsnm{{Hays}}, \binits{E.}},
\bauthor{\bsnm{{Hoffman}}, \binits{C.M.}},
\bauthor{\bsnm{{Kolterman}}, \binits{B.E.}},
\bauthor{\bsnm{{Kelley}}, \binits{L.A.}},
\bauthor{\bsnm{{Lansdell}}, \binits{C.P.}},
\bauthor{\bsnm{{Linnemann}}, \binits{J.T.}},
\bauthor{\bsnm{{McEnery}}, \binits{J.E.}},
\bauthor{\bsnm{{Mincer}}, \binits{A.I.}},
\bauthor{\bsnm{{Moskalenko}}, \binits{I.V.}},
\bauthor{\bsnm{{Nemethy}}, \binits{P.}},
\bauthor{\bsnm{{Noyes}}, \binits{D.}},
\bauthor{\bsnm{{Ryan}}, \binits{J.M.}},
\bauthor{\bsnm{{Samuelson}}, \binits{F.W.}},
\bauthor{\bsnm{{Saz Parkinson}}, \binits{P.M.}},
\bauthor{\bsnm{{Schneider}}, \binits{M.}},
\bauthor{\bsnm{{Shoup}}, \binits{A.}},
\bauthor{\bsnm{{Sinnis}}, \binits{G.}},
\bauthor{\bsnm{{Smith}}, \binits{A.J.}},
\bauthor{\bsnm{{Strong}}, \binits{A.W.}},
\bauthor{\bsnm{{Sullivan}}, \binits{G.W.}},
\bauthor{\bsnm{{Vasileiou}}, \binits{V.}},
\bauthor{\bsnm{{Walker}}, \binits{G.P.}},
\bauthor{\bsnm{{Williams}}, \binits{D.A.}},
\bauthor{\bsnm{{Xu}}, \binits{X.W.}},
\bauthor{\bsnm{{Yodh}}, \binits{G.B.}}:
\bjtitle{\apjl}
\bvolume{658}(\bissue{1}),
\bfpage{33}
(\byear{2007}).
\arxivurl{astro-ph/0611691}.
doi:\doiurl{10.1086/513696}
\end{barticle}
\endbibitem

\bibitem[\protect\citeauthoryear{Abdo and Abdo}{2010}]{Abdo:2010ht}
\begin{barticle}
\bauthor{\bsnm{Abdo}, \binits{A.A.}},
\bauthor{\bsnm{Abdo}, \binits{A.A.}}:
\bjtitle{Astrophys.J.}
\bvolume{711},
\bfpage{64}
(\byear{2010}).
\arxivurl{1001.0792}.
doi:\doiurl{10.1088/0004-637X/711/1/64}
\end{barticle}
\endbibitem

\bibitem[\protect\citeauthoryear{Abdo et~al.}{2007}]{Abdo:2007ad}
\begin{barticle}
\bauthor{\bsnm{Abdo}, \binits{A.A.}},
\bauthor{\bsnm{Allen}, \binits{B.T.}},
\bauthor{\bsnm{Berley}, \binits{D.}},
\bauthor{\bsnm{Casanova}, \binits{S.}},
\bauthor{\bsnm{Chen}, \binits{C.}}, \betal:
\bjtitle{Astrophys.J.}
\bvolume{664},
\bfpage{91}
(\byear{2007}).
\arxivurl{0705.0707}.
doi:\doiurl{10.1086/520717}
\end{barticle}
\endbibitem

\bibitem[\protect\citeauthoryear{Abdo et~al.}{2009}]{Abdo:2009ku}
\begin{barticle}
\bauthor{\bsnm{Abdo}, \binits{A.A.}},
\bauthor{\bsnm{Allen}, \binits{B.T.}},
\bauthor{\bsnm{Aune}, \binits{T.}},
\bauthor{\bsnm{Berley}, \binits{D.}},
\bauthor{\bsnm{Chen}, \binits{C.}}, \betal:
\bjtitle{Astrophys.J.}
\bvolume{700},
\bfpage{127}
(\byear{2009}).
\arxivurl{0904.1018}.
doi:\doiurl{10.1088/0004-637X/700/2/L127, 10.1088/0004-637X/703/2/L185}
\end{barticle}
\endbibitem

\bibitem[\protect\citeauthoryear{Abdo et~al.}{2012}]{Abdo:2012jg}
\begin{barticle}
\bauthor{\bsnm{Abdo}, \binits{A.A.}},
\bauthor{\bsnm{Abeysekara}, \binits{U.}},
\bauthor{\bsnm{Allen}, \binits{B.T.}},
\bauthor{\bsnm{Aune}, \binits{T.}},
\bauthor{\bsnm{Berley}, \binits{D.}}, \betal:
\bjtitle{Astrophys.J.}
\bvolume{753},
\bfpage{159}
(\byear{2012}).
\arxivurl{1202.0846}.
doi:\doiurl{10.1088/0004-637X/753/2/159}
\end{barticle}
\endbibitem

\bibitem[\protect\citeauthoryear{Abeysekara et~al.}{2016}]{Abeysekara:2015qba}
\begin{barticle}
\bauthor{\bsnm{Abeysekara}, \binits{A.U.}}, \betal:
\bjtitle{Astrophys. J.}
\bvolume{817}(\bissue{1}),
\bfpage{3}
(\byear{2016}).
\arxivurl{1509.05401}.
doi:\doiurl{10.3847/0004-637X/817/1/3}
\end{barticle}
\endbibitem

\bibitem[\protect\citeauthoryear{Abeysekara et~al.}{2017}]{Abeysekara:2017hyn}
\begin{barticle}
\bauthor{\bsnm{Abeysekara}, \binits{A.U.}}, \betal:
\bjtitle{Astrophys. J.}
\bvolume{843}(\bissue{1}),
\bfpage{40}
(\byear{2017}).
\arxivurl{1702.02992}.
doi:\doiurl{10.3847/1538-4357/aa7556}
\end{barticle}
\endbibitem

\bibitem[\protect\citeauthoryear{Abeysekara et~al.}{2020}]{Abeysekara:2019gov}
\begin{barticle}
\bauthor{\bsnm{Abeysekara}, \binits{A.U.}}, \betal:
\bjtitle{Phys. Rev. Lett.}
\bvolume{124}(\bissue{2}),
\bfpage{021102}
(\byear{2020}).
\arxivurl{1909.08609}.
doi:\doiurl{10.1103/PhysRevLett.124.021102}
\end{barticle}
\endbibitem

\bibitem[\protect\citeauthoryear{Abramowski et~al.}{2016}]{Abramowski:2016mir}
\begin{barticle}
\bauthor{\bsnm{Abramowski}, \binits{A.}}, \betal:
\bjtitle{Nature}
\bvolume{531},
\bfpage{476}
(\byear{2016}).
\arxivurl{1603.07730}.
doi:\doiurl{10.1038/nature17147}
\end{barticle}
\endbibitem

\bibitem[\protect\citeauthoryear{Acero et~al.}{2016}]{Acero:2016qlg}
\begin{barticle}
\bauthor{\bsnm{Acero}, \binits{F.}}, \betal:
\bjtitle{Astrophys. J. Suppl.}
\bvolume{223}(\bissue{2}),
\bfpage{26}
(\byear{2016}).
\arxivurl{1602.07246}.
doi:\doiurl{10.3847/0067-0049/223/2/26}
\end{barticle}
\endbibitem

\bibitem[\protect\citeauthoryear{Acharya et~al.}{2013}]{Acharya:2013sxa}
\begin{barticle}
\bauthor{\bsnm{Acharya}, \binits{B.S.}}, \betal:
\bjtitle{Astropart. Phys.}
\bvolume{43},
\bfpage{3}
(\byear{2013}).
doi:\doiurl{10.1016/j.astropartphys.2013.01.007}
\end{barticle}
\endbibitem

\bibitem[\protect\citeauthoryear{Acharya et~al.}{2018}]{Acharya:2017ttl}
\begin{bbook}
\bauthor{\bsnm{Acharya}, \binits{B.S.}}, \betal:
\bbtitle{{Science with the Cherenkov Telescope Array}}.
\bpublisher{WSP}, \blocation{???}
(\byear{2018}).
\arxivurl{1709.07997}.
doi:\doiurl{10.1142/10986}
\end{bbook}
\endbibitem

\bibitem[\protect\citeauthoryear{Achterberg et~al.}{2006}]{Achterberg:2006ik}
\begin{barticle}
\bauthor{\bsnm{Achterberg}, \binits{A.}}, \betal:
\bjtitle{Astropart. Phys.}
\bvolume{26},
\bfpage{282}
(\byear{2006}).
\arxivurl{astro-ph/0609534}.
doi:\doiurl{10.1016/J.ASTROPARTPHYS.2006.06.012}
\end{barticle}
\endbibitem

\bibitem[\protect\citeauthoryear{Ackermann et~al.}{2011}]{Ackermann:2011lfa}
\begin{barticle}
\bauthor{\bsnm{Ackermann}, \binits{M.}}, \betal:
\bjtitle{Science}
\bvolume{334}(\bissue{6059}),
\bfpage{1103}
(\byear{2011}).
doi:\doiurl{10.1126/science.1210311}
\end{barticle}
\endbibitem

\bibitem[\protect\citeauthoryear{Ackermann et~al.}{2012a}]{Ackermann:2012pya}
\begin{barticle}
\bauthor{\bsnm{Ackermann}, \binits{M.}}, \betal:
\bjtitle{Astrophys. J.}
\bvolume{750},
\bfpage{3}
(\byear{2012}a).
\arxivurl{1202.4039}.
doi:\doiurl{10.1088/0004-637X/750/1/3}
\end{barticle}
\endbibitem

\bibitem[\protect\citeauthoryear{Ackermann et~al.}{2012b}]{FermiLAT:2011lax}
\begin{barticle}
\bauthor{\bsnm{Ackermann}, \binits{M.}}, \betal:
\bjtitle{Astron. Astrophys.}
\bvolume{538},
\bfpage{71}
(\byear{2012}b).
\arxivurl{1110.6123}.
doi:\doiurl{10.1051/0004-6361/201117539}
\end{barticle}
\endbibitem

\bibitem[\protect\citeauthoryear{Ackermann et~al.}{2013}]{Ackermann:2013wqa}
\begin{barticle}
\bauthor{\bsnm{Ackermann}, \binits{M.}}, \betal:
\bjtitle{Science}
\bvolume{339},
\bfpage{807}
(\byear{2013}).
\arxivurl{1302.3307}.
doi:\doiurl{10.1126/science.1231160}
\end{barticle}
\endbibitem

\bibitem[\protect\citeauthoryear{Ackermann et~al.}{2014}]{Fermi-LAT:2014sfa}
\begin{barticle}
\bauthor{\bsnm{Ackermann}, \binits{M.}}, \betal:
\bjtitle{Astrophys.\ J.}
\bvolume{793}(\bissue{1}),
\bfpage{64}
(\byear{2014}).
\arxivurl{1407.7905}.
doi:\doiurl{10.1088/0004-637X/793/1/64}
\end{barticle}
\endbibitem

\bibitem[\protect\citeauthoryear{Ackermann et~al.}{2015}]{Ackermann:2014usa}
\begin{barticle}
\bauthor{\bsnm{Ackermann}, \binits{M.}}, \betal:
\bjtitle{Astrophys.\ J.}
\bvolume{799},
\bfpage{86}
(\byear{2015}).
\arxivurl{1410.3696}.
doi:\doiurl{10.1088/0004-637X/799/1/86}
\end{barticle}
\endbibitem

\bibitem[\protect\citeauthoryear{Adrian-Martinez
  et~al.}{2014}]{Adrian-Martinez:2014ito}
\begin{barticle}
\bauthor{\bsnm{Adrian-Martinez}, \binits{S.}}, \betal:
\bjtitle{JHEAp}
\bvolume{3-4},
\bfpage{9}
(\byear{2014}).
\arxivurl{1402.1600}.
doi:\doiurl{10.1016/j.jheap.2014.06.002}
\end{barticle}
\endbibitem

\bibitem[\protect\citeauthoryear{Adrian-Martinez
  et~al.}{2016}]{Adrian-Martinez:2016fdl}
\begin{barticle}
\bauthor{\bsnm{Adrian-Martinez}, \binits{S.}}, \betal:
\bjtitle{J. Phys.}
\bvolume{G43}(\bissue{8}),
\bfpage{084001}
(\byear{2016}).
\arxivurl{1601.07459}.
doi:\doiurl{10.1088/0954-3899/43/8/084001}
\end{barticle}
\endbibitem

\bibitem[\protect\citeauthoryear{Aharonian}{2009}]{Aharonian:2009je}
\begin{barticle}
\bauthor{\bsnm{Aharonian}, \binits{F.}}:
\bjtitle{Astron.Astrophys.}
\bvolume{499},
\bfpage{723}
(\byear{2009}).
\arxivurl{0904.3409}.
doi:\doiurl{10.1051/0004-6361/200811357}
\end{barticle}
\endbibitem

\bibitem[\protect\citeauthoryear{Aharonian et~al.}{2002}]{Aharonian:2002ij}
\begin{barticle}
\bauthor{\bsnm{Aharonian}, \binits{F.}}, \betal:
\bjtitle{Astron. Astrophys.}
\bvolume{393},
\bfpage{37}
(\byear{2002}).
\arxivurl{astro-ph/0207528}.
doi:\doiurl{10.1051/0004-6361:20021171}
\end{barticle}
\endbibitem

\bibitem[\protect\citeauthoryear{Aharonian et~al.}{2019}]{Aharonian:2018oau}
\begin{barticle}
\bauthor{\bsnm{Aharonian}, \binits{F.}},
\bauthor{\bsnm{Yang}, \binits{R.}},
\bauthor{\bparticle{de} \bsnm{Oña~Wilhelmi}, \binits{E.}}:
\bjtitle{Nat. Astron.}
\bvolume{3}(\bissue{6}),
\bfpage{561}
(\byear{2019}).
\arxivurl{1804.02331}.
doi:\doiurl{10.1038/s41550-019-0724-0}
\end{barticle}
\endbibitem

\bibitem[\protect\citeauthoryear{Aharonian et~al.}{2006}]{Aharonian:2005cx}
\begin{barticle}
\bauthor{\bsnm{Aharonian}, \binits{F.A.}},
\bauthor{\bsnm{Anchordoqui}, \binits{L.A.}},
\bauthor{\bsnm{Khangulyan}, \binits{D.}},
\bauthor{\bsnm{Montaruli}, \binits{T.}}:
\bjtitle{J. Phys. Conf. Ser.}
\bvolume{39},
\bfpage{408}
(\byear{2006}).
\arxivurl{astro-ph/0508658}.
doi:\doiurl{10.1088/1742-6596/39/1/106}
\end{barticle}
\endbibitem

\bibitem[\protect\citeauthoryear{Ahlers and Halzen}{2012}]{Ahlers:2012rz}
\begin{barticle}
\bauthor{\bsnm{Ahlers}, \binits{M.}},
\bauthor{\bsnm{Halzen}, \binits{F.}}:
\bjtitle{Phys.\ Rev.\ D}
\bvolume{86},
\bfpage{083010}
(\byear{2012}).
\arxivurl{1208.4181}.
doi:\doiurl{10.1103/PhysRevD.86.083010}
\end{barticle}
\endbibitem

\bibitem[\protect\citeauthoryear{Ahlers and Halzen}{2014}]{Ahlers:2014ioa}
\begin{barticle}
\bauthor{\bsnm{Ahlers}, \binits{M.}},
\bauthor{\bsnm{Halzen}, \binits{F.}}:
\bjtitle{Phys. Rev.}
\bvolume{D90}(\bissue{4}),
\bfpage{043005}
(\byear{2014}).
\arxivurl{1406.2160}.
doi:\doiurl{10.1103/PhysRevD.90.043005}
\end{barticle}
\endbibitem

\bibitem[\protect\citeauthoryear{Ahlers and Murase}{2014}]{Ahlers:2013xia}
\begin{barticle}
\bauthor{\bsnm{Ahlers}, \binits{M.}},
\bauthor{\bsnm{Murase}, \binits{K.}}:
\bjtitle{Phys. Rev.}
\bvolume{D90}(\bissue{2}),
\bfpage{023010}
(\byear{2014}).
\arxivurl{1309.4077}.
doi:\doiurl{10.1103/PhysRevD.90.023010}
\end{barticle}
\endbibitem

\bibitem[\protect\citeauthoryear{Ahlers et~al.}{2016}]{Ahlers:2015moa}
\begin{barticle}
\bauthor{\bsnm{Ahlers}, \binits{M.}},
\bauthor{\bsnm{Bai}, \binits{Y.}},
\bauthor{\bsnm{Barger}, \binits{V.}},
\bauthor{\bsnm{Lu}, \binits{R.}}:
\bjtitle{Phys. Rev.}
\bvolume{D93}(\bissue{1}),
\bfpage{013009}
(\byear{2016}).
\arxivurl{1505.03156}.
doi:\doiurl{10.1103/PhysRevD.93.013009}
\end{barticle}
\endbibitem

\bibitem[\protect\citeauthoryear{Aiello et~al.}{2019}]{Aiello:2018usb}
\begin{barticle}
\bauthor{\bsnm{Aiello}, \binits{S.}}, \betal:
\bjtitle{Astropart. Phys.}
\bvolume{111},
\bfpage{100}
(\byear{2019}).
\arxivurl{1810.08499}.
doi:\doiurl{10.1016/j.astropartphys.2019.04.002}
\end{barticle}
\endbibitem

\bibitem[\protect\citeauthoryear{Albert et~al.}{2017a}]{Albert:2017oba}
\begin{barticle}
\bauthor{\bsnm{Albert}, \binits{A.}}, \betal:
\bjtitle{Phys.\ Rev.\ D}
\bvolume{96}(\bissue{6}),
\bfpage{062001}
(\byear{2017}a).
\arxivurl{1705.00497}.
doi:\doiurl{10.1103/PhysRevD.96.062001}
\end{barticle}
\endbibitem

\bibitem[\protect\citeauthoryear{Albert et~al.}{2017b}]{Albert:2017bdv}
\begin{botherref}
\oauthor{\bsnm{Albert}, \binits{A.}}, et al.:
{The ANTARES Collaboration: Contributions to ICRC 2017 Part I: Neutrino
  astronomy (diffuse fluxes and point sources)}
(2017b).
\arxivurl{1711.01251}
\end{botherref}
\endbibitem

\bibitem[\protect\citeauthoryear{Albert et~al.}{2018}]{Albert:2018vxw}
\begin{barticle}
\bauthor{\bsnm{Albert}, \binits{A.}}, \betal:
\bjtitle{Astrophys. J.}
\bvolume{868}(\bissue{2}),
\bfpage{20}
(\byear{2018}).
\arxivurl{1808.03531}.
doi:\doiurl{10.3847/2041-8213/aaeecf}
\end{barticle}
\endbibitem

\bibitem[\protect\citeauthoryear{Albert et~al.}{2019}]{Albert:2019afb}
\begin{botherref}
\oauthor{\bsnm{Albert}, \binits{A.}}, et al.:
{Science Case for a Wide Field-of-View Very-High-Energy Gamma-Ray Observatory
  in the Southern Hemisphere}
(2019).
\arxivurl{1902.08429}
\end{botherref}
\endbibitem

\bibitem[\protect\citeauthoryear{Albert et~al.}{2008}]{Albert:2008yk}
\begin{barticle}
\bauthor{\bsnm{Albert}, \binits{J.}}, \betal:
\bjtitle{Astrophys. J.}
\bvolume{675},
\bfpage{25}
(\byear{2008}).
\arxivurl{0801.2391}.
doi:\doiurl{10.1086/529520}
\end{barticle}
\endbibitem

\bibitem[\protect\citeauthoryear{Aliu et~al.}{2014a}]{Aliu:2014rha}
\begin{barticle}
\bauthor{\bsnm{Aliu}, \binits{E.}}, \betal:
\bjtitle{Astrophys. J.}
\bvolume{788},
\bfpage{78}
(\byear{2014}a).
\arxivurl{1404.1841}.
doi:\doiurl{10.1088/0004-637X/788/1/78}
\end{barticle}
\endbibitem

\bibitem[\protect\citeauthoryear{Aliu et~al.}{2014b}]{Aliu:2014xra}
\begin{barticle}
\bauthor{\bsnm{Aliu}, \binits{E.}},
\bauthor{\bsnm{Archambault}, \binits{S.}},
\bauthor{\bsnm{Aune}, \binits{T.}},
\bauthor{\bsnm{Behera}, \binits{B.}},
\bauthor{\bsnm{Beilicke}, \binits{M.}}, \betal:
\bjtitle{Astrophys.J.}
\bvolume{787},
\bfpage{166}
(\byear{2014}b).
\arxivurl{1404.7185}.
doi:\doiurl{10.1088/0004-637X/787/2/166}
\end{barticle}
\endbibitem

\bibitem[\protect\citeauthoryear{Alvarez-Muniz and
  Halzen}{2002}]{AlvarezMuniz:2002yr}
\begin{barticle}
\bauthor{\bsnm{Alvarez-Muniz}, \binits{J.}},
\bauthor{\bsnm{Halzen}, \binits{F.}}:
\bjtitle{Astrophys.J.}
\bvolume{576},
\bfpage{33}
(\byear{2002}).
\arxivurl{astro-ph/0205408}.
doi:\doiurl{10.1086/342978}
\end{barticle}
\endbibitem

\bibitem[\protect\citeauthoryear{Amato and Arons}{2006}]{Amato:2006ts}
\begin{barticle}
\bauthor{\bsnm{Amato}, \binits{E.}},
\bauthor{\bsnm{Arons}, \binits{J.}}:
\bjtitle{Astrophys. J.}
\bvolume{653},
\bfpage{325}
(\byear{2006}).
\arxivurl{astro-ph/0609034}.
doi:\doiurl{10.1086/508050}
\end{barticle}
\endbibitem

\bibitem[\protect\citeauthoryear{Amato}{2014}]{Amato:2013fua}
\begin{barticle}
\bauthor{\bsnm{Amato}, \binits{E.}}:
\bjtitle{Int. J. Mod. Phys. Conf. Ser.}
\bvolume{28},
\bfpage{1460160}
(\byear{2014}).
\arxivurl{1312.5945}.
doi:\doiurl{10.1142/S2010194514601604}
\end{barticle}
\endbibitem

\bibitem[\protect\citeauthoryear{Amato et~al.}{2003}]{Amato:2003kw}
\begin{barticle}
\bauthor{\bsnm{Amato}, \binits{E.}},
\bauthor{\bsnm{Guetta}, \binits{D.}},
\bauthor{\bsnm{Blasi}, \binits{P.}}:
\bjtitle{Astron. Astrophys.}
\bvolume{402},
\bfpage{827}
(\byear{2003}).
\arxivurl{astro-ph/0302121}.
doi:\doiurl{10.1051/0004-6361:20030279}
\end{barticle}
\endbibitem

\bibitem[\protect\citeauthoryear{Anchordoqui}{2016}]{Anchordoqui:2016dcp}
\begin{barticle}
\bauthor{\bsnm{Anchordoqui}, \binits{L.A.}}:
\bjtitle{Phys. Rev.}
\bvolume{D94},
\bfpage{023010}
(\byear{2016}).
\arxivurl{1606.01816}.
doi:\doiurl{10.1103/PhysRevD.94.023010}
\end{barticle}
\endbibitem

\bibitem[\protect\citeauthoryear{Anchordoqui
  et~al.}{2014}]{Anchordoqui:2014rca}
\begin{barticle}
\bauthor{\bsnm{Anchordoqui}, \binits{L.A.}},
\bauthor{\bsnm{Goldberg}, \binits{H.}},
\bauthor{\bsnm{Paul}, \binits{T.C.}},
\bauthor{\bparticle{da} \bsnm{Silva}, \binits{L.H.M.}},
\bauthor{\bsnm{Vlcek}, \binits{B.J.}}:
\bjtitle{Phys. Rev.}
\bvolume{D90}(\bissue{12}),
\bfpage{123010}
(\byear{2014}).
\arxivurl{1410.0348}.
doi:\doiurl{10.1103/PhysRevD.90.123010}
\end{barticle}
\endbibitem

\bibitem[\protect\citeauthoryear{Antoni et~al.}{2005}]{Antoni:2005wq}
\begin{barticle}
\bauthor{\bsnm{Antoni}, \binits{T.}}, \betal:
\bjtitle{Astropart.\ Phys.}
\bvolume{24},
\bfpage{1}
(\byear{2005}).
\arxivurl{astro-ph/0505413}.
doi:\doiurl{10.1016/j.astropartphys.2005.04.001}
\end{barticle}
\endbibitem

\bibitem[\protect\citeauthoryear{Apel et~al.}{2013}]{Apel:2013uni}
\begin{barticle}
\bauthor{\bsnm{Apel}, \binits{W.D.}}, \betal:
\bjtitle{Astropart.\ Phys.}
\bvolume{47},
\bfpage{54}
(\byear{2013}).
\arxivurl{1306.6283}.
doi:\doiurl{10.1016/j.astropartphys.2013.06.004}
\end{barticle}
\endbibitem

\bibitem[\protect\citeauthoryear{Arg\"uelles et~al.}{2017}]{Arguelles:2017atb}
\begin{barticle}
\bauthor{\bsnm{Arg\"uelles}, \binits{C.A.}},
\bauthor{\bsnm{Kheirandish}, \binits{A.}},
\bauthor{\bsnm{Vincent}, \binits{A.C.}}:
\bjtitle{Phys. Rev. Lett.}
\bvolume{119}(\bissue{20}),
\bfpage{201801}
(\byear{2017}).
\arxivurl{1703.00451}.
doi:\doiurl{10.1103/PhysRevLett.119.201801}
\end{barticle}
\endbibitem

\bibitem[\protect\citeauthoryear{Arg\"uelles et~al.}{2019}]{Arguelles:2019ouk}
\begin{botherref}
\oauthor{\bsnm{Arg\"uelles}, \binits{C.A.}},
\oauthor{\bsnm{Diaz}, \binits{A.}},
\oauthor{\bsnm{Kheirandish}, \binits{A.}},
\oauthor{\bsnm{Olivares-Del-Campo}, \binits{A.}},
\oauthor{\bsnm{Safa}, \binits{I.}},
\oauthor{\bsnm{Vincent}, \binits{A.C.}}:
{Dark Matter Annihilation to Neutrinos: An Updated, Consistent \& Compelling
  Compendium of Constraints}
(2019).
\arxivurl{1912.09486}
\end{botherref}
\endbibitem

\bibitem[\protect\citeauthoryear{Avrorin et~al.}{2019}]{Avrorin:2019vfc}
\begin{bchapter}
\bauthor{\bsnm{Avrorin}, \binits{A.D.}}, \betal:
In: \bbtitle{{HAWC Contributions to the 36th International Cosmic Ray
  Conference (ICRC2019)}},
\byear{2019}.
\arxivurl{1908.05450}
\end{bchapter}
\endbibitem

\bibitem[\protect\citeauthoryear{Baade and Zwicky}{1934}]{BaadeAndZwicky}
\begin{barticle}
\bauthor{\bsnm{Baade}, \binits{W.}},
\bauthor{\bsnm{Zwicky}, \binits{F.}}:
\bjtitle{Proceedings of the National Academy of Sciences}
\bvolume{20}(\bissue{5}),
\bfpage{254}
(\byear{1934}).
doi:\doiurl{10.1073/pnas.20.5.254}
\end{barticle}
\endbibitem

\bibitem[\protect\citeauthoryear{{Baars} and
  {Wendker}}{1981}]{1981A&A...101...39B}
\begin{barticle}
\bauthor{\bsnm{{Baars}}, \binits{J.W.M.}},
\bauthor{\bsnm{{Wendker}}, \binits{H.J.}}:
\bjtitle{\aap}
\bvolume{101},
\bfpage{39}
(\byear{1981})
\end{barticle}
\endbibitem

\bibitem[\protect\citeauthoryear{Bai et~al.}{2016}]{Bai:2013nga}
\begin{barticle}
\bauthor{\bsnm{Bai}, \binits{Y.}},
\bauthor{\bsnm{Lu}, \binits{R.}},
\bauthor{\bsnm{Salvado}, \binits{J.}}:
\bjtitle{JHEP}
\bvolume{01},
\bfpage{161}
(\byear{2016}).
\arxivurl{1311.5864}.
doi:\doiurl{10.1007/JHEP01(2016)161}
\end{barticle}
\endbibitem

\bibitem[\protect\citeauthoryear{Bartoli et~al.}{2012}]{ARGO-YBJ:2012goa}
\begin{barticle}
\bauthor{\bsnm{Bartoli}, \binits{B.}}, \betal:
\bjtitle{Astrophys. J.}
\bvolume{760},
\bfpage{110}
(\byear{2012}).
\arxivurl{1207.6280}.
doi:\doiurl{10.1088/0004-637X/760/2/110}
\end{barticle}
\endbibitem

\bibitem[\protect\citeauthoryear{Bartoli et~al.}{2014a}]{Argo:2014tqa}
\begin{barticle}
\bauthor{\bsnm{Bartoli}, \binits{B.}}, \betal:
\bjtitle{Astrophys. J.}
\bvolume{790}(\bissue{2}),
\bfpage{152}
(\byear{2014}a).
\arxivurl{1406.6436}.
doi:\doiurl{10.1088/0004-637X/790/2/152}
\end{barticle}
\endbibitem

\bibitem[\protect\citeauthoryear{Bartoli et~al.}{2014b}]{Bartoli:2014irw}
\begin{barticle}
\bauthor{\bsnm{Bartoli}, \binits{B.}}, \betal:
\bjtitle{Astrophys. J.}
\bvolume{790}(\bissue{2}),
\bfpage{152}
(\byear{2014}b).
\arxivurl{1406.6436}.
doi:\doiurl{10.1088/0004-637X/790/2/152}
\end{barticle}
\endbibitem

\bibitem[\protect\citeauthoryear{Bartoli et~al.}{2015}]{Bartoli:2015era}
\begin{barticle}
\bauthor{\bsnm{Bartoli}, \binits{B.}}, \betal:
\bjtitle{Astrophys.\ J.}
\bvolume{806},
\bfpage{20}
(\byear{2015}).
\arxivurl{1507.06758}.
doi:\doiurl{10.1088/0004-637X/806/1/20}
\end{barticle}
\endbibitem

\bibitem[\protect\citeauthoryear{Bartoli et~al.}{2012}]{Bartoli:2012tj}
\begin{barticle}
\bauthor{\bsnm{Bartoli}, \binits{B.}},
\bauthor{\bsnm{Bernardini}, \binits{P.}},
\bauthor{\bsnm{Bi}, \binits{X.J.}},
\bauthor{\bsnm{Bleve}, \binits{C.}},
\bauthor{\bsnm{Bolognino}, \binits{I.}}, \betal:
\bjtitle{Astrophys.J.}
\bvolume{745},
\bfpage{22}
(\byear{2012}).
\arxivurl{1201.1973}.
doi:\doiurl{10.1088/2041-8205/745/2/L22}
\end{barticle}
\endbibitem

\bibitem[\protect\citeauthoryear{Beacom and Kistler}{2007}]{Beacom:2007yu}
\begin{barticle}
\bauthor{\bsnm{Beacom}, \binits{J.F.}},
\bauthor{\bsnm{Kistler}, \binits{M.D.}}:
\bjtitle{Phys. Rev.}
\bvolume{D75},
\bfpage{083001}
(\byear{2007}).
\arxivurl{astro-ph/0701751}.
doi:\doiurl{10.1103/PhysRevD.75.083001}
\end{barticle}
\endbibitem

\bibitem[\protect\citeauthoryear{Beacom et~al.}{2007}]{Beacom:2006tt}
\begin{barticle}
\bauthor{\bsnm{Beacom}, \binits{J.F.}},
\bauthor{\bsnm{Bell}, \binits{N.F.}},
\bauthor{\bsnm{Mack}, \binits{G.D.}}:
\bjtitle{Phys. Rev. Lett.}
\bvolume{99},
\bfpage{231301}
(\byear{2007}).
\arxivurl{astro-ph/0608090}.
doi:\doiurl{10.1103/PhysRevLett.99.231301}
\end{barticle}
\endbibitem

\bibitem[\protect\citeauthoryear{Becker~Tjus and
  Merten}{2020}]{BeckerTjus:2020xzg}
\begin{botherref}
\oauthor{\bsnm{Becker~Tjus}, \binits{J.}},
\oauthor{\bsnm{Merten}, \binits{L.}}:
{Closing in on the origin of cosmic rays using multimessenger information}
(2020).
\arxivurl{2002.00964}
\end{botherref}
\endbibitem

\bibitem[\protect\citeauthoryear{Bednarek}{2003}]{Bednarek:2003cv}
\begin{barticle}
\bauthor{\bsnm{Bednarek}, \binits{W.}}:
\bjtitle{Astron. Astrophys.}
\bvolume{407},
\bfpage{1}
(\byear{2003}).
\arxivurl{astro-ph/0305430}.
doi:\doiurl{10.1051/0004-6361:20030929}
\end{barticle}
\endbibitem

\bibitem[\protect\citeauthoryear{Bednarek and
  Protheroe}{1997}]{Bednarek:1997cn}
\begin{barticle}
\bauthor{\bsnm{Bednarek}, \binits{W.}},
\bauthor{\bsnm{Protheroe}, \binits{R.J.}}:
\bjtitle{Phys. Rev. Lett.}
\bvolume{79},
\bfpage{2616}
(\byear{1997}).
\arxivurl{astro-ph/9704186}.
doi:\doiurl{10.1103/PhysRevLett.79.2616}
\end{barticle}
\endbibitem

\bibitem[\protect\citeauthoryear{Berezinsky et~al.}{1993}]{Berezinsky:1992wr}
\begin{barticle}
\bauthor{\bsnm{Berezinsky}, \binits{V.S.}},
\bauthor{\bsnm{Gaisser}, \binits{T.K.}},
\bauthor{\bsnm{Halzen}, \binits{F.}},
\bauthor{\bsnm{Stanev}, \binits{T.}}:
\bjtitle{Astropart.\ Phys.}
\bvolume{1},
\bfpage{281}
(\byear{1993}).
doi:\doiurl{10.1016/0927-6505(93)90014-5}
\end{barticle}
\endbibitem

\bibitem[\protect\citeauthoryear{Bhattacharya
  et~al.}{2019}]{Bhattacharya:2019ucd}
\begin{barticle}
\bauthor{\bsnm{Bhattacharya}, \binits{A.}},
\bauthor{\bsnm{Esmaili}, \binits{A.}},
\bauthor{\bsnm{Palomares-Ruiz}, \binits{S.}},
\bauthor{\bsnm{Sarcevic}, \binits{I.}}:
\bjtitle{JCAP}
\bvolume{1905}(\bissue{05}),
\bfpage{051}
(\byear{2019}).
\arxivurl{1903.12623}.
doi:\doiurl{10.1088/1475-7516/2019/05/051}
\end{barticle}
\endbibitem

\bibitem[\protect\citeauthoryear{Bustard et~al.}{2017}]{Bustard:2016swa}
\begin{barticle}
\bauthor{\bsnm{Bustard}, \binits{C.}},
\bauthor{\bsnm{Zweibel}, \binits{E.G.}},
\bauthor{\bsnm{Cotter}, \binits{C.}}:
\bjtitle{Astrophys. J.}
\bvolume{835}(\bissue{1}),
\bfpage{72}
(\byear{2017}).
\arxivurl{1610.06565}.
doi:\doiurl{10.3847/1538-4357/835/1/72}
\end{barticle}
\endbibitem

\bibitem[\protect\citeauthoryear{Capanema et~al.}{2020}]{Capanema:2020rjj}
\begin{botherref}
\oauthor{\bsnm{Capanema}, \binits{A.}},
\oauthor{\bsnm{Esmaili}, \binits{A.}},
\oauthor{\bsnm{Murase}, \binits{K.}}:
{New Constraints on the Origin of Medium-Energy Neutrinos Observed by IceCube}
(2020).
\arxivurl{2002.07192}
\end{botherref}
\endbibitem

\bibitem[\protect\citeauthoryear{{Chen} et~al.}{1996}]{1996A&AS..120C.423C}
\begin{barticle}
\bauthor{\bsnm{{Chen}}, \binits{W.}},
\bauthor{\bsnm{{White}}, \binits{R.L.}},
\bauthor{\bsnm{{Bertsch}}, \binits{D.}}:
\bjtitle{\aaps}
\bvolume{120},
\bfpage{423}
(\byear{1996})
\end{barticle}
\endbibitem

\bibitem[\protect\citeauthoryear{Cheng et~al.}{1990}]{Cheng:1990au}
\begin{barticle}
\bauthor{\bsnm{Cheng}, \binits{K.S.}},
\bauthor{\bsnm{Cheung}, \binits{T.}},
\bauthor{\bsnm{Lau}, \binits{M.M.}},
\bauthor{\bsnm{Yu}, \binits{K.N.}},
\bauthor{\bsnm{Kwok}, \binits{P.W.}}:
\bjtitle{J. Phys.}
\bvolume{G16},
\bfpage{1115}
(\byear{1990}).
doi:\doiurl{10.1088/0954-3899/16/7/022}
\end{barticle}
\endbibitem

\bibitem[\protect\citeauthoryear{Crocker and Aharonian}{2011}]{Crocker:2010dg}
\begin{barticle}
\bauthor{\bsnm{Crocker}, \binits{R.M.}},
\bauthor{\bsnm{Aharonian}, \binits{F.}}:
\bjtitle{Phys.\ Rev.\ Lett.}
\bvolume{106},
\bfpage{101102}
(\byear{2011}).
\arxivurl{1008.2658}.
doi:\doiurl{10.1103/PhysRevLett.106.101102}
\end{barticle}
\endbibitem

\bibitem[\protect\citeauthoryear{Di~Sciascio}{2016}]{DiSciascio:2016rgi}
\begin{barticle}
\bauthor{\bsnm{Di~Sciascio}, \binits{G.}}:
\bjtitle{Nucl. Part. Phys. Proc.}
\bvolume{279-281},
\bfpage{166}
(\byear{2016}).
\arxivurl{1602.07600}.
doi:\doiurl{10.1016/j.nuclphysbps.2016.10.024}
\end{barticle}
\endbibitem

\bibitem[\protect\citeauthoryear{Domokos et~al.}{1993}]{Domokos:1991tt}
\begin{barticle}
\bauthor{\bsnm{Domokos}, \binits{G.}},
\bauthor{\bsnm{Elliott}, \binits{B.}},
\bauthor{\bsnm{Kovesi-Domokos}, \binits{S.}}:
\bjtitle{J.\ Phys.\ G}
\bvolume{19},
\bfpage{899}
(\byear{1993}).
doi:\doiurl{10.1088/0954-3899/19/6/010}
\end{barticle}
\endbibitem

\bibitem[\protect\citeauthoryear{Enberg et~al.}{2008}]{Enberg:2008te}
\begin{barticle}
\bauthor{\bsnm{Enberg}, \binits{R.}},
\bauthor{\bsnm{Reno}, \binits{M.H.}},
\bauthor{\bsnm{Sarcevic}, \binits{I.}}:
\bjtitle{Phys. Rev.}
\bvolume{D78},
\bfpage{043005}
(\byear{2008}).
\arxivurl{0806.0418}.
doi:\doiurl{10.1103/PhysRevD.78.043005}
\end{barticle}
\endbibitem

\bibitem[\protect\citeauthoryear{Evoli et~al.}{2007}]{Evoli:2007iy}
\begin{barticle}
\bauthor{\bsnm{Evoli}, \binits{C.}},
\bauthor{\bsnm{Grasso}, \binits{D.}},
\bauthor{\bsnm{Maccione}, \binits{L.}}:
\bjtitle{JCAP}
\bvolume{06},
\bfpage{003}
(\byear{2007}).
\arxivurl{astro-ph/0701856}.
doi:\doiurl{10.1088/1475-7516/2007/06/003}
\end{barticle}
\endbibitem

\bibitem[\protect\citeauthoryear{Fujita et~al.}{2017}]{Fujita:2016yvk}
\begin{barticle}
\bauthor{\bsnm{Fujita}, \binits{Y.}},
\bauthor{\bsnm{Murase}, \binits{K.}},
\bauthor{\bsnm{Kimura}, \binits{S.S.}}:
\bjtitle{JCAP}
\bvolume{04},
\bfpage{037}
(\byear{2017}).
\arxivurl{1604.00003}.
doi:\doiurl{10.1088/1475-7516/2017/04/037}
\end{barticle}
\endbibitem

\bibitem[\protect\citeauthoryear{Gaggero et~al.}{2017}]{Gaggero:2017jts}
\begin{barticle}
\bauthor{\bsnm{Gaggero}, \binits{D.}},
\bauthor{\bsnm{Grasso}, \binits{D.}},
\bauthor{\bsnm{Marinelli}, \binits{A.}},
\bauthor{\bsnm{Taoso}, \binits{M.}},
\bauthor{\bsnm{Urbano}, \binits{A.}}:
\bjtitle{Phys. Rev. Lett.}
\bvolume{119}(\bissue{3}),
\bfpage{031101}
(\byear{2017}).
\arxivurl{1702.01124}.
doi:\doiurl{10.1103/PhysRevLett.119.031101}
\end{barticle}
\endbibitem

\bibitem[\protect\citeauthoryear{Gaggero et~al.}{2015a}]{Gaggero:2014xla}
\begin{barticle}
\bauthor{\bsnm{Gaggero}, \binits{D.}},
\bauthor{\bsnm{Urbano}, \binits{A.}},
\bauthor{\bsnm{Valli}, \binits{M.}},
\bauthor{\bsnm{Ullio}, \binits{P.}}:
\bjtitle{Phys. Rev.}
\bvolume{D91}(\bissue{8}),
\bfpage{083012}
(\byear{2015}a).
\arxivurl{1411.7623}.
doi:\doiurl{10.1103/PhysRevD.91.083012}
\end{barticle}
\endbibitem

\bibitem[\protect\citeauthoryear{Gaggero et~al.}{2015b}]{Gaggero:2015xza}
\begin{barticle}
\bauthor{\bsnm{Gaggero}, \binits{D.}},
\bauthor{\bsnm{Grasso}, \binits{D.}},
\bauthor{\bsnm{Marinelli}, \binits{A.}},
\bauthor{\bsnm{Urbano}, \binits{A.}},
\bauthor{\bsnm{Valli}, \binits{M.}}:
\bjtitle{Astrophys. J.}
\bvolume{815}(\bissue{2}),
\bfpage{25}
(\byear{2015}b).
\arxivurl{1504.00227}.
doi:\doiurl{10.1088/2041-8205/815/2/L25}
\end{barticle}
\endbibitem

\bibitem[\protect\citeauthoryear{Gaisser}{1997}]{Gaisser1997}
\begin{botherref}
\oauthor{\bsnm{Gaisser}, \binits{T.K.}}:
{Neutrino astronomy: Physics goals, detector parameters}
(1997).
\arxivurl{astro-ph/9707283}
\end{botherref}
\endbibitem

\bibitem[\protect\citeauthoryear{Gallo et~al.}{2005}]{Gallo:2005tf}
\begin{barticle}
\bauthor{\bsnm{Gallo}, \binits{E.}},
\bauthor{\bsnm{Fender}, \binits{R.}},
\bauthor{\bsnm{Kaiser}, \binits{C.}},
\bauthor{\bsnm{Russell}, \binits{D.}},
\bauthor{\bsnm{Morganti}, \binits{R.}},
\bauthor{\bsnm{Oosterloo}, \binits{T.}},
\bauthor{\bsnm{Heinz}, \binits{S.}}:
\bjtitle{Nature}
\bvolume{436},
\bfpage{819}
(\byear{2005}).
\arxivurl{astro-ph/0508228}.
doi:\doiurl{10.1038/nature03879}
\end{barticle}
\endbibitem

\bibitem[\protect\citeauthoryear{Gonzalez-Garcia
  et~al.}{2014}]{Gonzalez-Garcia:2013iha}
\begin{barticle}
\bauthor{\bsnm{Gonzalez-Garcia}, \binits{M.C.}},
\bauthor{\bsnm{Halzen}, \binits{F.}},
\bauthor{\bsnm{Niro}, \binits{V.}}:
\bjtitle{Astropart. Phys.}
\bvolume{57-58},
\bfpage{39}
(\byear{2014}).
\arxivurl{1310.7194}.
doi:\doiurl{10.1016/j.astropartphys.2014.04.001}
\end{barticle}
\endbibitem

\bibitem[\protect\citeauthoryear{Gonzalez-Garcia
  et~al.}{2009}]{GonzalezGarcia:2009jc}
\begin{barticle}
\bauthor{\bsnm{Gonzalez-Garcia}, \binits{M.C.}},
\bauthor{\bsnm{Halzen}, \binits{F.}},
\bauthor{\bsnm{Mohapatra}, \binits{S.}}:
\bjtitle{Astropart. Phys.}
\bvolume{31},
\bfpage{437}
(\byear{2009}).
\arxivurl{0902.1176}.
doi:\doiurl{10.1016/j.astropartphys.2009.05.002}
\end{barticle}
\endbibitem

\bibitem[\protect\citeauthoryear{{Gottschalk}
  et~al.}{2012}]{2012A&A...541A..79G}
\begin{barticle}
\bauthor{\bsnm{{Gottschalk}}, \binits{M.}},
\bauthor{\bsnm{{Kothes}}, \binits{R.}},
\bauthor{\bsnm{{Matthews}}, \binits{H.E.}},
\bauthor{\bsnm{{Land ecker}}, \binits{T.L.}},
\bauthor{\bsnm{{Dent}}, \binits{W.R.F.}}:
\bjtitle{\aap}
\bvolume{541},
\bfpage{79}
(\byear{2012}).
\arxivurl{1202.0832}.
doi:\doiurl{10.1051/0004-6361/201118600}
\end{barticle}
\endbibitem

\bibitem[\protect\citeauthoryear{Guenduez et~al.}{2017}]{Guenduez:2017qrw}
\begin{botherref}
\oauthor{\bsnm{Guenduez}, \binits{M.}},
\oauthor{\bsnm{Becker~Tjus}, \binits{J.}},
\oauthor{\bsnm{Eichmann}, \binits{B.}},
\oauthor{\bsnm{Halzen}, \binits{F.}}:
{Identification of Gamma-Rays and Neutrinos from the Cygnus-X Complex
  Considering Radio Gamma Correlation}
(2017).
\arxivurl{1705.08337}
\end{botherref}
\endbibitem

\bibitem[\protect\citeauthoryear{Guetta et~al.}{2002}]{Guetta:2002hk}
\begin{bchapter}
\bauthor{\bsnm{Guetta}, \binits{D.}},
\bauthor{\bsnm{Distefano}, \binits{C.}},
\bauthor{\bsnm{Levinson}, \binits{A.}},
\bauthor{\bsnm{Waxman}, \binits{E.}}:
In: \bbtitle{{Proceedings, 4th Microquasar Workshop : Microquasars and their
  Relation to other Jet Sources in the Universe : New Views on Microquasars:
  Cargese, France, May 27-June 1, 2002}},
\byear{2002}.
\arxivurl{astro-ph/0207359}
\end{bchapter}
\endbibitem

\bibitem[\protect\citeauthoryear{Guetta and Amato}{2003}]{Guetta:2002hv}
\begin{barticle}
\bauthor{\bsnm{Guetta}, \binits{D.}},
\bauthor{\bsnm{Amato}, \binits{E.}}:
\bjtitle{Astropart. Phys.}
\bvolume{19},
\bfpage{403}
(\byear{2003}).
\arxivurl{astro-ph/0209537}.
doi:\doiurl{10.1016/S0927-6505(02)00221-9}
\end{barticle}
\endbibitem

\bibitem[\protect\citeauthoryear{Haack and Wiebusch}{2018}]{Haack:2017dxi}
\begin{barticle}
\bauthor{\bsnm{Haack}, \binits{C.}},
\bauthor{\bsnm{Wiebusch}, \binits{C.}}:
\bjtitle{PoS}
\bvolume{ICRC2017},
\bfpage{1005}
(\byear{2018}).
doi:\doiurl{10.22323/1.301.1005}
\end{barticle}
\endbibitem

\bibitem[\protect\citeauthoryear{Halzen and Klein}{2010}]{Halzen:2010yj}
\begin{barticle}
\bauthor{\bsnm{Halzen}, \binits{F.}},
\bauthor{\bsnm{Klein}, \binits{S.R.}}:
\bjtitle{Rev.\ Sci.\ Instrum.}
\bvolume{81},
\bfpage{081101}
(\byear{2010}).
\arxivurl{1007.1247}.
doi:\doiurl{10.1063/1.3480478}
\end{barticle}
\endbibitem

\bibitem[\protect\citeauthoryear{Halzen et~al.}{2008}]{Halzen:2008zj}
\begin{barticle}
\bauthor{\bsnm{Halzen}, \binits{F.}},
\bauthor{\bsnm{Kappes}, \binits{A.}},
\bauthor{\bsnm{O'Murchadha}, \binits{A.}}:
\bjtitle{Phys. Rev.}
\bvolume{D78},
\bfpage{063004}
(\byear{2008}).
\arxivurl{0803.0314}.
doi:\doiurl{10.1103/PhysRevD.78.063004}
\end{barticle}
\endbibitem

\bibitem[\protect\citeauthoryear{Halzen et~al.}{2017}]{Halzen:2016seh}
\begin{barticle}
\bauthor{\bsnm{Halzen}, \binits{F.}},
\bauthor{\bsnm{Kheirandish}, \binits{A.}},
\bauthor{\bsnm{Niro}, \binits{V.}}:
\bjtitle{Astropart. Phys.}
\bvolume{86},
\bfpage{46}
(\byear{2017}).
\arxivurl{1609.03072}.
doi:\doiurl{10.1016/j.astropartphys.2016.11.004}
\end{barticle}
\endbibitem

\bibitem[\protect\citeauthoryear{Heinz}{2005}]{Heinz:2005jc}
\begin{barticle}
\bauthor{\bsnm{Heinz}, \binits{S.}}:
\bjtitle{Astrophys. J.}
\bvolume{636},
\bfpage{316}
(\byear{2005}).
\arxivurl{astro-ph/0509777}.
doi:\doiurl{10.1086/497954}
\end{barticle}
\endbibitem

\bibitem[\protect\citeauthoryear{{Hillas}}{1984}]{1984ARA&A..22..425H}
\begin{barticle}
\bauthor{\bsnm{{Hillas}}, \binits{A.M.}}:
\bjtitle{\araa}
\bvolume{22},
\bfpage{425}
(\byear{1984}).
doi:\doiurl{10.1146/annurev.aa.22.090184.002233}
\end{barticle}
\endbibitem

\bibitem[\protect\citeauthoryear{Ingelman and Thunman}{1996}]{Ingelman:1996md}
\begin{botherref}
\oauthor{\bsnm{Ingelman}, \binits{G.}},
\oauthor{\bsnm{Thunman}, \binits{M.}}:
{Particle production in the interstellar medium}
(1996).
\arxivurl{hep-ph/9604286}
\end{botherref}
\endbibitem

\bibitem[\protect\citeauthoryear{Jogler}{2016}]{Jogler:2015ddc}
\begin{barticle}
\bauthor{\bsnm{Jogler}, \binits{T.}}:
\bjtitle{PoS}
\bvolume{ICRC2015},
\bfpage{888}
(\byear{2016}).
doi:\doiurl{10.22323/1.236.0888}
\end{barticle}
\endbibitem

\bibitem[\protect\citeauthoryear{Joshi et~al.}{2014}]{Joshi:2013aua}
\begin{barticle}
\bauthor{\bsnm{Joshi}, \binits{J.C.}},
\bauthor{\bsnm{Winter}, \binits{W.}},
\bauthor{\bsnm{Gupta}, \binits{N.}}:
\bjtitle{Mon. Not. Roy. Astron. Soc.}
\bvolume{439}(\bissue{4}),
\bfpage{3414}
(\byear{2014}).
\bcomment{[Erratum: Mon. Not. Roy. Astron. Soc.446,no.1,892(2014)]}.
\arxivurl{1310.5123}.
doi:\doiurl{10.1093/mnras/stu189, 10.1093/mnras/stu2132}
\end{barticle}
\endbibitem

\bibitem[\protect\citeauthoryear{Kachelrieß and
  Ostapchenko}{2014}]{Kachelriess:2014oma}
\begin{barticle}
\bauthor{\bsnm{Kachelrieß}, \binits{M.}},
\bauthor{\bsnm{Ostapchenko}, \binits{S.}}:
\bjtitle{Phys. Rev.}
\bvolume{D90}(\bissue{8}),
\bfpage{083002}
(\byear{2014}).
\arxivurl{1405.3797}.
doi:\doiurl{10.1103/PhysRevD.90.083002}
\end{barticle}
\endbibitem

\bibitem[\protect\citeauthoryear{Kappes et~al.}{2007}]{Kappes:2006fg}
\begin{barticle}
\bauthor{\bsnm{Kappes}, \binits{A.}},
\bauthor{\bsnm{Hinton}, \binits{J.}},
\bauthor{\bsnm{Stegmann}, \binits{C.}},
\bauthor{\bsnm{Aharonian}, \binits{F.A.}}:
\bjtitle{Astrophys. J.}
\bvolume{656},
\bfpage{870}
(\byear{2007}).
\bcomment{[Erratum: Astrophys. J.661,1348(2007)]}.
\arxivurl{astro-ph/0607286}.
doi:\doiurl{10.1086/508936, 10.1086/518161}
\end{barticle}
\endbibitem

\bibitem[\protect\citeauthoryear{Kargaltsev and
  Pavlov}{2010}]{Kargaltsev:2010jy}
\begin{barticle}
\bauthor{\bsnm{Kargaltsev}, \binits{O.}},
\bauthor{\bsnm{Pavlov}, \binits{G.}}:
\bjtitle{AIP Conf. Proc.}
\bvolume{1248},
\bfpage{25}
(\byear{2010}).
\arxivurl{1002.0885}.
doi:\doiurl{10.1063/1.3475228}
\end{barticle}
\endbibitem

\bibitem[\protect\citeauthoryear{Kelner et~al.}{2006}]{Kelner:2006tc}
\begin{barticle}
\bauthor{\bsnm{Kelner}, \binits{S.R.}},
\bauthor{\bsnm{Aharonian}, \binits{F.A.}},
\bauthor{\bsnm{Bugayov}, \binits{V.V.}}:
\bjtitle{Phys. Rev.}
\bvolume{D74},
\bfpage{034018}
(\byear{2006}).
\bcomment{[Erratum: Phys. Rev.D79,039901(2009)]}.
\arxivurl{astro-ph/0606058}.
doi:\doiurl{10.1103/PhysRevD.74.034018, 10.1103/PhysRevD.79.039901}
\end{barticle}
\endbibitem

\bibitem[\protect\citeauthoryear{Kheirandish and
  Wood}{2020}]{Kheirandish:2019bke}
\begin{barticle}
\bauthor{\bsnm{Kheirandish}, \binits{A.}},
\bauthor{\bsnm{Wood}, \binits{J.}}:
\bjtitle{PoS}
\bvolume{ICRC2019},
\bfpage{932}
(\byear{2020}).
\arxivurl{1908.08546}.
doi:\doiurl{10.22323/1.358.0932}
\end{barticle}
\endbibitem

\bibitem[\protect\citeauthoryear{{Kiminki} et~al.}{2015}]{2015ApJ...811...85K}
\begin{barticle}
\bauthor{\bsnm{{Kiminki}}, \binits{D.C.}},
\bauthor{\bsnm{{Kobulnicky}}, \binits{H.A.}},
\bauthor{\bsnm{{Vargas {\'A}lvarez}}, \binits{C.A.}},
\bauthor{\bsnm{{Alexander}}, \binits{M.J.}},
\bauthor{\bsnm{{Lundquist}}, \binits{M.J.}}:
\bjtitle{\apj}
\bvolume{811}(\bissue{2}),
\bfpage{85}
(\byear{2015}).
\arxivurl{1508.03108}.
doi:\doiurl{10.1088/0004-637X/811/2/85}
\end{barticle}
\endbibitem

\bibitem[\protect\citeauthoryear{{Kopper} and
  {Blaufuss}}{2017}]{2017GCN.21916....1K}
\begin{botherref}
\oauthor{\bsnm{{Kopper}}, \binits{C.}},
\oauthor{\bsnm{{Blaufuss}}, \binits{E.}}:
GRB Coordinates Network, Circular Service, No.~21916, \#1 (2017)
\textbf{21916}
(2017)
\end{botherref}
\endbibitem

\bibitem[\protect\citeauthoryear{Kopper et~al.}{2016}]{Kopper:2015vzf}
\begin{barticle}
\bauthor{\bsnm{Kopper}, \binits{C.}},
\bauthor{\bsnm{Giang}, \binits{W.}},
\bauthor{\bsnm{Kurahashi}, \binits{N.}}:
\bjtitle{PoS}
\bvolume{ICRC2015},
\bfpage{1081}
(\byear{2016})
\end{barticle}
\endbibitem

\bibitem[\protect\citeauthoryear{Koyama et~al.}{1996}]{Koyama:1996sj}
\begin{barticle}
\bauthor{\bsnm{Koyama}, \binits{K.}},
\bauthor{\bsnm{Maeda}, \binits{Y.}},
\bauthor{\bsnm{Sonobe}, \binits{T.}},
\bauthor{\bsnm{Takeshima}, \binits{T.}},
\bauthor{\bsnm{Tanaka}, \binits{Y.}},
\bauthor{\bsnm{Yamauchi}, \binits{S.}}:
\bjtitle{Publ.\ Astron.\ Soc.\ Jap.}
\bvolume{48},
\bfpage{249}
(\byear{1996})
\end{barticle}
\endbibitem

\bibitem[\protect\citeauthoryear{Lemoine et~al.}{2015}]{Lemoine:2014ala}
\begin{barticle}
\bauthor{\bsnm{Lemoine}, \binits{M.}},
\bauthor{\bsnm{Kotera}, \binits{K.}},
\bauthor{\bsnm{Pétri}, \binits{J.}}:
\bjtitle{JCAP}
\bvolume{1507},
\bfpage{016}
(\byear{2015}).
\arxivurl{1409.0159}.
doi:\doiurl{10.1088/1475-7516/2015/07/016}
\end{barticle}
\endbibitem

\bibitem[\protect\citeauthoryear{{Leung} and
  {Thaddeus}}{1992}]{1992ApJS...81..267L}
\begin{barticle}
\bauthor{\bsnm{{Leung}}, \binits{H.O.}},
\bauthor{\bsnm{{Thaddeus}}, \binits{P.}}:
\bjtitle{\apjs}
\bvolume{81},
\bfpage{267}
(\byear{1992}).
doi:\doiurl{10.1086/191693}
\end{barticle}
\endbibitem

\bibitem[\protect\citeauthoryear{Levinson and Waxman}{2001}]{Levinson:2001as}
\begin{barticle}
\bauthor{\bsnm{Levinson}, \binits{A.}},
\bauthor{\bsnm{Waxman}, \binits{E.}}:
\bjtitle{Phys. Rev. Lett.}
\bvolume{87},
\bfpage{171101}
(\byear{2001}).
\arxivurl{hep-ph/0106102}.
doi:\doiurl{10.1103/PhysRevLett.87.171101}
\end{barticle}
\endbibitem

\bibitem[\protect\citeauthoryear{Lipari and Vernetto}{2018}]{Lipari:2018gzn}
\begin{barticle}
\bauthor{\bsnm{Lipari}, \binits{P.}},
\bauthor{\bsnm{Vernetto}, \binits{S.}}:
\bjtitle{Phys. Rev.}
\bvolume{D98}(\bissue{4}),
\bfpage{043003}
(\byear{2018}).
\arxivurl{1804.10116}.
doi:\doiurl{10.1103/PhysRevD.98.043003}
\end{barticle}
\endbibitem

\bibitem[\protect\citeauthoryear{Liu and Kheirandish}{2020}]{Liu:2019iga}
\begin{barticle}
\bauthor{\bsnm{Liu}, \binits{Q.}},
\bauthor{\bsnm{Kheirandish}, \binits{A.}}:
\bjtitle{PoS}
\bvolume{ICRC2019},
\bfpage{944}
(\byear{2020}).
\arxivurl{1908.05279}.
doi:\doiurl{10.22323/1.358.0944}
\end{barticle}
\endbibitem

\bibitem[\protect\citeauthoryear{Lunardini et~al.}{2014}]{Lunardini:2013gva}
\begin{barticle}
\bauthor{\bsnm{Lunardini}, \binits{C.}},
\bauthor{\bsnm{Razzaque}, \binits{S.}},
\bauthor{\bsnm{Theodoseau}, \binits{K.T.}},
\bauthor{\bsnm{Yang}, \binits{L.}}:
\bjtitle{Phys.\ Rev.\ D}
\bvolume{90}(\bissue{2}),
\bfpage{023016}
(\byear{2014}).
\arxivurl{1311.7188}.
doi:\doiurl{10.1103/PhysRevD.90.023016}
\end{barticle}
\endbibitem

\bibitem[\protect\citeauthoryear{Manchester et~al.}{2005}]{Manchester:2004bp}
\begin{barticle}
\bauthor{\bsnm{Manchester}, \binits{R.N.}},
\bauthor{\bsnm{Hobbs}, \binits{G.B.}},
\bauthor{\bsnm{Teoh}, \binits{A.}},
\bauthor{\bsnm{Hobbs}, \binits{M.}}:
\bjtitle{Astron. J.}
\bvolume{129},
\bfpage{1993}
(\byear{2005}).
\arxivurl{astro-ph/0412641}.
doi:\doiurl{10.1086/428488}
\end{barticle}
\endbibitem

\bibitem[\protect\citeauthoryear{Mancina and Silva}{2020}]{Mancina:2019hsp}
\begin{barticle}
\bauthor{\bsnm{Mancina}, \binits{S.}},
\bauthor{\bsnm{Silva}, \binits{M.}}:
\bjtitle{PoS}
\bvolume{ICRC2019},
\bfpage{954}
(\byear{2020}).
\arxivurl{1908.04869}.
doi:\doiurl{10.22323/1.358.0954}
\end{barticle}
\endbibitem

\bibitem[\protect\citeauthoryear{Murase and Beacom}{2012}]{Murase:2012xs}
\begin{barticle}
\bauthor{\bsnm{Murase}, \binits{K.}},
\bauthor{\bsnm{Beacom}, \binits{J.F.}}:
\bjtitle{JCAP}
\bvolume{1210},
\bfpage{043}
(\byear{2012}).
\arxivurl{1206.2595}.
doi:\doiurl{10.1088/1475-7516/2012/10/043}
\end{barticle}
\endbibitem

\bibitem[\protect\citeauthoryear{Murase et~al.}{2013}]{Murase2013b}
\begin{barticle}
\bauthor{\bsnm{Murase}, \binits{K.}},
\bauthor{\bsnm{Ahlers}, \binits{M.}},
\bauthor{\bsnm{Lacki}, \binits{B.C.}}:
\bjtitle{Phys.Rev.}
\bvolume{D88}(\bissue{12}),
\bfpage{121301}
(\byear{2013}).
\arxivurl{1306.3417}.
doi:\doiurl{10.1103/PhysRevD.88.121301}
\end{barticle}
\endbibitem

\bibitem[\protect\citeauthoryear{Murase et~al.}{2016}]{Murase:2015xka}
\begin{barticle}
\bauthor{\bsnm{Murase}, \binits{K.}},
\bauthor{\bsnm{Guetta}, \binits{D.}},
\bauthor{\bsnm{Ahlers}, \binits{M.}}:
\bjtitle{Phys. Rev. Lett.}
\bvolume{116}(\bissue{7}),
\bfpage{071101}
(\byear{2016}).
\arxivurl{1509.00805}.
doi:\doiurl{10.1103/PhysRevLett.116.071101}
\end{barticle}
\endbibitem

\bibitem[\protect\citeauthoryear{Murase et~al.}{2015}]{Murase:2015gea}
\begin{barticle}
\bauthor{\bsnm{Murase}, \binits{K.}},
\bauthor{\bsnm{Laha}, \binits{R.}},
\bauthor{\bsnm{Ando}, \binits{S.}},
\bauthor{\bsnm{Ahlers}, \binits{M.}}:
\bjtitle{Phys.\ Rev.\ Lett.}
\bvolume{115}(\bissue{7}),
\bfpage{071301}
(\byear{2015}).
\arxivurl{1503.04663}.
doi:\doiurl{10.1103/PhysRevLett.115.071301}
\end{barticle}
\endbibitem

\bibitem[\protect\citeauthoryear{Nierstenhoefer
  et~al.}{2015}]{Nierstenhoefer:2015gta}
\begin{barticle}
\bauthor{\bsnm{Nierstenhoefer}, \binits{N.}},
\bauthor{\bsnm{Graeser}, \binits{P.}},
\bauthor{\bsnm{Schuppan}, \binits{F.}},
\bauthor{\bsnm{Becker~Tjus}, \binits{J.}}:
\bjtitle{J. Phys. Conf. Ser.}
\bvolume{632}(\bissue{1}),
\bfpage{012019}
(\byear{2015}).
\arxivurl{1501.06434}.
doi:\doiurl{10.1088/1742-6596/632/1/012019}
\end{barticle}
\endbibitem

\bibitem[\protect\citeauthoryear{Niro et~al.}{2019}]{Niro:2019mzw}
\begin{botherref}
\oauthor{\bsnm{Niro}, \binits{V.}},
\oauthor{\bsnm{Neronov}, \binits{A.}},
\oauthor{\bsnm{Fusco}, \binits{L.}},
\oauthor{\bsnm{Gabici}, \binits{S.}},
\oauthor{\bsnm{Semikoz}, \binits{D.}}:
{Neutrinos from the gamma-ray source eHWC J1825-134: predictions for Km$^3$
  detectors}
(2019).
\arxivurl{1910.09065}
\end{botherref}
\endbibitem

\bibitem[\protect\citeauthoryear{Palma et~al.}{2017}]{DiPalma:2016yfy}
\begin{barticle}
\bauthor{\bsnm{Palma}, \binits{I.D.}},
\bauthor{\bsnm{Guetta}, \binits{D.}},
\bauthor{\bsnm{Amato}, \binits{E.}}:
\bjtitle{Astrophys. J.}
\bvolume{836}(\bissue{2}),
\bfpage{159}
(\byear{2017}).
\arxivurl{1605.01205}.
doi:\doiurl{10.3847/1538-4357/836/2/159}
\end{barticle}
\endbibitem

\bibitem[\protect\citeauthoryear{Romero et~al.}{2005}]{Romero:2005fr}
\begin{barticle}
\bauthor{\bsnm{Romero}, \binits{G.E.}},
\bauthor{\bsnm{Christiansen}, \binits{H.R.}},
\bauthor{\bsnm{Orellana}, \binits{M.}}:
\bjtitle{Astrophys. J.}
\bvolume{632},
\bfpage{1093}
(\byear{2005}).
\arxivurl{astro-ph/0506735}.
doi:\doiurl{10.1086/444446}
\end{barticle}
\endbibitem

\bibitem[\protect\citeauthoryear{Romero et~al.}{2003}]{Romero:2003td}
\begin{barticle}
\bauthor{\bsnm{Romero}, \binits{G.E.}},
\bauthor{\bsnm{Torres}, \binits{D.F.}},
\bauthor{\bsnm{Bernado}, \binits{M.M.K.}},
\bauthor{\bsnm{Mirabel}, \binits{I.F.}}:
\bjtitle{Astron. Astrophys.}
\bvolume{410},
\bfpage{1}
(\byear{2003}).
\arxivurl{astro-ph/0309123}.
doi:\doiurl{10.1051/0004-6361:20031314-1}
\end{barticle}
\endbibitem

\bibitem[\protect\citeauthoryear{Ryu et~al.}{2013}]{Ryu:2012ib}
\begin{barticle}
\bauthor{\bsnm{Ryu}, \binits{S.G.}},
\bauthor{\bsnm{Nobukawa}, \binits{M.}},
\bauthor{\bsnm{Nakashima}, \binits{S.}},
\bauthor{\bsnm{Tsuru}, \binits{T.G.}},
\bauthor{\bsnm{Koyama}, \binits{K.}},
\bauthor{\bsnm{Uchiyama}, \binits{H.}}:
\bjtitle{Publ.\ Astron.\ Soc.\ Jap.}
\bvolume{65},
\bfpage{33}
(\byear{2013}).
\arxivurl{1211.4529}.
doi:\doiurl{10.1093/pasj/65.2.33}
\end{barticle}
\endbibitem

\bibitem[\protect\citeauthoryear{Schneider}{2020}]{Schneider:2019ayi}
\begin{barticle}
\bauthor{\bsnm{Schneider}, \binits{A.}}:
\bjtitle{PoS}
\bvolume{ICRC2019},
\bfpage{1004}
(\byear{2020}).
\arxivurl{1907.11266}.
doi:\doiurl{10.22323/1.358.1004}
\end{barticle}
\endbibitem

\bibitem[\protect\citeauthoryear{{Schneider}
  et~al.}{2006}]{2006A&A...458..855S}
\begin{barticle}
\bauthor{\bsnm{{Schneider}}, \binits{N.}},
\bauthor{\bsnm{{Bontemps}}, \binits{S.}},
\bauthor{\bsnm{{Simon}}, \binits{R.}},
\bauthor{\bsnm{{Jakob}}, \binits{H.}},
\bauthor{\bsnm{{Motte}}, \binits{F.}},
\bauthor{\bsnm{{Miller}}, \binits{M.}},
\bauthor{\bsnm{{Kramer}}, \binits{C.}},
\bauthor{\bsnm{{Stutzki}}, \binits{J.}}:
\bjtitle{\aap}
\bvolume{458}(\bissue{3}),
\bfpage{855}
(\byear{2006}).
doi:\doiurl{10.1051/0004-6361:20065088}
\end{barticle}
\endbibitem

\bibitem[\protect\citeauthoryear{{Schneider}
  et~al.}{2016}]{2016A&A...591A..40S}
\begin{barticle}
\bauthor{\bsnm{{Schneider}}, \binits{N.}},
\bauthor{\bsnm{{Bontemps}}, \binits{S.}},
\bauthor{\bsnm{{Motte}}, \binits{F.}},
\bauthor{\bsnm{{Blazere}}, \binits{A.}},
\bauthor{\bsnm{{Andr{\'e}}}, \binits{P.}},
\bauthor{\bsnm{{Anderson}}, \binits{L.D.}},
\bauthor{\bsnm{{Arzoumanian}}, \binits{D.}},
\bauthor{\bsnm{{Comer{\'o}n}}, \binits{F.}},
\bauthor{\bsnm{{Didelon}}, \binits{P.}},
\bauthor{\bsnm{{Di Francesco}}, \binits{J.}},
\bauthor{\bsnm{{Duarte-Cabral}}, \binits{A.}},
\bauthor{\bsnm{{Guarcello}}, \binits{M.G.}},
\bauthor{\bsnm{{Hennemann}}, \binits{M.}},
\bauthor{\bsnm{{Hill}}, \binits{T.}},
\bauthor{\bsnm{{K{\"o}nyves}}, \binits{V.}},
\bauthor{\bsnm{{Marston}}, \binits{A.}},
\bauthor{\bsnm{{Minier}}, \binits{V.}},
\bauthor{\bsnm{{Rygl}}, \binits{K.L.J.}},
\bauthor{\bsnm{{R{\"o}llig}}, \binits{M.}},
\bauthor{\bsnm{{Roy}}, \binits{A.}},
\bauthor{\bsnm{{Spinoglio}}, \binits{L.}},
\bauthor{\bsnm{{Tremblin}}, \binits{P.}},
\bauthor{\bsnm{{White}}, \binits{G.J.}},
\bauthor{\bsnm{{Wright}}, \binits{N.J.}}:
\bjtitle{\aap}
\bvolume{591},
\bfpage{40}
(\byear{2016}).
\arxivurl{1604.03967}.
doi:\doiurl{10.1051/0004-6361/201628328}
\end{barticle}
\endbibitem

\bibitem[\protect\citeauthoryear{Silva and Mancina}{2020}]{Silva:2019fnq}
\begin{barticle}
\bauthor{\bsnm{Silva}, \binits{M.}},
\bauthor{\bsnm{Mancina}, \binits{S.}}:
\bjtitle{PoS}
\bvolume{ICRC2019},
\bfpage{1010}
(\byear{2020}).
\arxivurl{1908.06586}.
doi:\doiurl{10.22323/1.358.1010}
\end{barticle}
\endbibitem

\bibitem[\protect\citeauthoryear{Smith}{2010}]{Smith:2010yn}
\begin{botherref}
\oauthor{\bsnm{Smith}, \binits{A.J.}}:
{A Survey of Fermi Catalog Sources using Data from the Milagro Gamma-Ray
  Observatory }
(2010).
\arxivurl{1001.3695}
\end{botherref}
\endbibitem

\bibitem[\protect\citeauthoryear{Spengler}{2020}]{Spengler:2019sde}
\begin{barticle}
\bauthor{\bsnm{Spengler}, \binits{G.}}:
\bjtitle{Astron. Astrophys.}
\bvolume{633},
\bfpage{138}
(\byear{2020}).
\arxivurl{1912.05221}.
doi:\doiurl{10.1051/0004-6361/201936632}
\end{barticle}
\endbibitem

\bibitem[\protect\citeauthoryear{Spiering}{2012}]{Spiering:2012xe}
\begin{barticle}
\bauthor{\bsnm{Spiering}, \binits{C.}}:
\bjtitle{Eur. Phys. J.}
\bvolume{H37},
\bfpage{515}
(\byear{2012}).
\arxivurl{1207.4952}.
doi:\doiurl{10.1140/epjh/e2012-30014-2}
\end{barticle}
\endbibitem

\bibitem[\protect\citeauthoryear{Stecker}{1979}]{Stecker:1978ah}
\begin{barticle}
\bauthor{\bsnm{Stecker}, \binits{F.W.}}:
\bjtitle{Astrophys.\ J.}
\bvolume{228},
\bfpage{919}
(\byear{1979}).
doi:\doiurl{10.1086/156919}
\end{barticle}
\endbibitem

\bibitem[\protect\citeauthoryear{Stettner}{2020}]{Stettner:2019tok}
\begin{barticle}
\bauthor{\bsnm{Stettner}, \binits{J.}}:
\bjtitle{PoS}
\bvolume{ICRC2019},
\bfpage{1017}
(\byear{2020}).
\arxivurl{1908.09551}.
doi:\doiurl{10.22323/1.358.1017}
\end{barticle}
\endbibitem

\bibitem[\protect\citeauthoryear{Tanabashi et~al.}{2018}]{Tanabashi:2018oca}
\begin{barticle}
\bauthor{\bsnm{Tanabashi}, \binits{M.}}, \betal:
\bjtitle{Phys. Rev.}
\bvolume{D98}(\bissue{3}),
\bfpage{030001}
(\byear{2018}).
doi:\doiurl{10.1103/PhysRevD.98.030001}
\end{barticle}
\endbibitem

\bibitem[\protect\citeauthoryear{Tchernin et~al.}{2013}]{Tchernin:2013wfa}
\begin{barticle}
\bauthor{\bsnm{Tchernin}, \binits{C.}},
\bauthor{\bsnm{Aguilar}, \binits{J.A.}},
\bauthor{\bsnm{Neronov}, \binits{A.}},
\bauthor{\bsnm{Montaruli}, \binits{T.}}:
\bjtitle{Astron. Astrophys.}
\bvolume{560},
\bfpage{67}
(\byear{2013}).
\arxivurl{1305.4113}.
doi:\doiurl{10.1051/0004-6361/201321801}
\end{barticle}
\endbibitem

\bibitem[\protect\citeauthoryear{Torres and Halzen}{2007}]{Torres:2006ub}
\begin{barticle}
\bauthor{\bsnm{Torres}, \binits{D.F.}},
\bauthor{\bsnm{Halzen}, \binits{F.}}:
\bjtitle{Astropart. Phys.}
\bvolume{27},
\bfpage{500}
(\byear{2007}).
\arxivurl{astro-ph/0607368}.
doi:\doiurl{10.1016/j.astropartphys.2007.02.004}
\end{barticle}
\endbibitem

\bibitem[\protect\citeauthoryear{{Vladimirov}
  et~al.}{2011}]{2011CoPhC.182.1156V}
\begin{barticle}
\bauthor{\bsnm{{Vladimirov}}, \binits{A.E.}},
\bauthor{\bsnm{{Digel}}, \binits{S.W.}},
\bauthor{\bsnm{{J{\'o}hannesson}}, \binits{G.}},
\bauthor{\bsnm{{Michelson}}, \binits{P.F.}},
\bauthor{\bsnm{{Moskalenko}}, \binits{I.V.}},
\bauthor{\bsnm{{Nolan}}, \binits{P.L.}},
\bauthor{\bsnm{{Orland o}}, \binits{E.}},
\bauthor{\bsnm{{Porter}}, \binits{T.A.}},
\bauthor{\bsnm{{Strong}}, \binits{A.W.}}:
\bjtitle{Computer Physics Communications}
\bvolume{182}(\bissue{5}),
\bfpage{1156}
(\byear{2011}).
\arxivurl{1008.3642}.
doi:\doiurl{10.1016/j.cpc.2011.01.017}
\end{barticle}
\endbibitem

\bibitem[\protect\citeauthoryear{Wakely and Horan}{2007}]{Wakely:2007qpa}
\begin{bchapter}
\bauthor{\bsnm{Wakely}, \binits{S.P.}},
\bauthor{\bsnm{Horan}, \binits{D.}}:
In: \bbtitle{{Proceedings, 30th International Cosmic Ray Conference (ICRC
  2007): Merida, Yucatan, Mexico, July 3-11, 2007}},
\byear{2007}
\end{bchapter}
\endbibitem

\bibitem[\protect\citeauthoryear{{Wright} et~al.}{2015}]{2015MNRAS.449..741W}
\begin{barticle}
\bauthor{\bsnm{{Wright}}, \binits{N.J.}},
\bauthor{\bsnm{{Drew}}, \binits{J.E.}},
\bauthor{\bsnm{{Mohr-Smith}}, \binits{M.}}:
\bjtitle{\mnras}
\bvolume{449}(\bissue{1}),
\bfpage{741}
(\byear{2015}).
\arxivurl{1502.05718}.
doi:\doiurl{10.1093/mnras/stv323}
\end{barticle}
\endbibitem

\bibitem[\protect\citeauthoryear{Yoast-Hull et~al.}{2013}]{Yoast-Hull:2013wwa}
\begin{barticle}
\bauthor{\bsnm{Yoast-Hull}, \binits{T.M.}},
\bauthor{\bsnm{Everett}, \binits{J.E.}},
\bauthor{\bsnm{Gallagher}, \binits{J.S.}},
\bauthor{\bsnm{Zweibel}, \binits{E.G.}}:
\bjtitle{Astrophys. J.}
\bvolume{768},
\bfpage{53}
(\byear{2013}).
\arxivurl{1303.4305}.
doi:\doiurl{10.1088/0004-637X/768/1/53}
\end{barticle}
\endbibitem

\bibitem[\protect\citeauthoryear{Yoast-Hull et~al.}{2017}]{Yoast-Hull:2017gaj}
\begin{barticle}
\bauthor{\bsnm{Yoast-Hull}, \binits{T.M.}},
\bauthor{\bsnm{Gallagher}, \binits{J.S.}},
\bauthor{\bsnm{Halzen}, \binits{F.}},
\bauthor{\bsnm{Kheirandish}, \binits{A.}},
\bauthor{\bsnm{Zweibel}, \binits{E.G.}}:
\bjtitle{Phys. Rev.}
\bvolume{D96}(\bissue{4}),
\bfpage{043011}
(\byear{2017}).
\arxivurl{1703.02590}.
doi:\doiurl{10.1103/PhysRevD.96.043011}
\end{barticle}
\endbibitem

\bibitem[\protect\citeauthoryear{Zanin and Holder}{2018}]{Zanin:2017rao}
\begin{barticle}
\bauthor{\bsnm{Zanin}, \binits{R.}},
\bauthor{\bsnm{Holder}, \binits{J.}}:
\bjtitle{PoS}
\bvolume{ICRC2017},
\bfpage{740}
(\byear{2018}).
\bcomment{[35,740(2017)]}.
\arxivurl{1709.04354}.
doi:\doiurl{10.22323/1.301.0740}
\end{barticle}
\endbibitem

\end{thebibliography}
\end{document}